\documentclass[a4paper,11pt]{article}
\usepackage[utf8]{inputenc}
\usepackage{jcappub} 
\usepackage{lineno}
\usepackage{multirow}
\usepackage{indentfirst}

\title{Topological Defects in Cosmic Shear}

\author[a]{Jiachen Bai}
\author[a]{Benjamin Giblin}
\author[a]{Andy Taylor}
\affiliation[a]{Institute for Astronomy, University of Edinburgh,\\
Royal Observatory, Blackford Hill, Edinburgh, EH9 3HJ, U.K.}
\emailAdd{J.Bai@ed.ac.uk}

\abstract{
  In this work, we present a novel approach for extracting cosmological information from cosmic shear maps based on topological defects in the shear orientation field. 
These defects are defined purely by the phase structure of the shear field and are therefore complementary to conventional weak lensing statistics that primarily probe amplitude information.
We characterise their statistical properties, including number densities, spatial clustering, and the convergence distribution at defect locations, and demonstrate with Fisher analyses under idealised and realistic survey conditions that topological defect statistics provide a valuable and complementary source of cosmological information.
With Euclid-like simulated data, the inclusion of defect statistics leads to constraints on $\Omega_{\rm m}$ and $\sigma_8$ that are up to 3.9 times tighter than those obtained from the convergence two-point correlation function alone.
}

\begin{document}
\maketitle
\flushbottom

\section{Introduction}

Cosmology has entered an era of precision measurements, especially in the field of large-scale structure (LSS) of the universe.
Stage-III LSS surveys have made significant progress in measuring the matter distribution of the Universe, such as the Dark Energy Survey (DES) \cite{2022PhRvD.105b3520A}, the Kilo-Degree Survey (KiDS) \cite{2025A&A...703A.158W}, and the Hyper Suprime-Cam Survey (HSC) \cite{2023PhRvD.108l3519D}.
The primary observable in these surveys is weak gravitational lensing. The underlying matter distribution can be inferred from the coherent distortions of galaxy images induced by gravitational lensing, a phenomenon known as cosmic shear \cite{2001PhR...340..291B}.

Stage-IV surveys, such as Euclid \cite{2025A&A...697A...1E} and the Vera C. Rubin Legacy Survey of Space and Time (LSST) \cite{2019ApJ...873..111I}, are expected to provide massive amounts of data by improving the depth of the observations, and increasing the survey area.
This data revolution provides an unprecedented opportunity to test models of cosmology and gravity, which have recently been challenged by the enigmatic possibility of dynamical dark energy \cite{2025JCAP...02..021A, 2025PhRvD.112h3515A} and the persistent Hubble tension \cite{2021CQGra..38o3001D}.

However, the improvement in the precision of measurements also demands commensurate advances in our data-analysis methodologies.
The most common approach to extracting information from cosmological data relies on two-point correlation functions (2PCFs) and power spectra, which provide a powerful statistical characterisation of such fields.
In the context of weak lensing, the 2PCFs of cosmic shear has been widely used to extract information from the observed lensing field, and it has been used to place $10\%$ constraints on $\Omega_{\rm m}$ and $\sigma_8$ \cite{2022PhRvD.105b3520A, 2025A&A...703A.158W, 2023PhRvD.108l3519D}.

Although the 2PCF is a powerful tool, it only captures the second-order Gaussian information of the field, while for non-Gaussian fields, there is non-negligible information encoded in higher-order correlation functions.
Considering the observed cosmic shear maps are highly non-Gaussian on small scales, it is necessary to extract higher-order information from the data, which means we need to develop methods beyond the traditional 2PCFs.

In recent years, a variety of higher-order statistics have been proposed, such as the convergence probability distribution function ($\kappa$-PDF) \cite{2023MNRAS.520.1721G, 2023A&A...675A.120E}, the Minkowski functionals \cite{2012PhRvD..85j3513K, 2023A&A...675A.120E}, and the peak counts \cite{2017A&A...599A..79P, 2023A&A...675A.120E}.
These statistics have been proven to provide higher precision constraints on cosmological parameters than the 2PCF in parameter inference from real observational data \cite{2021MNRAS.506.1623H, 2024MNRAS.534.3305H, 2025PhRvD.111h3510N}.

Currently, most statistics used to extract information from cosmic shear focus on either the convergence field or the shear amplitude, while the shear orientation has not yet been thoroughly investigated.
In this work, we focus on the shear orientation field and identify a previously less explored feature that may encode additional information, namely topological defects \cite{2023EPJST.232.1813F, 2023arXiv231102449C, 2009JCAP...09..034V}. 

Topological defects are singular points in the cosmic shear field where the curl of the shear orientation gradient is nonzero and quantised. At these locations, the shear amplitude drops to zero, rendering the shear orientation undefined and the image distortion locally isotropic \cite{2009JCAP...09..034V, 2003JPhA...36.6611D}.
These defects are also called phase singularities or umbilic points \cite{MVBerry_1977, 2003JPhA...36.6611D}, and have been extensively studied in mathematics and physics.
For example, they have been applied to random vector waves to understand their polarisation \cite{2001RSPSA.457..141B, 2004NJPh....6..162B}, and also appear in quantum field theory as non-trivial field configurations \cite{2023EPJST.232.1813F,  2023arXiv231102449C}.

In cosmology, topological defects continue to be an under-used metric, having been measured from polarisation of the cosmic microwave background (CMB) \cite{2005PhRvD..72d3004H}, but never having been used to constrain cosmological parameters. In the context of weak lensing, topological defects were only investigated preliminarily in a coarse way without looking into their properties and cosmological dependencies  \cite{2009JCAP...09..034V}.
In this work, we therefore investigate topological defects in cosmic shear maps in detail, establishing them as a viable statistical tool for weak lensing observations.
We characterise the statistical properties of these defects, including their number density, spatial distribution, and the convergence distribution at their locations, and investigate the sensitivity to the underlying cosmology. We demonstrate, therefore, the application of cosmic shear defects as a novel probe for constraining cosmological parameters.

The structure of this paper is as follows. In Sec.~\ref{sec:theory}, we introduce the theoretical background of topological defects in cosmic shear maps, describe our defect-finding algorithm, and validate its performance. In Sec.~\ref{sec:sims}, we present the weak lensing simulations employed in this study, encompassing various datasets. Sec.~\ref{sec:stats} is devoted to the statistical characterisation of defects. In Sec.~\ref{sec:fisher}, we perform a Fisher analysis to forecast the constraining power of defect statistics on cosmology, and in Sec.~\ref{sec:results}, we present the results of this analysis. Finally, in Sec.~\ref{sec:conclusion}, we summarise our findings and discuss potential directions for future research.

\section{Theory and Methods}\label{sec:theory}

\subsection{Topological Defects in Cosmic Shear Maps}\label{sec:defect_theory}

To establish the theoretical basis for our analysis, we briefly review the relevant theoretical background. 
Cosmic shear is a phenomenon where the images of distant galaxies are distorted due to the gravitational lensing effect caused by the large-scale structure of the universe. 
This causes the observed angular position of a galaxy, $\boldsymbol{\theta}$, to differ from the true value, $\boldsymbol{\beta}$, by the deflection angle, $\boldsymbol{\alpha}$, as given by the lens equation \cite{2001PhR...340..291B}
\begin{equation}
    \boldsymbol{\beta} = \boldsymbol{\theta} - \boldsymbol{\alpha}.
\end{equation}
The deflection angle can be expressed as
\begin{equation}
    \boldsymbol{\alpha} = \frac{2}{c^2} \int_0^{\chi_s} d\chi \frac{f_K(\chi_s - \chi)}{f_K(\chi_s) f_K(\chi)} \nabla_\perp \Phi(\chi, \boldsymbol{\theta}),
\end{equation}
where $\chi$ is the comoving distance, $\chi_s$ is the comoving distance to the source, $f_K(\chi)$ is the comoving angular diameter distance, and $\Phi$ is the gravitational potential.
Here we have applied the Born approximation, which assumes that light rays travel along unperturbed paths.
The approximation might include some bias in very high-precision measurements \cite{2024MNRAS.533.3209F}, but for simplicity we will not consider this effect in this work.
We can define a lens potential, $\psi$, as
\begin{equation}
    \psi(\boldsymbol{\theta}) = \frac{2}{c^2} \int_0^{\chi_s} d\chi \frac{f_K(\chi_s - \chi)}{f_K(\chi_s) f_K(\chi)} \Phi(\chi, \boldsymbol{\theta}).
\end{equation}
Thus the deflection angle can be written as the gradient of the lens potential,
\begin{equation}
    \boldsymbol{\alpha} = \nabla_\theta \psi(\boldsymbol{\theta}).
\end{equation}
This makes the deflection angle a conservative field.

The deformation of galaxy images by weak gravitational lensing is described by the Jacobian matrix $\mathbf{A} \equiv \frac{\partial \boldsymbol{\beta}}{\partial \boldsymbol{\theta}}$, which can be expressed in terms of the Hessian matrix of the lens potential as
\begin{equation}
\mathbf{A} = \delta_{ij} - \frac{\partial^2\psi}{\partial \theta_i \partial \theta_j} =
 \begin{pmatrix}1 - \kappa - \gamma_1 & -\gamma_2 \\
-\gamma_2 & 1 - \kappa + \gamma_1\end{pmatrix},
\end{equation}
where $\delta_{ij}$ is the Kronecker delta, and $\kappa$ and $\gamma_1, \gamma_2$ denote the isotropic and anisotropic components of the lensing-induced deformation, commonly referred to as the convergence and the two components of the cosmic shear, respectively.
We note that the convergence, $\kappa$ is the Laplacian of the lens potential. Using the Poisson equation, one can derive
\begin{equation}
    \kappa(\boldsymbol{\theta}) = \frac{4\pi G}{c^2} \int_0^{\chi_s} d\chi \frac{f_K(\chi) f_K(\chi_s - \chi)}{f_K(\chi_s)} a^2 \rho_m \delta_m(\chi, \boldsymbol{\theta}),
\end{equation}
where the factor $a$ is the scale factor evaluated at comoving distance $\chi$, $\rho_m$ is the mean matter density, and $\delta_m(\chi,\boldsymbol{\theta})$ is the matter density contrast at comoving distance $\chi$ and angular position $\boldsymbol{\theta}$. The integral is taken along the line of sight from the observer to the source plane.

The other quantity is cosmic shear, which can be represented as a complex field
\begin{equation}
\mathbf{\gamma}({\bf \theta}) = \gamma_1({\bf \theta}) + i\gamma_2({\bf \theta}) = |\mathbf{\gamma}({\bf \theta})| e^{2i\varphi({\bf \theta})},
\end{equation}
where $|\mathbf{\gamma}| = \sqrt{\gamma_1^2 + \gamma_2^2}$ is the shear amplitude, and $\varphi = \frac{1}{2}\tan^{-1}\left(\frac{\gamma_2}{\gamma_1}\right)$ is the shear orientation angle.
The factor of 2 in the exponent reflects the spin-2 nature of the shear field, meaning that the shear is invariant under a $180^\circ$ global rotation of the coordinate frame.

We note that the second equality is only well-defined when the shear amplitude is non-zero, as the shear orientation becomes undefined at points where the amplitude vanishes.
These points correspond to locations where the eigenvalues of the Hessian matrix of the lens potential are equal, leading to isotropic image distortion.
We refer to these points as topological defects, which are also called umbilic points or phase singularities in the literature \cite{MVBerry_1977, 2003JPhA...36.6611D}.

At these points, one can prove \cite{2003JPhA...36.6611D}
\begin{equation}
\nabla \times \nabla \varphi |_{\mathbf{\gamma}=0} = 2 \pi m,
\end{equation}
where $m$ is called the topological charge of the defect, which can take integer multiples of $1/2$, i.e. $m = \pm 1/2, \pm 1, \pm 3/2, \ldots$, for shear fields.
With Stokes' theorem, we can express the topological charge as a line integral around a closed loop enclosing the defect
\begin{equation}
m = \frac{1}{2\pi} \oint \nabla \varphi \cdot d\mathbf{l}.
\end{equation}

\begin{figure*}[htp]
    \centering
    \includegraphics[width=0.98\textwidth]{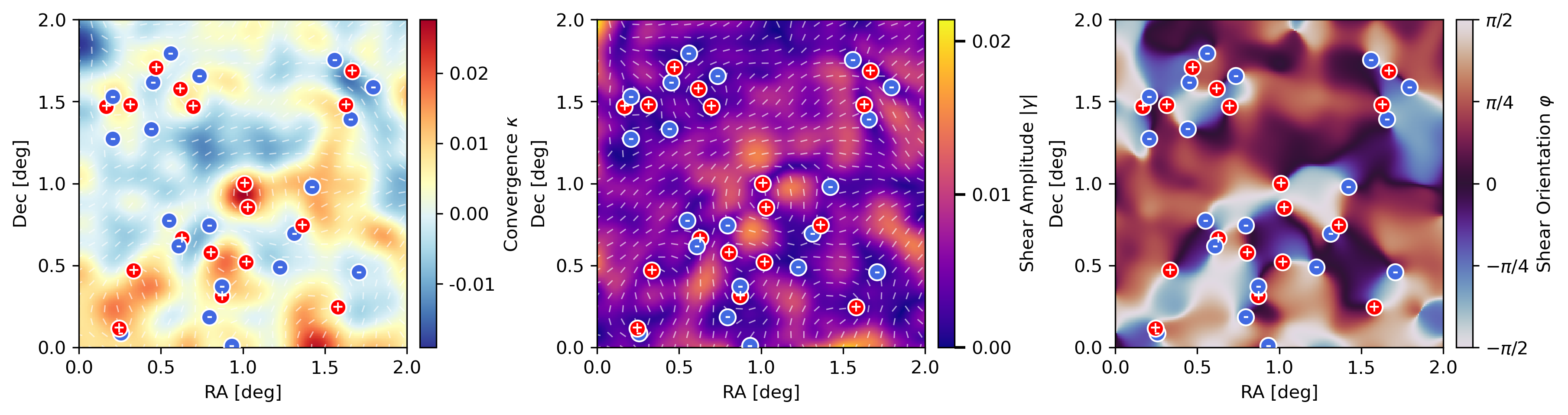}
    \caption{An example of topological defects in a cosmic shear map. The left panel shows the convergence map, with shear orientation represented by white lines and defects marked by red and blue points for positive and negative charges respectively. The middle panel shows the shear amplitude map, where defects correspond to points with near zero shear amplitude. The right panel shows the shear orientation map, with defects located at points where the orientations are singular.}
    \label{fig:defect_example}
\end{figure*}

Fig.~\ref{fig:defect_example} visualises the defects with positive ($m=1/2$) and negative ($m=-1/2$) charges in a simulated map. Here, we select a representative cosmic shear map from the SLICS (Scinet LIghtCone Simulations) \cite{2018MNRAS.481.1337H} and mark the identified defects with red and blue dots, corresponding to positive and negative topological charges, respectively. To reduce noise, and for improved visualisation, the shear map has been smoothed with a Gaussian kernel of 4.69 arcmin.
From the left panel, it is evident that defects predominantly appear near the peaks and minima of the convergence map, indicating that defect locations are associated with distinct local convergence environments. At these locations, as shown in the middle panel, the shear amplitude approaches zero, consistent with the defining property of defects. The right panel further illustrates that defects occur precisely at points where the shear orientation becomes undefined, i.e., at singularities.

\subsection{Density and Correlation Functions of Topological Defects} \label{sec:defect_stats}

According to the definition of topological defects, their local number density can be expressed formally as
\begin{equation}
    d(\theta) = \langle \delta_{\rm 2D}[\mathbf{\gamma(\theta)}] |\mathcal{J}[\gamma(\theta)]| \rangle,
\end{equation}
where $\delta_{\rm 2D}[\mathbf{\gamma(\theta)}]$ is the 2-D Dirac delta function, which ensures that we only count points where the shear amplitude is zero, and $\mathcal{J}(\gamma)$ is the Jacobian determinant of the shear field, which assigns a geometric weight to each defect based on their area in the shear space.
We note that the Jacobian of the shear field is composed of derivatives of the shear components, which are related to cosmic flexion \cite{2006MNRAS.365..414B, 2022PhRvD.105l3521A}.
This is a very interesting correspondence, which we examine in detail in App.~\ref{app:flexion}.

Another interesting property of defects can be discovered by this formula, that is the topological defects are independent of global amplitude rescaling of the shear field.
If we multiply the shear field by a globally constant factor $A$, one can prove the local number density stays the same as
\begin{equation}
    d(\theta) = \langle \delta_{\rm 2D}[A\mathbf{\gamma(\theta)}] |\mathcal{J}[A\gamma(\theta)]| \rangle = \langle \frac{1}{A^2} \delta_{\rm 2D}[\mathbf{\gamma(\theta)}] A^2 |\mathcal{J}[\gamma(\theta)]| \rangle = \langle \delta_{\rm 2D}[\mathbf{\gamma(\theta)}] |\mathcal{J}[\gamma(\theta)]| \rangle,
\end{equation}
in which we have used the property of the Dirac delta function that $\delta_{\rm 2D}(Ax) = \frac{1}{A^2} \delta_{\rm 2D}(x)$.

For a 2-D Gaussian random field, the number density of topological defects can be derived analytically. The expected number density of defects, $d$, is given by \cite{2009JCAP...09..034V, 2003JPhA...36.6611D}
\begin{equation}
    d = \left\vert \frac{3\xi^{(6)}_{\psi,0}}{10 \pi \xi^{(4)}_{\psi,0}} \right\vert + \frac{1}{\pi}, \label{eq:defect_density}
\end{equation}
where the correlation function of the lens potential is given by
\begin{equation}
    \xi_\psi(\theta) = \sum_\ell \frac{2\ell + 1}{4\pi} C_\ell^\psi P_\ell(\cos \theta) \approx \int \frac{\ell d\ell}{2\pi} C_\ell^\psi J_0(\ell \theta), \label{eq:lens_potential_2pcf}
\end{equation}
and $\xi^{(n)}_{\psi,0}$ is the $n$-th derivative of $\xi_\psi(\theta)$ at $\theta=0$. 
$C_\ell^\psi$ is the angular power spectrum of the lens potential at angular scale $\ell$, and $P_\ell$ is the Legendre polynomial of order $\ell$.
The second equality is the flat-sky approximation, which is valid for small angular scales, where $J_0$ is the Bessel function of the first kind of order zero.
We find that computing the defect number density using derivatives of Legendre polynomials rather than the Bessel function will only involve combinations of $\ell$ and $C_\ell^\psi$, which largely eases the numerical computation.
We show the exact formula in App.~\ref{app:theories} for reference.

The lens potential power spectrum can be computed from the convergence power spectrum, $C_\ell^\kappa$, as
\begin{equation}
    C_\ell^\psi = \frac{4}{\ell^2 (\ell + 1)^2} C_\ell^\kappa.\label{eq:Cl_psi}
\end{equation}
where $C_\ell^\kappa$ is the angular power spectrum of the convergence field, which can be computed from the matter power spectrum.

We note that the original formula in \cite{2009JCAP...09..034V} does not include the second term $\frac{1}{\pi}$, which is a correction accounting for the spherical topology of the sky.
Specifically, the Poincaré--Hopf theorem \cite{poincare1881memoire, hopf1926vektorfelder} states that the total topological charge of a vector field on a closed surface must equal the Euler characteristic of that surface. For a sphere, the Euler characteristic is 2, and as cosmic shear is a spin-2 field, each defect carries a charge of $\pm 1/2$, leading to a difference of 4 between the number of positive and negative defects. This implies that there must be 4 more positive defects than negative defects in the full-sky maps, which corresponds to an additional contribution of $\frac{1}{\pi}$ to the defect number density.

We can also define the 2PCFs of topological defects, which can be formulated as

\begin{align}
    g_{\rm Q}(\theta) &= \frac{\langle\delta_{\rm 2D}(\mathbf{\gamma}_A) \delta_{\rm 2D}(\mathbf{\gamma}_B) \mathcal{J}(\mathbf{\gamma_A}) \mathcal{J}(\mathbf{\gamma_B})\rangle}{d_A(\mathbf{\theta}_A) d_B(\mathbf{\theta}_B)},\label{eq:charge_2pcf}\\
    g_{\rm N}(\theta) &= \frac{\langle\delta_{\rm 2D}(\mathbf{\gamma}_A) \delta_{\rm 2D}(\mathbf{\gamma}_B) |\mathcal{J}(\mathbf{\gamma_A})| |\mathcal{J}(\mathbf{\gamma_B})|\rangle}{d_A(\mathbf{\theta}_A) d_B(\mathbf{\theta}_B)}, 
\end{align}
where $g_{\rm Q}(\theta)$ and $g_{\rm N}(\theta)$ are the charge density and number density 2PCFs of defects, respectively.
The footnote $A$ and $B$ denote two different locations separated by an angle $\theta$, and $d_A$ and $d_B$ are the local number densities of defects at these two locations.
From the formula above, we can see that the clustering of defects is also independent of the amplitude of the shear field, since multiplying the shear field by a constant factor will not change the final mathematical form of the 2PCFs.
We note that computing the charge density correlation function $g_{\rm Q}(\theta)$ is much easier than computing the number density correlation function $g_{\rm N}(\theta)$, since the latter involves the modulus of the Jacobian determinant, which is more difficult to handle analytically.
The exact expression for $g_{\rm Q}(\theta)$ in Gaussian random fields has already been derived in \cite{2003JPhA...36.6611D}.

In practice, we can measure 3 types of 2PCFs for defects which are called partial correlation functions: the auto-correlation of positive defects, $g_{++}(\theta)$, the auto-correlation of negative defects $g_{--}(\theta)$, and the cross-correlation between positive and negative defects $g_{+-}(\theta)$.
They are defined as
\begin{equation}
    g_{++}(\theta) = \langle d_+ d_+\rangle, \quad g_{--}(\theta) = \langle d_- d_-\rangle, \quad g_{+-}(\theta) = \langle d_+ d_-\rangle,
\end{equation}
where $d_+$ and $d_-$ are the local number densities of positive and negative defects, respectively.
The partial correlation functions can be related to the charge density and number density correlation functions as
\begin{align}
    &g_{\rm Q}(\theta) = \langle \frac{1}{2} (d_+ - d_-) \frac{1}{2} (d_+ - d_-)\rangle = 
    \frac{g_{++}(\theta) + g_{--}(\theta) - 2 g_{+-}(\theta)}{4}, \\
    &g_{\rm N}(\theta) = \langle \frac{1}{2} (d_+ + d_-) \frac{1}{2} (d_+ + d_-)\rangle = 
    \frac{g_{++}(\theta) + g_{--}(\theta) + 2 g_{+-}(\theta)}{4}.
\end{align}

It has been proved that the charge density 2PCF of defects provides a clear theoretical understanding of an effect known as ``topological charge screening'' \cite{2003JPhA...36.6611D}. 
The notion of screening here refers to a global constraint on the correlations of topological charges, which arises from topological charge neutrality.
Specifically, when considering the charge density 2PCF, the integral over all separations is constrained to yield a total contribution of $-1$, indicating that the net topological charge surrounding any given defect must exactly compensate its own charge.
This integral constraint implies that the presence of a defect induces an excess of opposite-charge defects in its neighbourhood, such that the total charge is screened on large scales.
As a direct consequence of this screening condition, defects of opposite charges exhibit positive correlations, while defects carrying the same charge are anti-correlated, reflecting an effective attraction between opposite charges and a repulsion between like charges.

\subsection{Defect Finder}\label{sec:defect_finder}
With the definition of the topological charge, we can easily identify defects in discrete shear maps. For a given pixel, we calculate the change of shear angle $\Delta \varphi$ along 8 neighbouring pixels in a counter-clockwise manner. 
This procedure effectively measures the winding number of the shear orientation around the pixel, which quantifies how many times the orientation rotates by $2\pi$ (or $\pi$ for spin-2 fields) along a closed loop
\begin{equation}
    m = \frac{1}{2\pi} \sum_{i=1}^{8} \Delta \varphi_i,
\end{equation}
where $\Delta \varphi_i = \varphi_{i+1} - \varphi_i$ is the change of shear angle between neighbouring pixels, and $\varphi_9 = \varphi_1$ to close the loop.
If the total change $\sum \Delta \varphi = \pm \pi$, then there is a defect with charge $\pm 1/2$ inside the pixel. This method was first used to identify defects in CMB polarisation maps \cite{2005PhRvD..72d3004H}. 

We can validate the performance of our defect finder by applying it to Gaussian random fields, where the statistical properties of defects can be computed analytically.
We find defects with our defect finder, and compute their number density and 2PCF, and then compare them with the theoretical predictions.

One of the key factors that can affect the performance of the defect finder is the pixelisation of the maps, as we are using discrete pixels to approximate a continuous field.
A commonly used way to mitigate pixelisation effects is to smooth the maps with some smoothing scale, which effectively suppresses small-scale power.
Our first test is to find the smoothing scale needed to ensure that the defect finder can yield accurate results without biases from pixelisation effects.
To do this, we generate full-sky Gaussian convergence maps using the \texttt{glass} package \cite{2023OJAp....6E..11T}, setting the resolution parameter $N_{\rm side} = 256$ (corresponding to a pixel size of 13.7 arcmin), $\ell_{\rm max} = 3\times N_{\rm side}$, and varying the Gaussian smoothing kernel size from 1 pixel to 16 pixels. 
The cosmological parameters are fixed at $\Omega_{\rm m} = 0.3$ and $\sigma_8 = 0.8$, and the input power spectrum is computed using \texttt{camb} \cite{2011ascl.soft02026L, 2000ApJ...538..473L}.
We generate 10 realisations of the random maps to showcase the possible fluctuations of defect number density.
To include smoothing in the theoretical prediction, we apply a Gaussian smoothing kernel to the lens potential power spectrum
\begin{equation}
    C_{\ell, \rm smoothed}^{\psi} = C_\ell^\psi W_\ell^2,
\end{equation}
where $W_\ell = \exp\left(-\ell(\ell+1)\theta_G^2/2\right)$ is the Gaussian smoothing kernel in harmonic space, and $\theta_G$ is the smoothing scale in radians.

We then measure the defect number density and their charge density 2PCF ($g_{\rm Q}(\theta)$) in the following 3 ways:
\begin{enumerate}
    \item Using the input power spectrum directly: we apply the smoothing kernel and then compute the defect number density and 2PCF via the theoretical formula. This approach yields the theoretical prediction without contributions from cosmic variance or pixelisation effects. We denote this prediction as ``Theory''.
    \item Using convergence power specta computed from the smoothed simulated maps: we measure the convergence power spectrum from the smoothed maps, and then compute the defect number density and 2PCF via the theoretical formula. This procedure accounts for cosmic variance and minor pixelisation effects. We denote this prediction as ``Expected''.
    \item Using the defect finder: we apply our defect finder to the smoothed maps to directly measure the defect number density and 2PCF. This procedure includes all effects, such as cosmic variance, pixelisation, and any potential biases from the defect finder. We denote this measurement as ``Measured''.
\end{enumerate}

\begin{figure}[htp]
    \centering
    \includegraphics[width=0.6\textwidth]{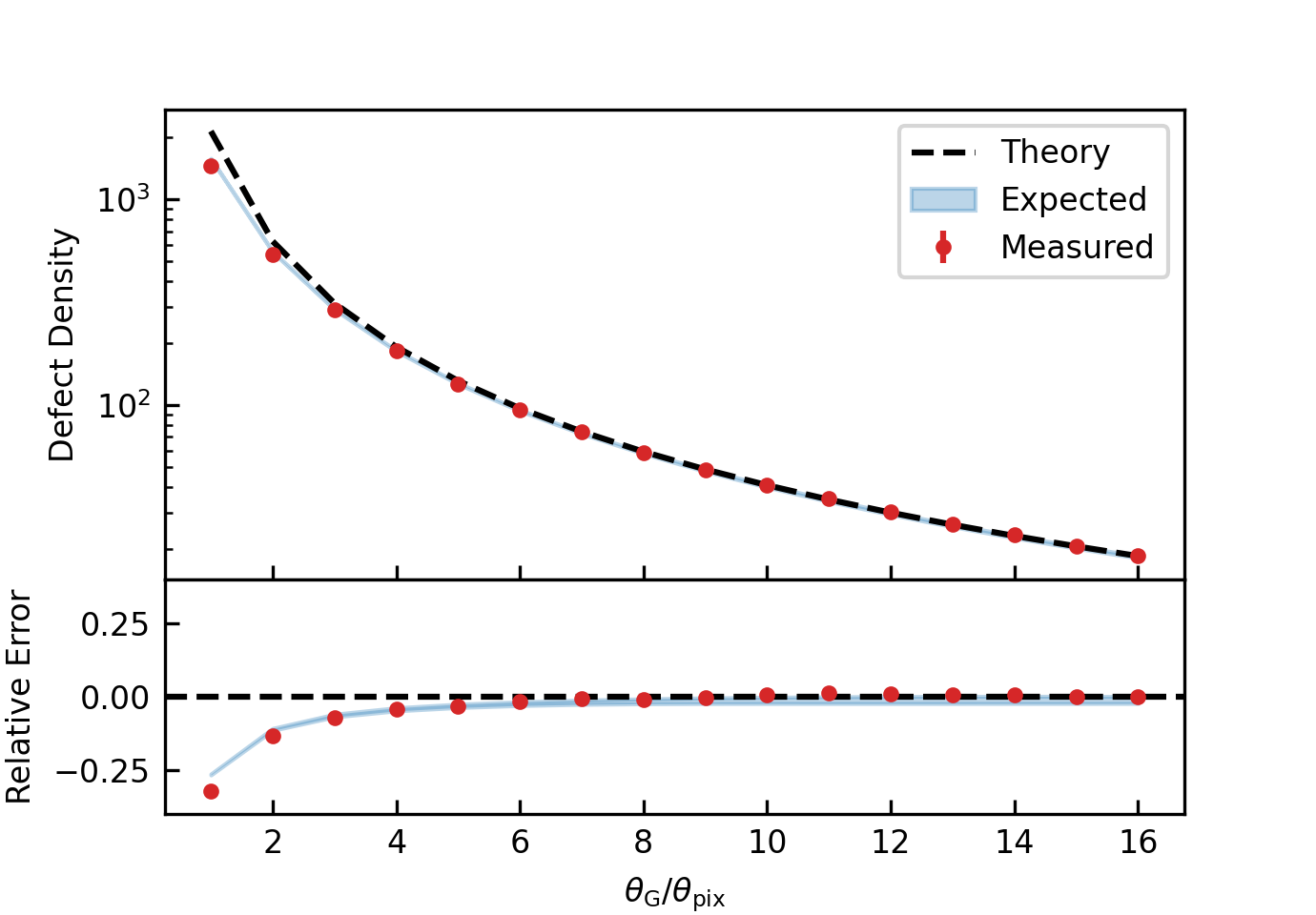}
    \caption{Defect number density in Gaussian maps as a function of smoothing scale.
    The black dashed line corresponds to the theoretical prediction directly computed from the input power spectrum (Eq.~\ref{eq:defect_density}), while the blue shaded region corresponds to the theoretical prediction computed from the power spectrum measured from the smoothed maps.
    The red points represent the defect number densities measured by our defect finder, and the error bars (which are smaller than the size of the data points) are estimated uncertainties on the mean densities from 10 realisations.}
    \label{fig:smoothing}   
\end{figure}

The results for defect number densities are shown in Fig.~\ref{fig:smoothing}. The defect number density measured by our defect finder (red points) is in good agreement with the analytical prediction computed from the smoothed power spectrum of the map (blue shaded region), while both deviate from the prediction (\ref{eq:defect_density}) based on the input power spectrum (black dashed line) when the smoothing scale is small. As the smoothing scale increases, this discrepancy gradually decreases and the three estimates become consistent once the smoothing scale reaches approximately 10 times the pixel size.

\begin{figure}[htp]
    \centering
    \includegraphics[width=0.6\textwidth]{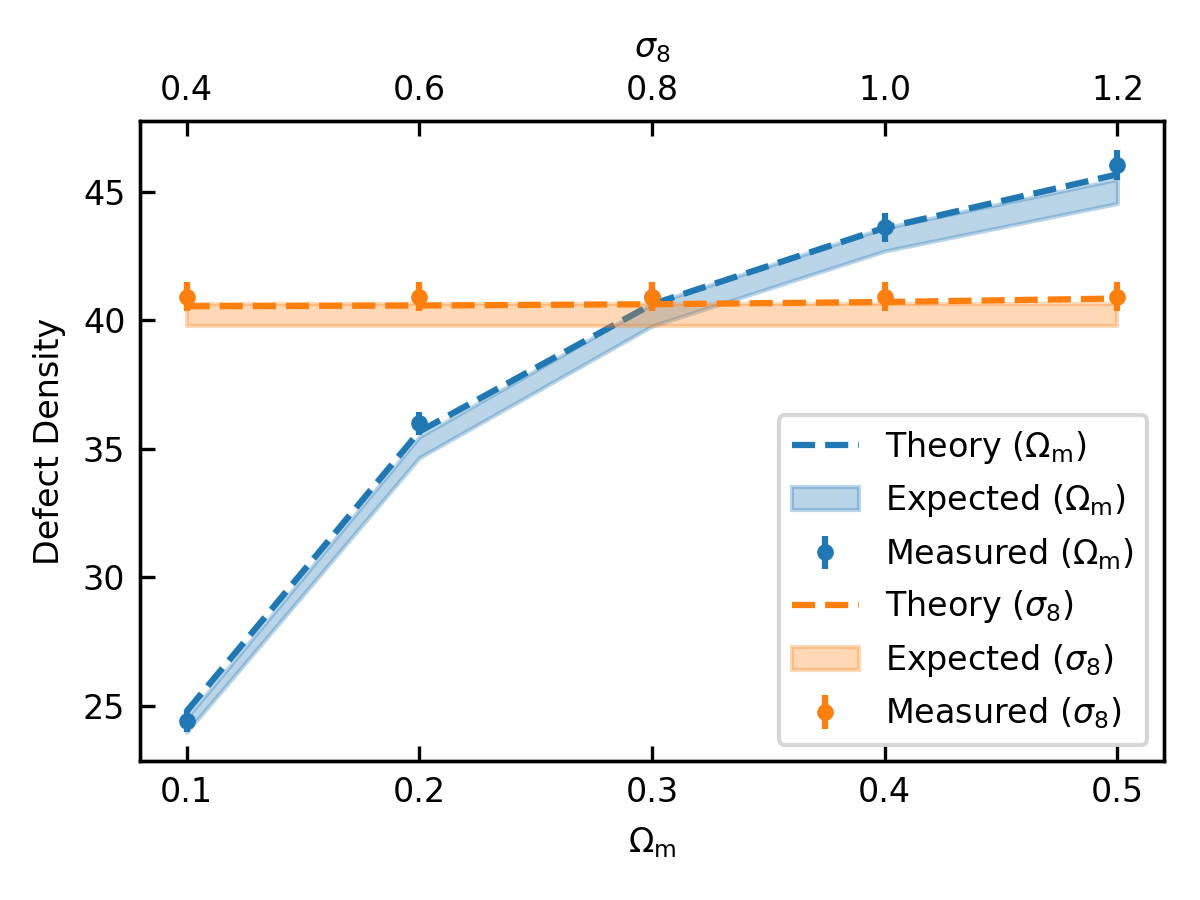}
    \caption{Defect number density in Gaussian maps as a function of cosmology. 
The blue points correspond to variations in $\Omega_{\rm m}$ while keeping $\sigma_8$ fixed at its fiducial value (0.8), whereas the orange points correspond to variations in $\sigma_8$ with $\Omega_{\rm m}$ held fixed to 0.3. 
The dashed lines represent the theoretical predictions computed from the input power spectrum, while the shaded regions correspond to the theoretical predictions computed from the power spectrum measured from the smoothed maps.
The error bars are estimated uncertainties of the mean densities from 10 realisations.}
    \label{fig:density}
\end{figure}

We also validate our defect finder by checking its performance across different cosmologies, as the defect number density is expected to vary with cosmology through the power spectrum.
Again, we generate full-sky Gaussian convergence maps with the same settings as before, but now we vary $\Omega_{\rm m}$ from 0.1 to 0.5 and $\sigma_8$ from 0.4 to 1.2.
When varying one parameter, we keep the other one fixed at $\Omega_{\rm m} = 0.3$ and $\sigma_8 = 0.8$ respectively.
Still, we generate 10 realisations for each cosmology.
We smooth the maps with a Gaussian kernel of 10 pixels, which is a ``safe'' smoothing scale according to the previous test, and apply the same smoothing kernel to the theoretical prediction.
We then measure the defect number density in the same 3 ways as before, and compare them with each other.
We demonstrate the results in Fig.~\ref{fig:density}, which shows a good agreement between the theoretical expectation and the measured defect number density from our defect finder.
Although we expect the defect number density computed from the power spectrum measured from the smoothed maps to be in the middle of the theoretical prediction and the measurement, we find that it is slightly lower than both, which likely arises from residual bias in the measured power spectrum.

Moreover, we can glimpse the potential cosmological sensitivity of the defect number density from the figure. The number density varies noticeably with changes in $\Omega_{\rm m}$, but remains largely unchanged when $\sigma_8$ is varied. This behaviour arises because the maps considered here are extremely smooth, meaning that the structures are effectively in the linear regime where variations in $\sigma_8$ only alter the amplitude of the field fluctuations. 
As we explained in Sec.~\ref{sec:defect_stats}, the existence of defects is independent of the amplitude rescaling of the shear field, which explains why the defect number density is insensitive to $\sigma_8$ here.
We will also show later that even the clustering properties of defects remain insensitive to $\sigma_8$ in linear fields.
While for $\Omega_{\rm m}$, it affects the shape (break scale) of the power spectrum, which leads to different distributions of critical points in the maps, and thus different number densities of defects.

As a further validation of our defect-finding algorithm, we explicitly checked the net topological charge constraint discussed in Sec.~\ref{sec:defect_stats}. For the full-sky maps, we find that, when the smoothing is applied appropriately, the number of positive defects is exactly 4 greater than the number of negative defects. This confirms the expectation from the Poincaré–Hopf theorem and provides an additional sanity check on the robustness of our defect-finding procedure.

\begin{figure}[htp]
    \centering
    \includegraphics[width=0.8\textwidth]{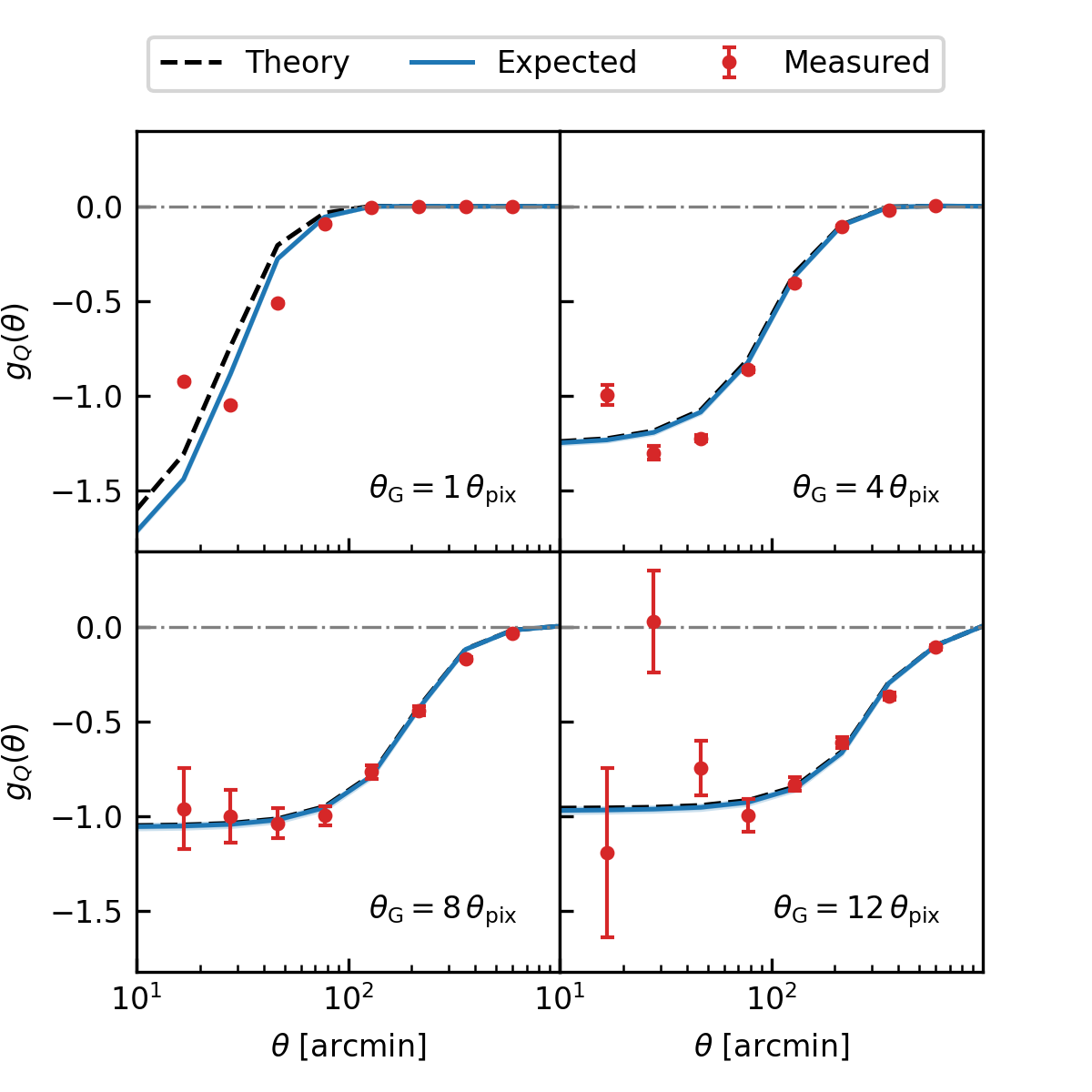}
    \caption{Charge density 2PCF of defects in Gaussian maps as a function of smoothing scale. The black dashed lines correspond to the theoretical prediction computed from the input power spectrum, while the blue shaded regions correspond to the theoretical prediction computed from the power spectrum measured from the smoothed maps. 
    The red points represent the charge density 2PCF measured by our defect finder, and the error bars are estimated uncertainties of the mean values from 10 realisations.}
    \label{fig:defect_correlation}
\end{figure}

We then validate our defect finder by checking the charge density 2PCF ($g_{\rm Q}(\theta)$) of defects in Gaussian random fields, and comparing it with the theoretical prediction. 
As mentioned, the charge density 2PCF of defects is easier to compute analytically than the number density 2PCF, so we only focus on validating this statistic.
The basic settings and 3 methods for measuring the 2PCF are the same as before, and we test the performance of the defect finder across 4 smoothing scales, $\theta_{\rm G}/\theta_{\rm pix} = 1, 4, 8, 12$, where $\theta_{\rm pix}=13.7$ arcmin.
The results are shown in Fig.~\ref{fig:defect_correlation}.

As observed, when increasing the smoothing scale, the measured charge density 2PCF of defects gets closer to the theoretical prediction.
The improvement of the agreement at small angular scales is particularly significant, which is  due to the fact that smoothing effectively suppresses small-scale pixelisation effects.
The results with smoothing scale of 12 pixels seem worse than those with smoothing scale of 8 pixels, which is because the number of defects decreases significantly when the smoothing scale is too large, which leads to lack of defect pairs in specific separation bins (e.g. the second bin in the bottom-right panel).

\section{Weak Lensing Simulations} \label{sec:sims}
We investigate the properties of topological defects with the help of simulations, which are separated into three types: idealised low-resolution full-sky noise-free Gaussian/lognormal simulations (``idealised'' hereafter), 
high-resolution full-sky noisy lognormal simulations (``noisy'' hereafter), and high-resolution Euclid-like noisy simulations (``Euclid-like'' hereafter).
The purposes of these simulations are different: the idealised simulations are used for validating our numerical implementation and understanding the basic properties of defects, while the noisy case serves a step-stone towards more realistic settings, and the Euclid-like simulations are used for forecasting cosmological constraints from defects in Stage-IV lensing surveys.

We generate full-sky maps using the \texttt{glass} package \cite{2023OJAp....6E..11T} to generate Gaussian and lognormal random fields on the full sky with given angular power spectra.
For the idealised case, we set the resolution parameter $N_{\rm side} = 512$, $\ell_{\rm max} = 3 \times N_{\rm side}$ (corresponding to a resolution of about 7 arcmin per pixel), and place all source galaxies at redshift $z_s = 1$.
We choose $\Omega_{\rm m} = 0.3$, $\sigma_8 = 0.8$ for the fiducial cosmology, and generate 1000 Gaussian and lognormal realisations with different random seeds respectively for covariance estimation.
We also vary $\Omega_{\rm m}$ and $\sigma_8$ around the fiducial values by $\Delta \Omega_{\rm m} = \Delta \sigma_8 = 0.1$ to compute their derivatives as a preparation for Fisher analysis, and generate 100 Gaussian and lognormal realisations for each cosmology to reduce the statistical uncertainty.
We justify our choice of the parameter step size in Sec.~\ref{sec:fisher_derivs}.

In practice, we smooth the maps with 3 smoothing scales, $\theta_G = 15, 30, 90$ arcmin respectively, using a Gaussian kernel, implemented with the \texttt{healpy} package \cite{2005ApJ...622..759G}.
We are particularly interested in whether the defect statistics encode different information at various smoothing scales, as this may prove useful for breaking parameter degeneracies \cite{2023MNRAS.520.1721G}.
Smoothing can be interpreted as a low-pass filter, effectively suppressing power on small-scales. In our settings, the smoothing scales are high enough to suppress nonlinear structures.
We adopt this approach for two reasons. First, it mitigates pixelisation effects in defect identification. Second, it ensures that the maps are dominated by the linear matter power spectrum, where the impact of cosmological parameters is more transparent: $\Omega_{\rm m}$ affects both the amplitude and shape of the power spectrum, whereas $\sigma_8$ primarily alters the amplitude. As we will show later, this separation facilitates a clearer understanding of the cosmological sensitivities of defect statistics.

For the noisy case, we improve the resolution to $N_{\rm side} = 1024$ (corresponding to a resolution of about 3.5 arcmin per pixel), and add shape noise to the maps by assuming a KiDS-like noise level of $\sigma_{\rm \epsilon} = 0.3$ and a galaxy number density of $n_{\rm gal} = 6 \; \text{arcmin}^{-2}$ .
Considering the computational cost of generating high-resolution maps, we only generate 300 lognormal realisations for the fiducial cosmology, and 50 realisations for each cosmology with varied parameters.
We then smooth the maps with a Gaussian kernel of 2 pixels, as this analysis aims to investigate the impact of shape noise on the defect statistics rather than to ensure agreement with theoretical predictions. This relatively small smoothing scale retains the influence of noise and small-scale features on the identified defects, while, as we show later, being sufficient to capture the cosmological information encoded in the defect statistics.

The more realistic Euclid-like shear maps are taken from the non-tomographic Euclid HOWLS (Higher Order Weak Lensing Statistics) analysis \cite{2023A&A...675A.120E}, which consist of two sets of simulations: the DUSTGRAIN-\textit{pathfinder} simulations \cite{2018MNRAS.481.2813G} and the SLICS \cite{2018MNRAS.481.1337H}.
The simulations are designed to match the expected specifications of the Euclid survey, including the redshift distribution and number density of source galaxies with the specified density of 30 galaxies per square arcminute.
Moreover, unlike the idealised simulations, these maps incorporate shape noise, rendering them more realistic for forecasting cosmological constraints.
Further details on the Euclid HOWLS simulations can be found in \cite{2023A&A...675A.120E}.
We only introduce some of the essential details here.

The DUSTGRAIN-\textit{pathfinder} simulations are used for computing the derivatives of our statistics with respect to cosmological parameters.
This suite contains four cosmologies varying $\Omega_{\rm m}$ and $\sigma_8$ respectively, and each cosmology has 128 independent realisations.
While the SLICS are used for covariance matrix estimation, it contains 924 independent realisations for a fiducial cosmology.
The parameters of these simulations are summarised in Tab.~\ref{tab:howls}.

\begin{table}[t]
    \centering
    \begin{tabular}{c c c c c}
    \hline
    Name & $\Omega_{\rm m}$ & $\sigma_8$  & N \\
    \hline
    DUSTGRAIN-path $\Omega_{\rm m}-$ & 0.200 & 0.842  & 128 \\
    DUSTGRAIN-path $\Omega_{\rm m}+$ & 0.400 & 0.842  & 128 \\
    DUSTGRAIN-path $\sigma_8-$ & 0.313 & 0.707  & 128 \\
    DUSTGRAIN-path $\sigma_8+$ & 0.313 & 0.977  & 128 \\
    SLICS Fiducial & 0.291 & 0.826  & 924 \\
    \hline
    \end{tabular}
    \caption{The input cosmologies and the number of realisations in the Euclid HOWLS simulations.
            DUTGRAIN-$\textit{pathfinder}$ are used to compute the derivatives with respect to $\Omega_{\rm m}$ and $\sigma_8$ whilst the SLICS Fiducial is used for covariance estimation.}
    \label{tab:howls}
\end{table}

The resolution of these simulations is 0.59 arcmin, and we also apply smoothing to these maps, but only at a single scale, consistent with the approach adopted in the Euclid HOWLS analysis.
For higher-order statistics, we apply a top-hat filter with a smoothing radius of $\theta_{\rm s} = 4.69$ arcmin, which is equivalent to 8 pixels in the maps.
Here we use $\theta_{\rm s}$ to denote the top-hat smoothing scale for the Euclid-like maps, to distinguish it from $\theta_{\rm G}$ used for the Gaussian smoothing scale.
We choose the top-hat filter to make our results more comparable to the $\kappa$-PDF results in the Euclid HOWLS analysis, and a smoothing scale of 8 pixels to ensure a robust detection of defects (see Sec.~\ref{sec:defect_finder}).
The smoothing is implemented by convolving the shear maps with a top-hat kernel using the \texttt{scipy} package \cite{2020NatMe..17..261V}.

The two simulation suites also have different survey areas; the DUSTGRAIN-\textit{pathfinder} maps span $5 \times 5$ deg$^2$, while the SLICS maps are $10 \times 10$ deg$^2$.
Our preliminary tests reveal that our defect statistics, like angular lensing power spectra and many HOWLS, are affected by masking \cite{2013ApJ...774..111S, 2014ApJ...784...31L}. In order to keep consistent masking between the SLICS and DUSTGRAIN maps, we divide the former into four $5 \times 5$ deg$^2$ sub-maps, and treat them as independent realisations in our analysis. This is valid because our data vectors only contain information on angular scales much smaller than 5 deg. Specifically, they probe up to approximately 100 arcmin.
This means we have 3696 pseudo-independent realisations from SLICS for covariance matrix estimation. 
This factor 4 increase in the number of realisations means a greater number of bins can be used in estimating the covariance of our statistics.

\section{Statistics of Defects} \label{sec:stats}
Following the procedure outlined in the previous section, we apply the defect finder to these simulated shear maps to construct a ``defect catalogue'' for each simulation.
In this section, we focus on the properties of defects, including their clustering, and convergence distribution at their locations. 
Specifically, we will investigate the defect-2PCFs (D-2PCFs), the cumulative distribution of k-th nearest neighbours (kNN-CDF) of defects, and the defect-conditioned convergence distribution (DCC-CDF).

\subsection{Two-point Correlation Function of Defects}
We first consider the two-point correlation function (2PCF), denoted by $g(\theta)$ here, which quantifies the excess probability of finding a pair of objects separated by an angular distance $\theta$ compared to a random distribution.
Here we choose to measure the partial 2PCFs of defects (D-2PCFs) $g_{++}(\theta)$, $g_{--}(\theta)$, and $g_{+-}(\theta)$, using the Landy-Szalay estimator \cite{1993ApJ...412...64L}
\begin{equation}
g(\theta) = \frac{DD(\theta) - 2DR(\theta) + RR(\theta)}{RR(\theta)},
\end{equation}
where $DD(\theta)$, $DR(\theta)$, and $RR(\theta)$ represent the binned pair counts of defect-defect, defect-random, and random-random pairs at angular separation $\theta$, respectively.
We use the \texttt{TreeCorr} package \cite{2015ascl.soft08007J} to compute the 2PCFs.

\begin{figure*}[htp]
    \centering
    \includegraphics[width=0.6\textwidth]{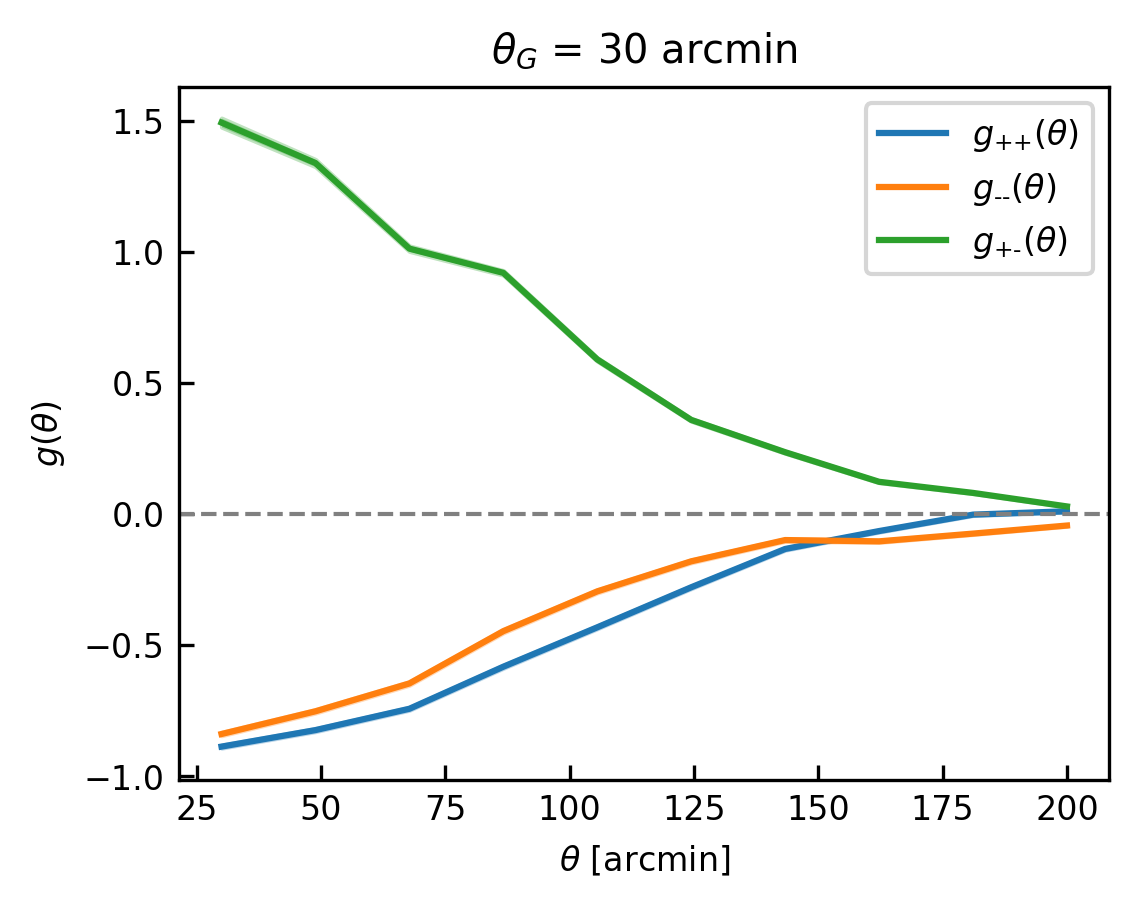}
    \caption{D-2PCFs in Gaussian maps, with smoothing scale $\theta_G = 30$ arcmin. The blue and orange lines are the 2PCFs for defects with the same charges, while the green line is for defects with opposite charges. The lines are the mean values from 1000 realisations, while the shaded regions indicate the uncertainty of the mean, though their magnitude is so small ($\lesssim 0.005$) that they are barely visible.}
    \label{fig:2pcf_gauss}
\end{figure*}

In Fig.~\ref{fig:2pcf_gauss}, we show the D-2PCFs in the idealised Gaussian convergence maps, with smoothing scale $\theta_G = 30$ arcmin. 
We observe a clear manifestation of the defect screening effect: defects with opposite charges exhibit a strong positive correlation, whereas defects with the same charge show negative correlations, indicating a tendency to repel one another.
Also, with decreasing angular scale, the absolute values of the 2PCFs increase, indicating a stronger clustering/repulsion effect at smaller scales.
The figure also reveals noticeable differences between the 2PCFs of positive and negative defects, suggesting that positive defects are more strongly repulsive than negative defects on small-scales.

\subsection{Cumulative Distribution of k-th Nearest Neighbours}

For Gaussian random fields, the 2PCF is a complete statistic to describe the statistical properties of the field.
However, even for Gaussian weak lensing maps, the defect clustering might not be Gaussian.
Calculating higher-order correlation functions is computationally expensive.
Therefore, in this work, we choose to focus on the k-th nearest neighbour (kNN) statistics of defects, which is proved to be an efficient way to extract non-Gaussian information from discrete point sets \cite{2021MNRAS.500.5479B}.

Here we define the cumulative distribution function of k-th nearest neighbours (hereafter referred to as the kNN-CDF), 
which is defined as the cumulative number density of defects whose k-th nearest neighbour is within an angular distance $\theta$.
We note, to keep defect density information, this function is not normalised to one, but normalised with a factor of total solid angle of the survey, so that it has the dimension of number density (${\rm sr}^{-1}$).

As the defects are marked with topological charges, we can measure the kNN distributions for defects with different charges separately, yielding 4 different kNN distributions in total:
\begin{itemize}
    \item $C_{\rm kNN}^{\rm pos-pos}(\theta)$: kNN-CDF for defects with charge $+1/2$ only,
    \item $C_{\rm kNN}^{\rm neg-neg}(\theta)$: kNN-CDF for defects with charge $-1/2$ only,
    \item $C_{\rm kNN}^{\rm pos-neg}(\theta)$: kNN-CDF for defects with charge $+1/2$, but counting the distance to the nearest defect with charge $-1/2$,
    \item $C_{\rm kNN}^{\rm neg-pos}(\theta)$: kNN-CDF for defects with charge $-1/2$, but counting the distance to the nearest defect with charge +$1/2$.
\end{itemize}

We use the \texttt{scikit-learn}'s ``ball-tree" algorithm \cite{2011JMLR...12.2825P} to efficiently find the k-th nearest neighbours for each defect, before counting the cumulative number of defects whose k-th nearest neighbour is within an angular distance $\theta$ to get the kNN-CDFs.
\begin{figure}[htp]
    \centering
    \includegraphics[width=0.6\textwidth]{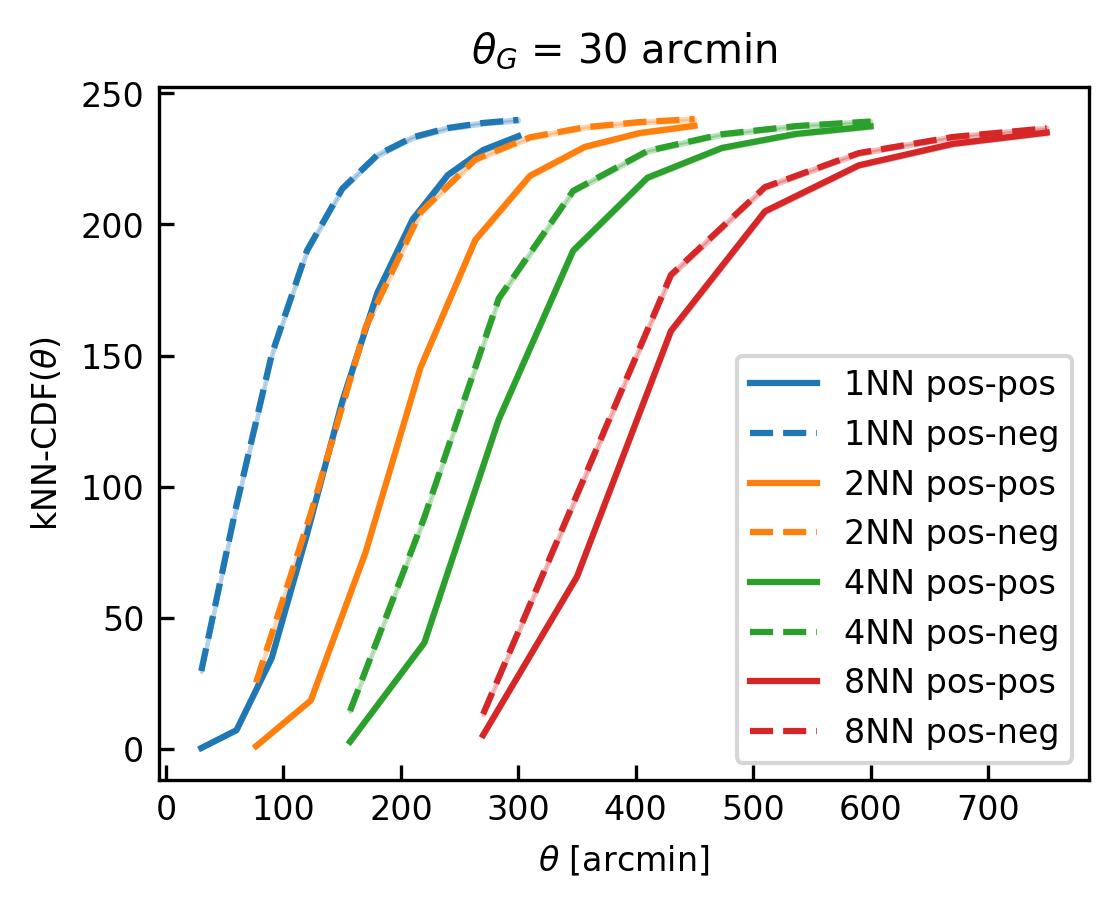}
    \caption{kNN-CDFs for positive defects in Gaussian convergence maps, with smoothing scale $\theta_G = 30$ arcmin. The different colours represent different k values, i.e., 1st, 2nd, 4th, and 8th nearest neighbours respectively. Solid lines are for same-charge neighbours, while dashed lines are for opposite-charge neighbours. The lines are the mean values from 1000 realisations, while the shaded regions indicate the uncertainty of the mean ($\lesssim 0.5$, thus barely visible).}
    \label{fig:nns_example}
\end{figure}
In Fig.~\ref{fig:nns_example}, we show an example of 1st, 2nd, 4th and 8th NN-CDFs for positive defects in idealised Gaussian convergence maps, with smoothing scale $\theta_G = 30$ arcmin.
As expected, as we increase k, the CDFs shift to larger angular scales, as the k-th nearest neighbour is further away.
For NNs with opposite charges, the CDFs are always higher than those with same charges, which is a manifestation of the defect screening effect.
We note that, due to the general symmetry between positive and negative defects, the kNN-CDFs for negative defects are similar to those for positive defects, thus we do not show them here for brevity.
We will also see in Sec.~\ref{sec:results} that their cosmological sensitivities are similar as well.

\subsection{Cumulative Distribution of Defect-conditioned Convergence}\label{sec:DCC-CDF}

\begin{figure*}[htp]
    \centering
    \includegraphics[width=0.9\textwidth]{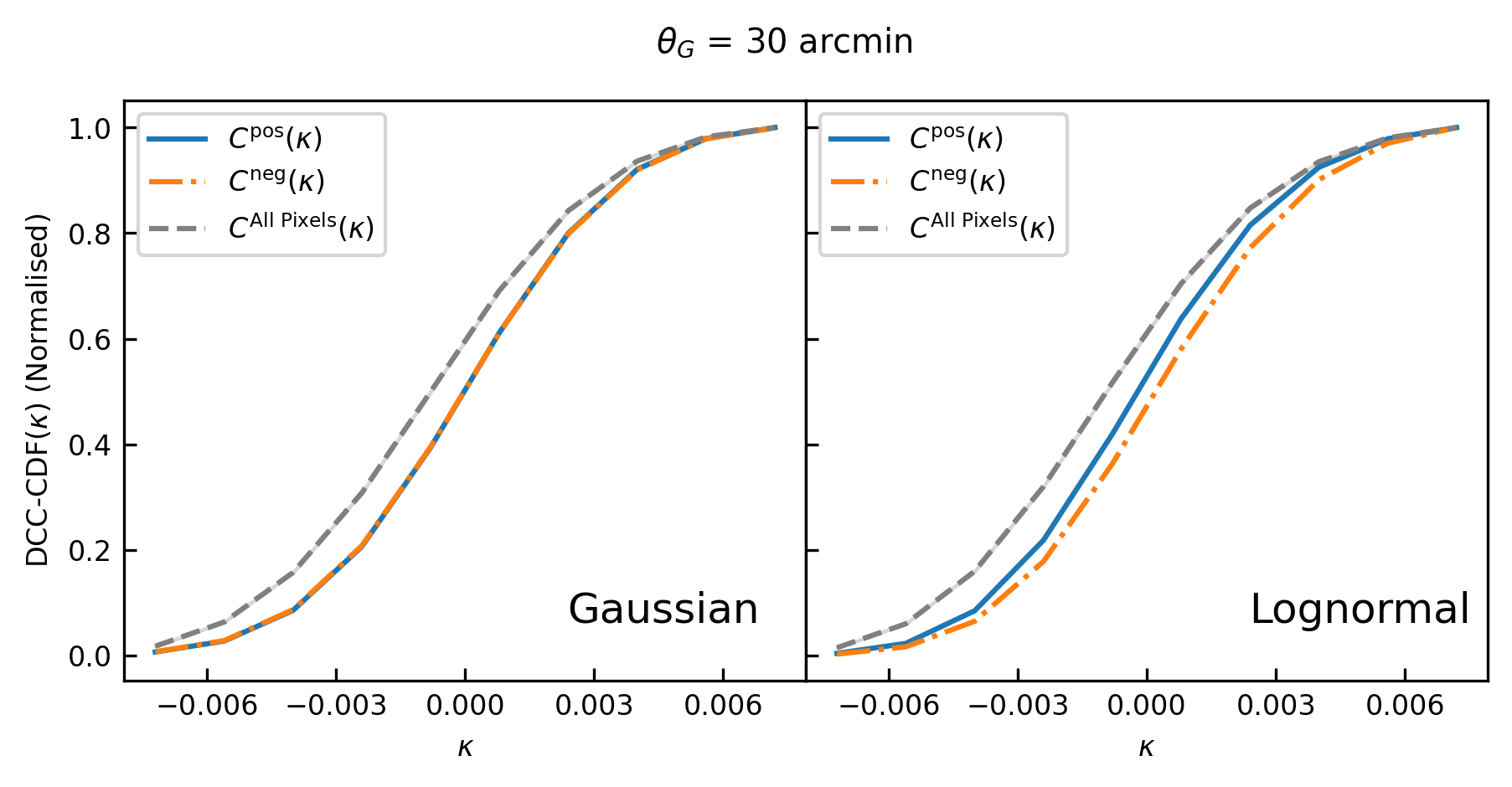}
    \caption{Comparison between the DCC-CDFs (orange/blue lines for positive/negative defects) and the overall convergence CDF (grey lines) for Gaussian (left panel) and lognormal (right panel) maps. All maps are smoothed with a Gaussian kernel with $\theta_G = 30$ arcmin. The shaded regions on $C^{\rm pos}(\kappa)$ and $C^{\text{All Pixels}}(\kappa)$ indicate the uncertainty of the mean over 1000 realisations ($\lesssim 0.002$, thus barely visible).} 
    \label{fig:dcdf_example}
\end{figure*}

The kNN-CDF we introduced above contains the clustering and number density information of defects, i.e., the phase information of the shear field.
To incorporate additional information about the field amplitude, we supplement the analysis by measuring the cumulative distribution function of convergence values at defect locations, which we refer to as the defect-conditioned CDF (DCC-CDF hereafter).

The DCC-CDF can be interpreted as a selection of special samples in the convergence field, where the deformation of images is isotropic. Also it gives us information on the matter density environment for defects to exist.
We expect this statistic to deviate from the overall convergence distribution, as defects are special points encoding both amplitude and phase information of the shear field.

For different defect charges, we can measure the convergence CDFs separately, resulting in two different statistics:
\begin{itemize}
    \item $C^{\rm pos}(\kappa)$: CDF of convergence values at defects with charge $+1/2$,
    \item $C^{\rm neg}(\kappa)$: CDF of convergence values at defects with charge $-1/2$.
\end{itemize}
We normalise the CDFs by the area of the maps; hence this statistic quantifies the cumulative counts of defect-conditioned convergence per solid angle (${\rm sr}^{-1}$).

Fig.~\ref{fig:dcdf_example} demonstrates how the DCC-CDFs differ from the overall convergence CDFs using all pixels in our maps, for both positive and negative defects, in Gaussian and lognormal density fields, with a smoothing scale of $\theta_G = 30$ arcmin.
For this figure only, we normalise the DCC-CDFs to unity for comparison with the overall convergence CDF with the same normalisation. 
The slope of the CDF provides insight into the distribution of convergence at defect locations: a steeper slope indicates a higher concentration of defects within that specific convergence range.

The DCC-CDFs differ significantly from the overall convergence CDF. 
We observe that defects preferentially occur at moderate convergence values, avoiding both the high and low extremes.
Moreover, for Gaussian maps, the positive and negative defects have the same convergence distributions, while for lognormal maps, they differ.
Specifically, positive defects tend to be more concentrated in low-convergence regions, while negative defects tend to be more concentrated in high-convergence regions.
Here arises one important feature of defects in non-Gaussian fields: the asymmetry environment between positive and negative defects.

\section{Fisher Analysis}\label{sec:fisher}
In this section, we investigate how much cosmological information can be extracted from topological defect statistics by performing a Fisher analysis \cite{Tegmark1996KarhunenLoveEP}.
We will introduce the general formalism of Fisher analysis, then test our defect statistics with various survey settings.
We note that we only analyse non-tomographic cases in this work, while tomographic analysis will be left for more realistic analyses in the future.

While the Fisher formalism is adopted for this initial proof-of-concept study of defects, it is well known that such Gaussian approximations to the parameter space can lead to biased constraints \cite{2006JCAP...10..013P}.
Therefore, in this work, we focus on the relative performance of defect statistics compared to other statistics, rather than the absolute values of the forecasted constraints.

\subsection{Fisher Formalism}\label{sec:fisher_formalism}
Assuming a multivariate Gaussian likelihood with parameter-independent covariance, the Fisher information matrix is given by
\begin{equation}
F_{ij}
=
\left(\frac{\partial \mathbf{D}}{\partial \theta_i}\right)^{\mathrm{T}}
\boldsymbol{\Sigma}_{\rm d}^{-1}
\left(\frac{\partial \mathbf{D}}{\partial \theta_j}\right).
\end{equation}
The parameter covariance matrix is then obtained from the inverse Fisher matrix,
\begin{equation}
\mathbf{C}_{\rm p} = \mathbf{F}^{-1},
\end{equation}
and the forecasted parameter likelihood is approximated as a multivariate Gaussian centred on the fiducial parameters $\boldsymbol{\theta}_0$:
\begin{equation}
\mathcal{L}(\boldsymbol{\theta}|\boldsymbol{D})
=
\frac{1}{\sqrt{(2\pi)^n |\mathbf{C}_{\rm p}|}}
\exp\left[
-\frac{1}{2}
(\boldsymbol{\theta} - \boldsymbol{\theta}_0)^{\mathrm{T}}
\mathbf{C}_{\rm p}^{-1}
(\boldsymbol{\theta} - \boldsymbol{\theta}_0)
\right].
\end{equation}

Since the data covariance matrix is estimated from a finite number of simulations, we correct its inverse for the resulting bias \cite{2007A&A...464..399H}:
\begin{equation}
    \boldsymbol{\Sigma}_{\rm unbiased}^{-1}
    =
    \frac{N_{\rm sims} - N_{\rm bins} - 2}
    {N_{\rm sims} - 1}
    \boldsymbol{\Sigma}_{\rm biased}^{-1},
\end{equation}
where $N_{\rm sims}$ is the number of simulations and $N_{\rm bins}$ is the length of the data vector.

Following the Euclid HOWLS analysis, we further rescale the covariance matrix by the ratio of the simulation area to the target survey area, assuming that the covariance scales inversely with survey area. We use this rescaling to forecast constraints for a Euclid-like survey with an area of $15000\,{\rm deg}^2$.

\subsection{Binning Scheme and Data Selection}
To ensure the robustness of our Fisher analysis, we carefully select the bins used in each data vector.
We first set initial binning schemes for different statistics, then apply several criteria to select the bins that are included in the final analysis.

The statistics we consider in the Fisher analysis include the D-2PCFs (only for the idealised case), the kNN-CDFs, and the DCC-CDFs.
In all cases, we also measure the $\kappa$-PDF as a reference, which is one of the most constraining statistics in weak lensing analyses \cite{2023A&A...675A.120E}.
Also by comparing DCC-CDFs to this, we gain insight into the impact of down-sampling the field to select defects (using only about 1\% of the pixels instead of all).
The $\kappa$-2PCF is also included, as it is one of the most commonly used statistic in weak lensing analyses, and if the field is Gaussian, it is a complete statistic, so it serves as a useful benchmark for the information content of our defect statistics.

As we have a finite number of simulations for full-sky setups (1000 in the idealised cases and 300 in the noisy case), we carefully choose the binning schemes for different statistics by visually inspecting the data vectors and their derivatives, ensuring that the most prominent features in the statistics and their derivatives receive sufficient sampling while keeping the number of bins manageable for covariance estimation.
For higher-resolution, Euclid-like cases, as we have more simulation realisations (3696) and only one smoothing scale, we can afford to assign more bins in the data vectors. We therefore choose the binning schemes in a nearly agnostic way, with larger ranges and more bins.
We list the binning schemes for these cases in Tab.~\ref{tab:binning} and Tab.~\ref{tab:howls_binning}.

\begin{table}[t]
\centering
\small
\setlength{\tabcolsep}{4pt}

\resizebox{\textwidth}{!}{
\begin{tabular}{|c|c|c|c|c|c|c|}
\hline
\multirow{2}{*}{Statistic} 
& \multirow{2}{*}{$N_{\rm bins}$} 
& \multicolumn{4}{c|}{Bin Edges at smoothing scales} 
& \multirow{2}{*}{Spacing} \\
\cline{3-6}

& 
& $\theta_G = 15'$ 
& $\theta_G = 30'$ 
& $\theta_G = 90'$ 
& $\theta_G = 7'$ (Noisy) 
& \\
\hline

$g_{\rm defects}(\theta)$ 
& 10 
& $[15, 100]'$ 
& $[30, 200]'$ 
& $[90, 600]'$ 
& N/A 
& Logarithmic \\

$C_{\rm 1NN}(\theta)$ 
& 10 
& $[30, 100]'$ 
& $[30, 300]'$ 
& $[30, 750]'$ 
& $[7, 40]'$ 
& Linear \\

$C_{\rm 2NN}(\theta)$ 
& 10 
& $[30, 200]'$ 
& $[30, 450]'$ 
& $[30, 1125]'$ 
& $[15, 60]'$ 
& Linear \\

$C_{\rm 4NN}(\theta)$ 
& 10 
& $[30, 300]'$ 
& $[30, 600]'$ 
& $[30, 1500]'$ 
& $[15, 120]'$ 
& Linear \\

$C_{\rm 8NN}(\theta)$ 
& 10 
& $[30, 400]'$ 
& $[30, 750]'$ 
& $[30, 1875]'$ 
& $[30, 200]'$ 
& Linear \\

$C^{\rm pos/neg}(\kappa)$ 
& 10 
& $[-0.01, 0.01]$ 
& $[-0.008, 0.008]$ 
& $[-0.005, 0.005]$ 
& $[-0.01, 0.01]$ 
& Linear \\

${\rm PDF}(\kappa)$ 
& 10 
& $[-0.01, 0.01]$ 
& $[-0.008, 0.008]$ 
& $[-0.005, 0.005]$ 
& $[-0.01, 0.01]$ 
& Linear \\

$\xi^{\kappa\kappa}(\theta)$ 
& 30 
& $[15, 600]'$ 
& $[30, 600]'$ 
& $[90, 600]'$ 
& $[7, 200]'$ 
& Logarithmic \\
\hline
\end{tabular}
}

\caption{Binning schemes for different defect statistics in the idealised and noisy case.
The bin edges are given in arcmin for angular statistics.}
\label{tab:binning}
\end{table}

\begin{table}[t]
    \centering
    \begin{tabular}{|c|c|c|c|}
    \hline
    Statistic & $N_{\rm bins}$ & Bin Edges & Spacing \\
    \hline
    $C_{\rm kNN}(\theta)$ & 100 & $[0.59, 100]'$ & Logarithmic \\
    $C^{\rm pos/neg}(\kappa)$ & 100 & $[-0.05, 0.05]$ & Linear \\
    ${\rm PDF}(\kappa)$ & 100 & $[-0.05, 0.05]$ & Linear \\
    $\xi^{\kappa\kappa}(\theta)$ & 100 & $[0.59, 100]'$ & Logarithmic \\
    \hline
    \end{tabular}
    \caption{Binning schemes for different defect statistics in the Euclid-like Fisher analysis.
    The bin edges are given in arcmin for angular statistics.}
    \label{tab:howls_binning}
\end{table}

For several of the statistics considered in this work, most notably the CDF-based observables, the tails of the distributions approach regions of near-zero variance. In particular, at the extreme ends of the CDFs, the values become effectively deterministic, leading to covariance matrix elements with vanishing variance. These nearly singular modes give rise to ill-defined or unstable elements in the inverse covariance matrix, thereby significantly amplifying noise in its estimation from a finite number of simulations.
To mitigate this issue, we only select the bins where the average value of the statistic across all simulations is larger than $N\times\sigma$ of the statistic, where $\sigma$ is the square root of the diagonal elements of the covariance matrix.
We therefore retain only bins with a signal-to-noise ratio greater than $N$, adopting $N=1$ for the idealised and noisy cases and $N=2$ for the Euclid-like case. In the first two cases, the near-zero tails of the distributions already have low signal-to-noise, so a mild cut of $N=1$ is sufficient. In contrast, the Euclid-like maps contain many more defects and stronger bin correlations, requiring a stricter threshold to ensure numerical stability of the precision matrix.

We also remove highly-correlated bins from our defect statistics which are located primarily in the tails; this helps to make the inversion of the covariance matrices more stable. We identify 99.99\% correlated pairs of bins in the correlation matrix and retain only the bin with the higher signal-to-noise ratio.
This criterion is selected by experimentation to balance the trade-off between retaining informative bins and ensuring the stability of the covariance matrix inversion.

We check the stability of our selected covariance matrices by computing the condition number of the covariance matrix.
Most of the statistics considered in this work have condition numbers of order $10^7$ or smaller, while \texttt{NumPy} \cite{2020Natur.585..357H} linear algebra package can safely handle matrices with condition numbers up to $10^{12}$, so these covariance matrices are well-conditioned and suitable for inversion.
The only exceptions are those combining higher-order statistics with $\kappa$-2PCFs in the Euclid-like case, where the condition numbers reach up to $10^{23}$, which is too large for stable inversion. This is primarily due to values in combined vectors can span several orders of magnitude.
For these cases, we perform whitening transformations on the covariances matrices by introducing a transformation matrix $\mathcal{T}$ \cite{2013MNRAS.432.1928T}, which is defined as
\begin{equation}
    \mathcal{T}_{ij} = \frac{1}{\sqrt{\mathbf{\Sigma_{\rm d}}_{ii}}} \delta_{ij},
\end{equation}
where $\mathbf{\Sigma_{\rm d}}_{ii}$ is the diagonal element of the covariance matrix for the $i$-th bin, and $\delta_{ij}$ is the Kronecker delta function. 

We then invert the whitened matrix $\Sigma_{\rm w} \equiv \mathcal{T} \boldsymbol{\mathbf{\Sigma_{\rm d}}} \mathcal{T}$, and transform it back to get the inverse covariance matrix
\begin{equation}
    \boldsymbol{\Sigma_{\rm d}}^{-1} = \mathcal{T} \left(\Sigma_{\rm w}\right)^{-1} \mathcal{T}.
\end{equation}
The whitening transformation effectively reduces the condition number to the order of $10^7$, making the matrix suitable for inversion.

We find, however, that the whitening transformation or not does not significantly change the forecasted constraints, which might be due to the fact that the directions corresponding to the smallest singular values of the covariance matrix contribute very little to the overall signal. 
These modes, although numerically unstable, carry negligible weight in the likelihood because their associated signal-to-noise is low. 
As a result, stabilizing the inversion through whitening primarily affects only these weakly constrained directions, leaving the dominant modes and their constraints unchanged.

Fisher analysis assumes Gaussian-distributed data. As this assumption can, in general, be violated for beyond-two-point metrics like our defect statistics, we remove the data elements which are least-consistent with a Gaussian distribution. This is similar in principle to the approach used in the Euclid HOWLS analysis \cite{2023A&A...675A.120E}, but with some key differences. They test the Gaussianity of the likelihood by evaluating the quadratic form
\begin{equation}
y_i = \left(\mathbf{D}_i-\langle \mathbf{D}\rangle\right)\,\mathbf{\Sigma_{\rm d}}^{-1}\,\left(\mathbf{D}_i-\langle \mathbf{D}\rangle\right)^{\mathrm T},
\end{equation}  
for each realisation, where $\mathbf{D}_i$ is the data vector for the $i$-th realisation, $\langle \mathbf{D}\rangle$ is the mean data vector across all realisations, and $\mathbf{\Sigma}_{\rm d}$ is the covariance matrix of the data vector estimated from all realisations. If the likelihood is Gaussian, the resulting $y_i$ values should follow a $\chi^2$ distribution with $N_{\rm d}$ degrees of freedom, where $N_{\rm d}$ is the length of the data vector. They therefore compare the measured distribution $P_{\rm obs}(y_i)$ to the expected $\chi^2$ distribution and quantify their agreement using a weighted SMAPE (Symmetrized Mean Absolute Percentage Deviation) statistic \cite{2023MNRAS.521.1152R} defined as
\begin{align}
\mathcal{S} &= \frac{\sum_i \omega_i \mathcal{S}_i}{\sum_i \omega_i}, \\
\text{with} \quad \mathcal{S}_i &= \frac{\left| P_{\mathrm{obs}}(y_i) - P_{\chi^2}(y_i) \right|}
{\left| P_{\mathrm{obs}}(y_i) \right| + \left| P_{\chi^2}(y_i) \right|}, \\
\text{and} \quad \omega_i &= P_{\chi^2}(y_i).
\end{align}
where $P_{\chi^2}(y_i)$ is the expected $\chi^2$ distribution with $N_{\rm d}$ degrees of freedom, and $P_{\rm obs}(y_i)$ is the observed distribution of $y_i$ values across all simulations. A smaller SMAPE value indicates better agreement between the observed and expected distributions, suggesting that the Gaussian likelihood assumption is more valid for the data vector under consideration.
In the Euclid HOWLS analysis, they minimise the SMAPE by iteratively varying the range and number of bins in the data vector, to find the optimal binning scheme whose $\mathcal{S}$ value is below a chosen threshold (set to 0.15 in their work).

\begin{figure*}[htp]
    \centering
    \includegraphics[width=0.9\textwidth]{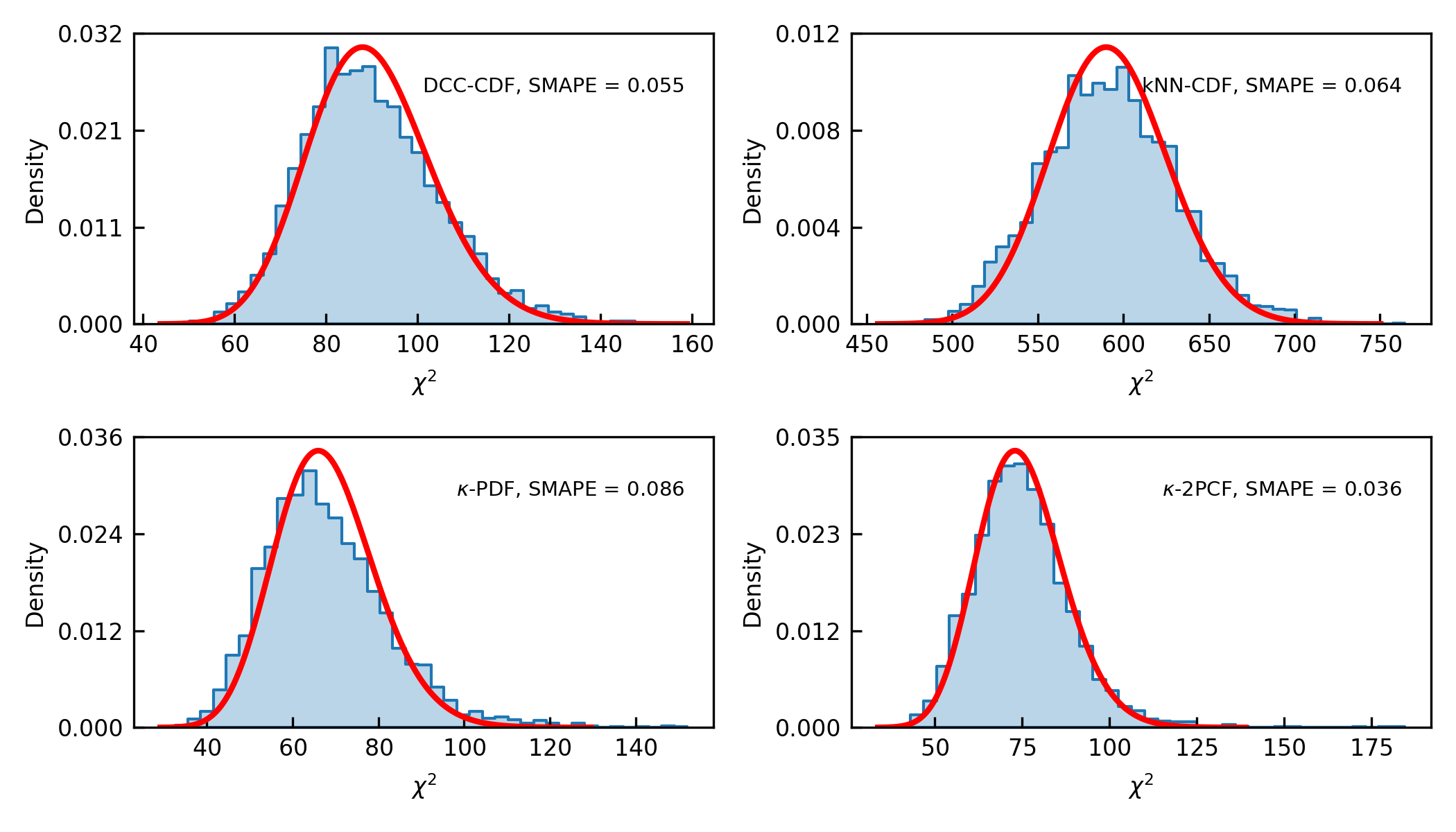}
    \caption{The distribution of $\chi^2$ values computed from the selected data vector across all Euclid-like simulations (blue histogram) compared to the expected $\chi^2$ distribution with the same number of degrees of freedom (red curve).}
    \label{fig:smape}
\end{figure*}

Since this method only assesses the overall Gaussianity of the multidimensional distribution, in this analysis, we examine each bin of the data vector individually by comparing its distribution to a Gaussian and computing the Kullback-Leibler (K-L) divergence \cite{Kullback:1951zyt} between them
\begin{equation}
    D_{\rm KL}(P||Q) = \int_{-\infty}^{+\infty} P(x) \log \frac{P(x)}{Q(x)} dx,
\end{equation}
where $P(x)$ is the probability distribution of the bin, estimated from all simulations, and $Q(x)$ is the Gaussian distribution with the same mean and variance as $P(x)$.

If the K-L divergence for a bin exceeds a chosen threshold (set to 0.1 in this work), we discard that bin from the data vector. This threshold is selected to balance two considerations: it is small enough to remove bins that contribute negligible independent information or are dominated by noise, yet large enough to retain bins that carry meaningful cosmological signal. In practice, 0.1 provides a conservative cutoff that stabilises the covariance matrix while preserving most of the informative content. 

To illustrate the consistency of the data selection criteria with the Gaussian likelihood assumption, we show in Fig.~\ref{fig:smape} the distribution of $\chi^2$ values computed from the selected data vector across all Euclid-like simulations, compared to the expected $\chi^2$ distribution with the same number of degrees of freedom. The SMAPE measured are all below 0.1, indicating good agreement between the observed and expected distributions, and thus supporting the validity of the Gaussian likelihood assumption for the selected data vector.

Hence, our data vector optimisation strategy is summarised as follows:
\begin{enumerate}
    \item For each statistic, estimate the covariance matrix from all available simulations in the fiducial cosmology.
    \item Select bins with SNR $ > N$ (where $N=1$ for the idealised and noisy cases and $N=2$ for the Euclid-like case).
    \item Remove highly correlated bins (correlation coefficient $> 99.99\%$) by retaining only the bin with the higher SNR.
    \item Remove non-Gaussian distributed bins according to the 0.1 K-L divergence criterion.
\end{enumerate}

\subsection{Fisher Derivatives}\label{sec:fisher_derivs}
As described in Sec.~\ref{sec:fisher_formalism}, the Fisher matrix requires the derivatives of the data vector with respect to the parameters.
We estimate these derivatives using a finite difference method, by varying each parameter around the fiducial cosmology while keeping the other parameter fixed.
Specifically, we compute the derivatives using the symmetric (central) finite difference:

\begin{equation}
\frac{\partial \mathbf{D}}{\partial \theta_i}
\approx \frac{\mathbf{D}(\boldsymbol{\theta}_0 + h_i) - \mathbf{D}(\boldsymbol{\theta}_0 - h_i)}{2h_i},
\end{equation}
where $\boldsymbol{\theta}_0$ is the fiducial parameter vector, and $h_i$ is a small step size along the $i$-th parameter. 

The choice of step size $h_i$ is crucial for accurately estimating the derivative. 
The step size adopted for the Euclid-like simulations has been validated in \cite{2023A&A...675A.120E}, and we follow the same choice, with the parameters summarised in Tab.~\ref{tab:howls}. while, for the idealised case, we estimate the optimal step size using the following formula, which balances the truncation error of the finite difference approximation and the numerical noise in the data \cite{10.1145/2168773.2168777}
\begin{equation}
h_{i, {\rm opt}} = \left( \frac{3\sigma}{\left|\frac{\partial^3 \mathbf{D}}{\partial \theta_i^3}\right|} \right)^{1/3},
\end{equation}
where $\sigma$ is the uncertainty of the mean data vector across all simulations in the fiducial cosmology, and $\frac{\partial^3 \mathbf{D}}{\partial \theta_i^3}$ is the third derivative of the data vector with respect to the $i$-th parameter.

In order to simplify the estimation of the optimal step size, we use only the defect number density (total number of defects per solid angle) part of the data vector, because most of the statistics we consider have the same order of magnitude as the defect number density. Then the optimal step size is applied to all defect statistics.
We estimate the third derivative of the density using a second-order finite difference method. The formula is given by
\begin{equation}
\frac{\partial^3 \mathbf{D}}{\partial \theta_i^3}
\approx \frac{\mathbf{D}(\boldsymbol{\theta}_0 + 2h_i) - 2\mathbf{D}(\boldsymbol{\theta}_0 + h_i) + 2\mathbf{D}(\boldsymbol{\theta}_0 - h_i) - \mathbf{D}(\boldsymbol{\theta}_0 - 2h_i)}{2h_i^3}.
\end{equation}

We then only focus on the step size for $\Omega_{\rm m}$, as the derivatives with respect to $\sigma_8$ are close to zero.
In this case, the uncertainty of the mean defect number density across 1000 realisations at the fiducial cosmology is $\sigma_{\rm d} \approx 1 \; {\rm sr}^{-1}$ . 
Since we only have 100 realisations for the derivative estimation, we scale this uncertainty to $\sigma_{\rm d} \approx \sqrt{10} \; {\rm sr}^{-1}$. 
The third derivative of the defect number density with respect to $\Omega_{\rm m}$ is estimated using a small step size of $h_{\Omega_{\rm m}} = 0.01$, and the theoretical density is computed with \texttt{pyCCL} \cite{2019ApJS..242....2C} at different $\Omega_{\rm m}$ values. 
Following the setup of the idealised case, we estimate $\left|\frac{\partial^3 \mathbf{D}}{\partial \Omega_{\rm m}^3}\right| \approx 10^4 \; {\rm sr}^{-1}$. 
Combining these numbers, we obtain an optimal step size of $h_{\Omega_{\rm m, opt}} \approx 0.1$ for the idealised case. 
We adopt the same step size for $\sigma_8$, noting that the derivatives with respect to $\sigma_8$ are close to zero and therefore largely insensitive to the choice of $h_{\sigma_{8}}$. 

\subsection{Theory Versus Numerical Forecasts}
The only available analytical statistics we have for defects are the charge density 2PCFs in Gaussian fields. We therefore validate our numerical Fisher forecast for $g_{\rm Q}(\theta)$ in Gaussian maps by comparing it to the analytical forecast computed with \texttt{jax} and \texttt{jax-cosmo} \cite{2019arXiv191204232S, 2023OJAp....6E..15C}.
We only compare the results in the idealised case and taking the largest smoothing scale $\theta_G = 90$ arcmin to avoid the impact of pixelisation.

\begin{figure*}[htp]
    \centering
    \includegraphics[width=0.9\textwidth]{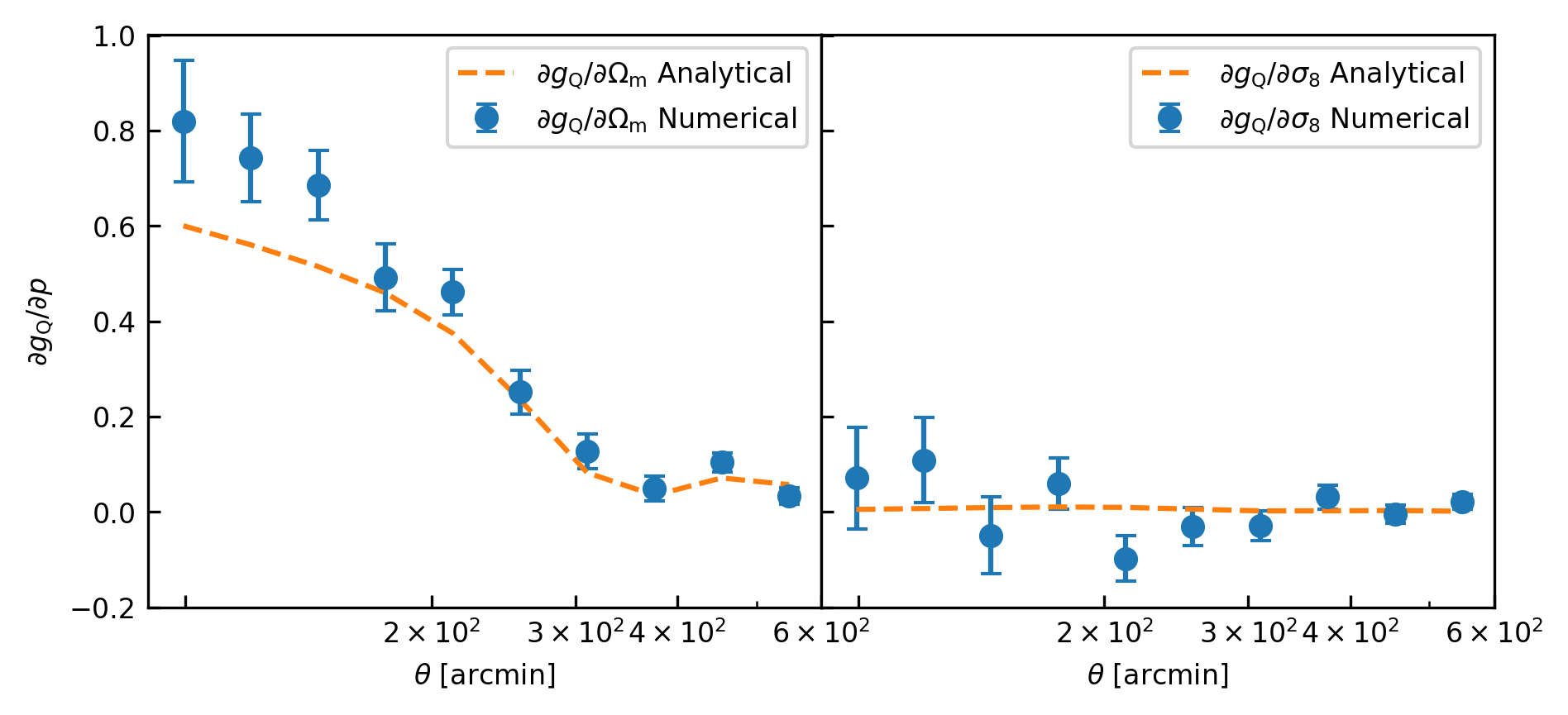}
    \caption{Comparison of the numerical (blue data points) and analytical (orange dashed line) Fisher derivatives of $g_{\rm Q}(\theta)$ with respect to $\Omega_{\rm m}$ (left) and $\sigma_8$ (right) in Gaussian maps with a smoothing scale of $\theta_G = 90$ arcmin. 
    The error bars indicate the uncertainty of the mean across 100 realisations.
    }
    \label{fig:theory_vs_numerical}
\end{figure*}

\begin{figure*}[htp]
    \centering
    \includegraphics[width=0.6\textwidth]{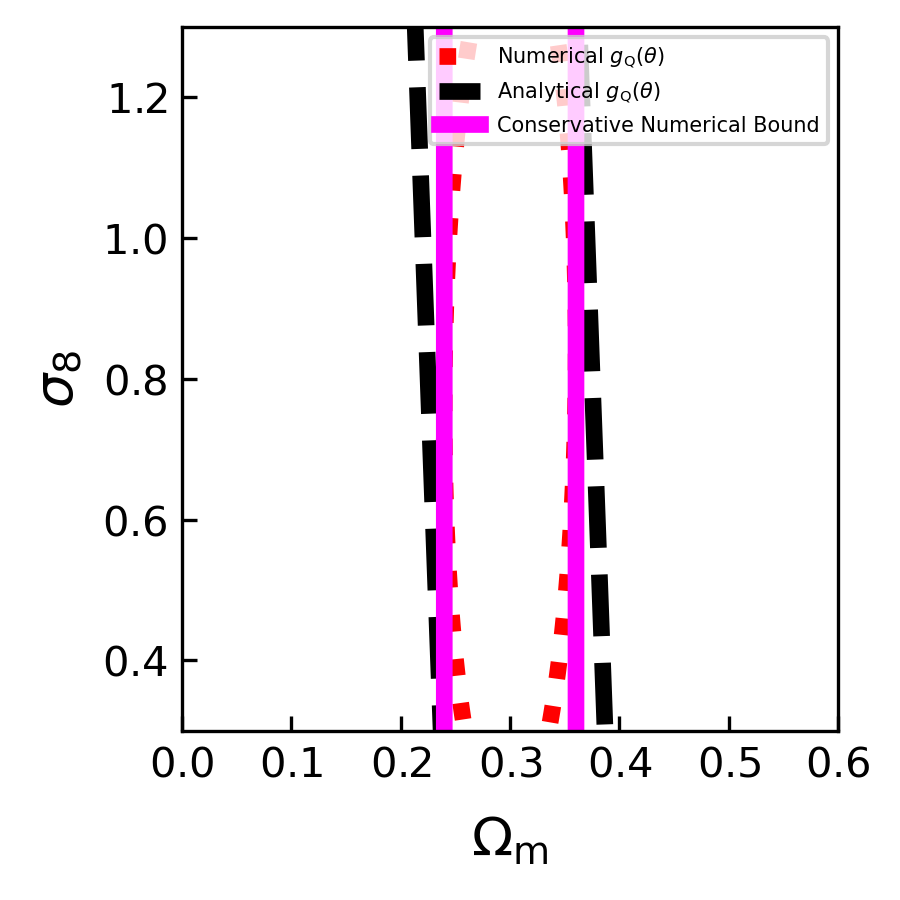}
    \caption{Comparison of the numerical and analytical Fisher forecasted constraints on $\Omega_{\rm m}$ and $\sigma_8$ from $g_{\rm Q}(\theta)$ in Gaussian maps with a smoothing scale of $\theta_G = 90$ arcmin. The black dashed contour represents the analytical forecast while the red dotted contour represents the numerical forecast estimated from simulations. 
    The magenta solid lines correspond to the most conservative case of numerical forecast.
    Both contours correspond to the 68\% confidence level.}
    \label{fig:theory_vs_numerical_contours}
\end{figure*}

The numerical and analytical Fisher derivatives of $g_{\rm Q}(\theta)$ with respect to $\Omega_{\rm m}$ and $\sigma_8$ are shown in Fig.~\ref{fig:theory_vs_numerical}, and the Fisher forecasted constraints are shown in Fig.~\ref{fig:theory_vs_numerical_contours}.
We find that the numerical derivatives are consistent with the analytical ones, while there are some small discrepancies at small scales, which might be due to the impact of pixelisation and the finite difference method used for derivative estimation.
The forecasted constraints from these two methods (dashed red and solid black contours) are consistent for $\Omega_{\rm m}$, while the numerical forecast for $\sigma_8$ is slightly tighter than the analytical one, which is due to the noisy numerical derivatives with respect to $\sigma_8$ having non-zero flucuations that can artificially tighten the constraints.
We thus obtain a conservative bound for the numerical forecast by setting the derivatives with respect to $\sigma_8$ to those closest to zero within a $1\sigma$ range. As shown by the magenta contour, this choice largely reconciles the difference between the analytical and numerical forecasts, suggesting that the discrepancy is primarily driven by noise in the numerical derivatives.
We note that the deviations between the numerical and analytical forecasts are only at extreme values of $\sigma_8$, which are out of the range of any cosmological priors, so they do not affect the conclusions of our analysis.
We will also see later in Sec.~\ref{sec:idealised_results} that the constraints on $\sigma_8$ from the defect-2PCFs are negligible once restricted to practical cosmological ranges.

\section{Results}\label{sec:results}

In this section, we present the results of our Fisher analysis, including the Fisher derivatives as tests for cosmological sensitivities of different statistics, and the forecasted constraints on parameters from them.
We will first present the results for the idealised case, then move on to the noisy case, and finally the Euclid-like case.
We note all the contours we present in this section 68\% confidence intervals.
In the end of this section, we also discuss about how these three cases differentiate from each other, and what we can learn from the comparison between them.

\subsection{Idealised Case}\label{sec:idealised_results}

\begin{figure*}[htp]
\centering
    \includegraphics[width=0.98\textwidth]{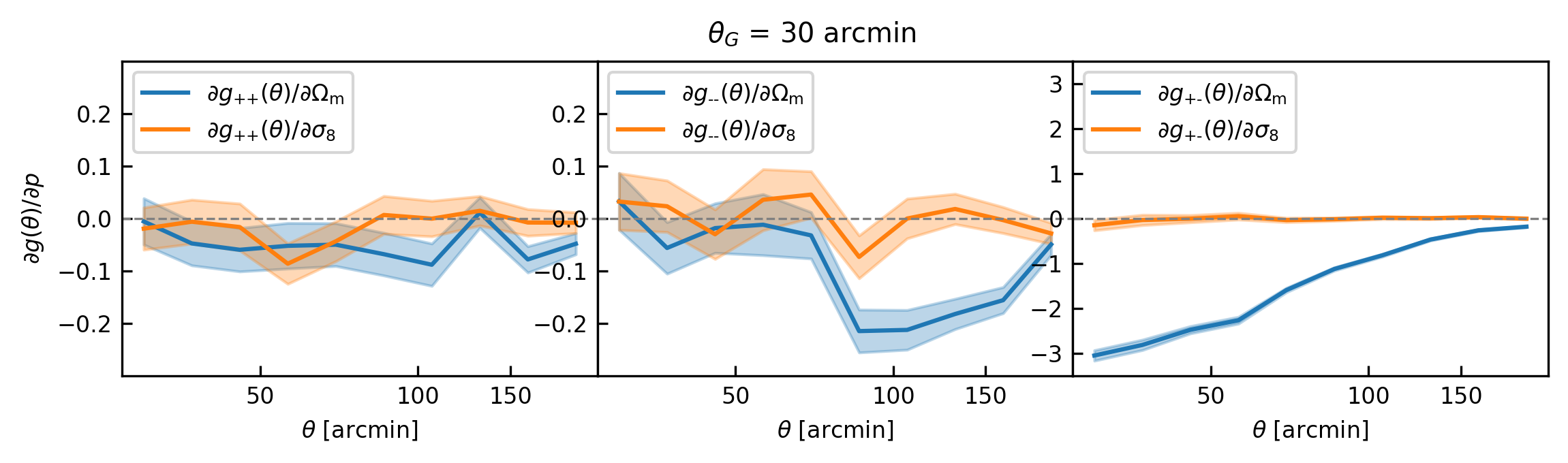}
    \caption{Derivatives of D-2PCFs in Gaussian maps with respect to $\Omega_{\rm m}$ and $\sigma_8$ (blue and orange respectively). 
    The shaded regions indicate the uncertainty of the mean over 100 realisations.
    }
    \label{fig:derivs_2pcf}
\end{figure*}

In Fig.~\ref{fig:derivs_2pcf}, we show the Fisher derivatives of D-2PCFs in Gaussian maps with respect to $\Omega_{\rm m}$ and $\sigma_8$ at smoothing scale $\theta_G = 30$ arcmin.
A clear distinction emerges between the two sets of derivatives: those with respect to $\sigma_8$ are consistent with zero, indicating that defect clustering is largely insensitive to $\sigma_8$; in contrast, the derivatives with respect to $\Omega_{\rm m}$ are substantial, demonstrating that defect clustering is sensitive to $\Omega_{\rm m}$.

We then focus on the dependence of the D-2PCFs on $\Omega_{\rm m}$. 
We find that the positive-negative correlation exhibits the strongest response, especially on small scales. 
In contrast, the positive--positive and negative--negative correlations vary more mildly, as their amplitudes are bounded between $-1$ and $0$. 
As $\Omega_{\rm m}$ increases, the positive-negative 2PCF decreases significantly, while both same-charge correlations show a more modest decline, with the negative-negative term displaying a slightly stronger suppression than the positive-positive one.

\begin{figure*}[htp]
    \centering
    \includegraphics[width=0.9\textwidth]{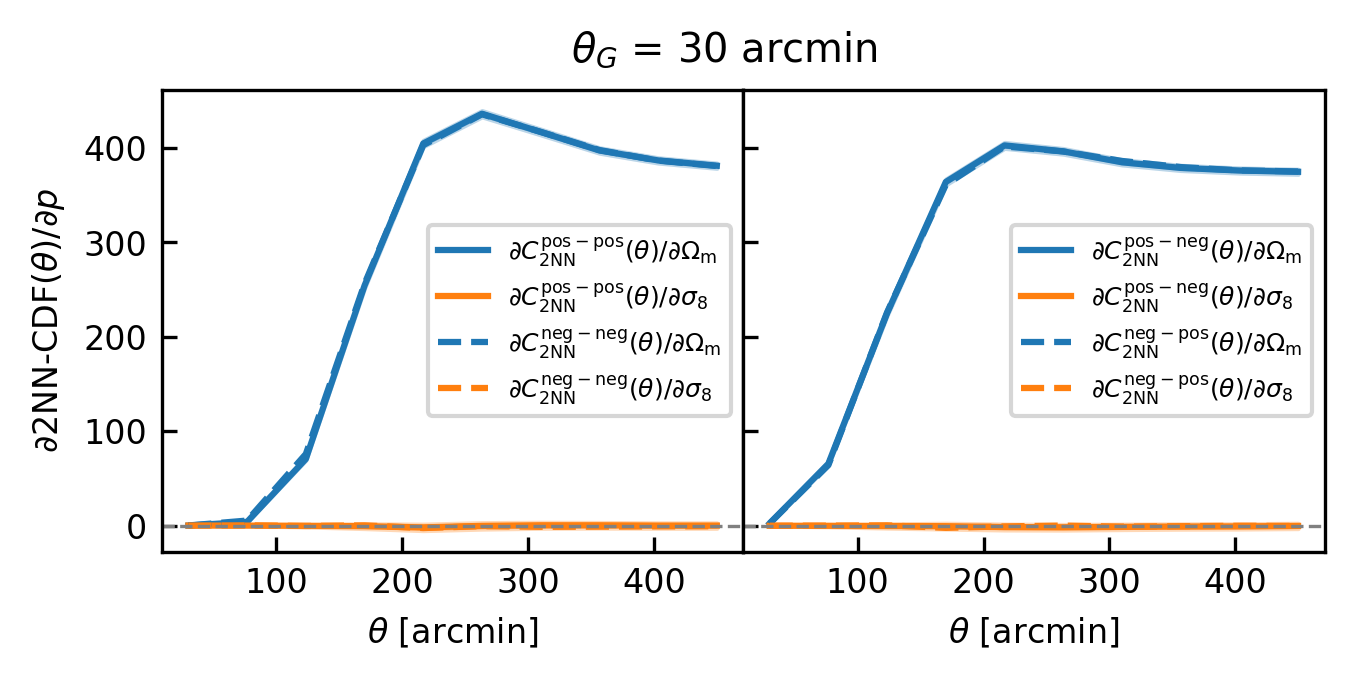}
    \caption{Derivatives of the 2NN-CDFs of defects with respect to $\Omega_{\rm m}$ (blue) and $\sigma_8$ (orange). The left and right panels show the same-charge and opposite-charge 2NN-CDFs, respectively. Shaded regions indicate the uncertainty on the mean across 100 realisations ($\lesssim 5$, thus barely visible). Positive and negative 2NN results are distinguished by solid and dashed lines, respectively.}
    \label{fig:derivs_nns}
\end{figure*}

In Fig.~\ref{fig:derivs_nns}, we show the derivatives of kNN-CDFs of defects with respect to $\Omega_{\rm m}$ and $\sigma_8$.
We here only show the positive 2NN results as examples, while other kNN statistics show similar behaviours.  
As with the D-2PCFs, the kNN-CDFs also show a clear sensitivity to $\Omega_{\rm m}$ and little dependence to $\sigma_8$. Again, we attribute this to $\Omega_{\rm m}$ alone impacting the large-scale shear phase information which the defects are sensitive to.
We also see the kNN-CDFs for positive and negative defects have the same cosmological sensitivities, this feature also holds for lognormal maps.

\begin{figure*}[htp]
    \centering
    \includegraphics[width=0.9\textwidth]{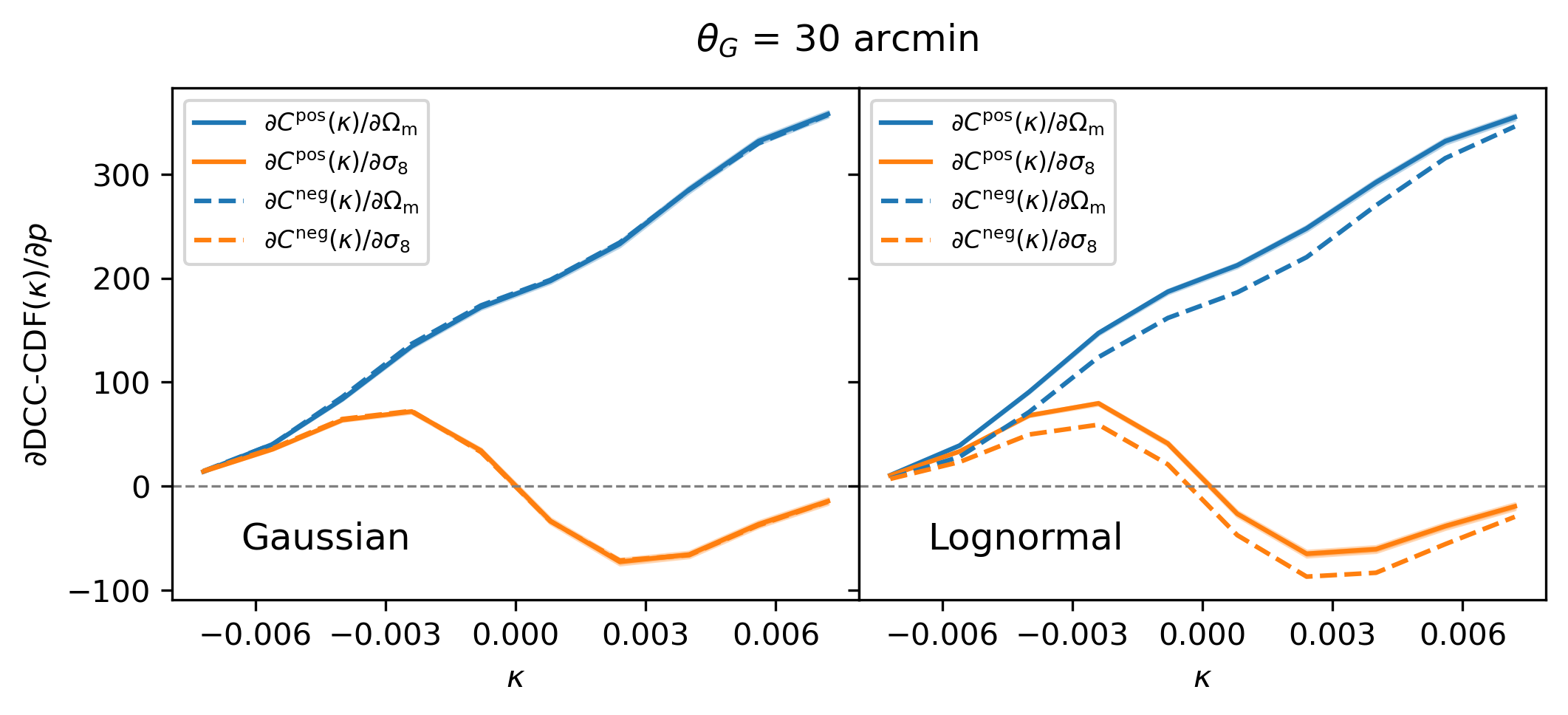}
    \caption{Derivatives of the DCC-CDFs with respect to $\Omega_{\rm m}$ (blue) and $\sigma_8$ (orange). The left and right panels show results for Gaussian and lognormal maps, respectively. Shaded regions indicate the uncertainty on the mean over 100 realisations ($\lesssim 3$, thus barely visible). Positive and negative defect results are distinguished by solid and dashed lines, respectively.}
    \label{fig:derivs_dcdf}
\end{figure*}

In Fig.~\ref{fig:derivs_dcdf}, we show the derivatives of DCC-CDFs with respect to $\Omega_{\rm m}$ and $\sigma_8$.
As we learnt from Sec.~\ref{sec:DCC-CDF}, the DCC-CDFs are different between positive and negative charges in non-Gaussian fields, the figure shows that their cosmological sensitivities are also different.
Interestingly, the cosmology dependencies of the DCC-CDFs are significant for both $\Omega_{\rm m}$ and $\sigma_8$, while the dependencies differ between the two parameters.
For $\Omega_{\rm m}$, the DCC-CDFs increase across all convergence values as $\Omega_{\rm m}$ rises, reflecting the overall increase in the total number of defects. For $\sigma_8$, the DCC-CDFs also change noticeably; however, this variation is primarily driven by the overall convergence distribution, since $\sigma_8$ mainly affects the fluctuation amplitude of the field. The defects themselves, being determined by the shear orientation, remain largely insensitive to $\sigma_8$, so the observed changes in the DCC-CDF mainly reflect the shift in the underlying convergence field rather than a direct effect on defect locations.

\begin{figure*}[htp]
    \centering
    \includegraphics[width=0.48\textwidth]{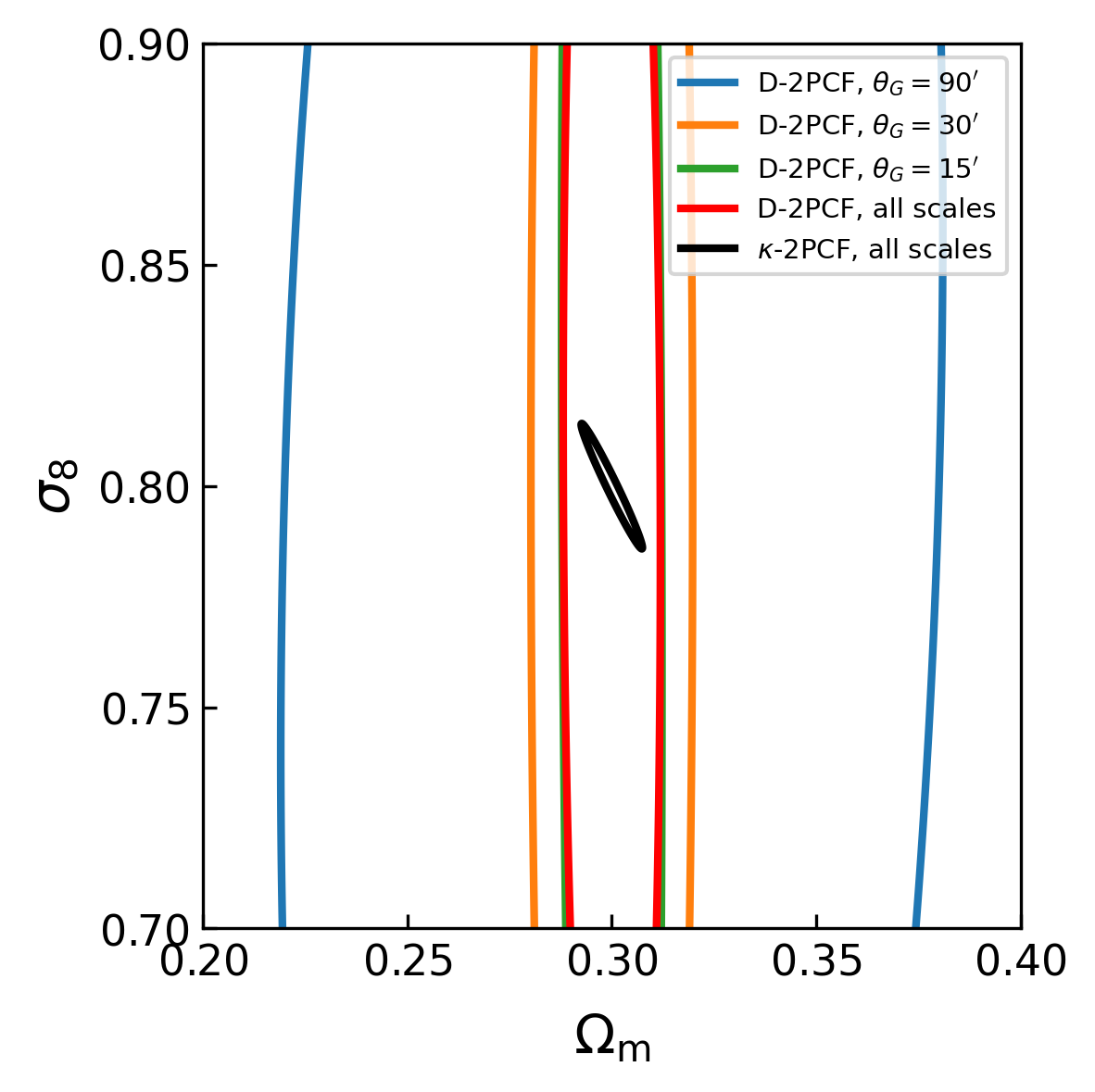}
    \includegraphics[width=0.48\textwidth]{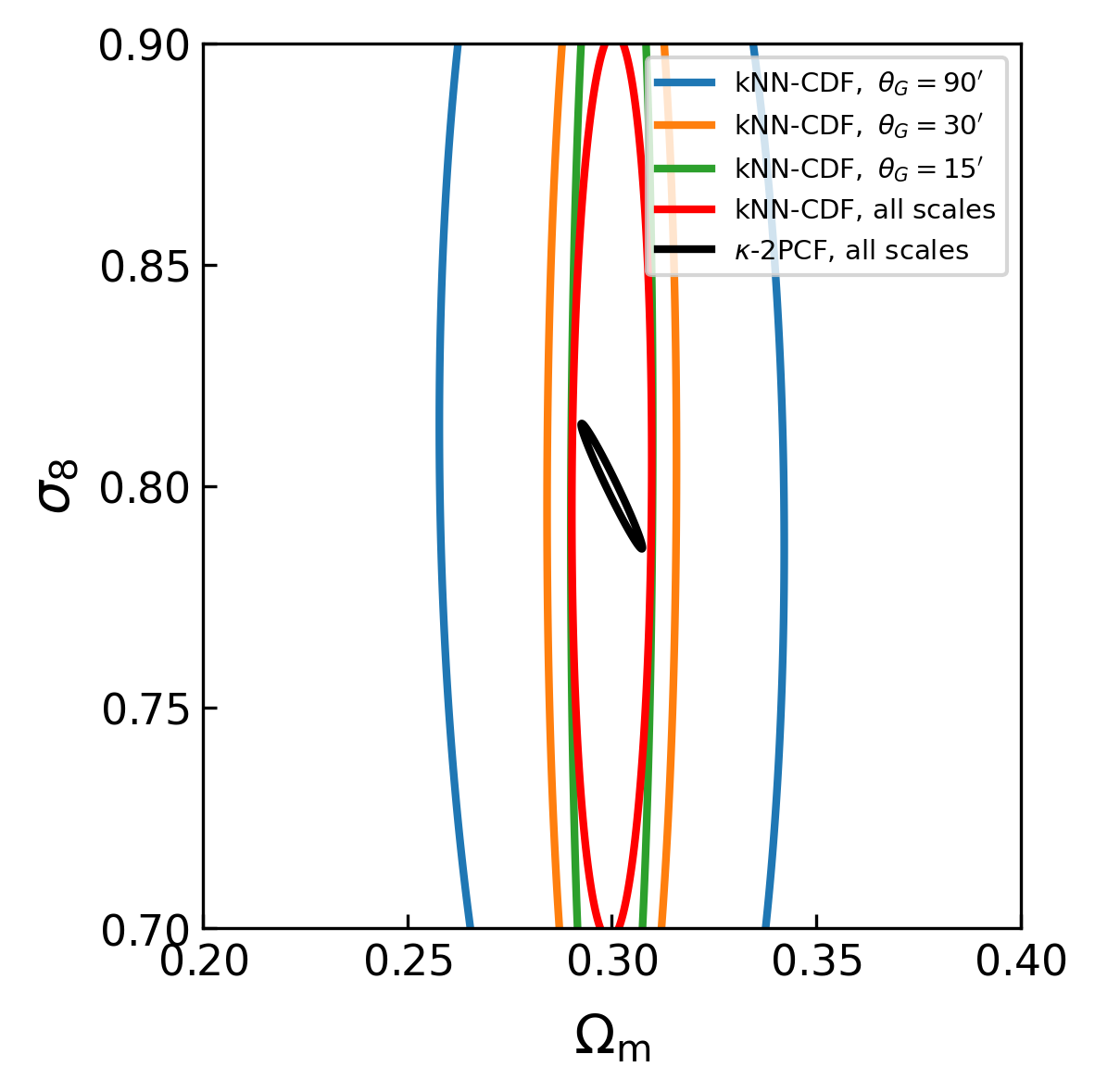}
    \includegraphics[width=0.48\textwidth]{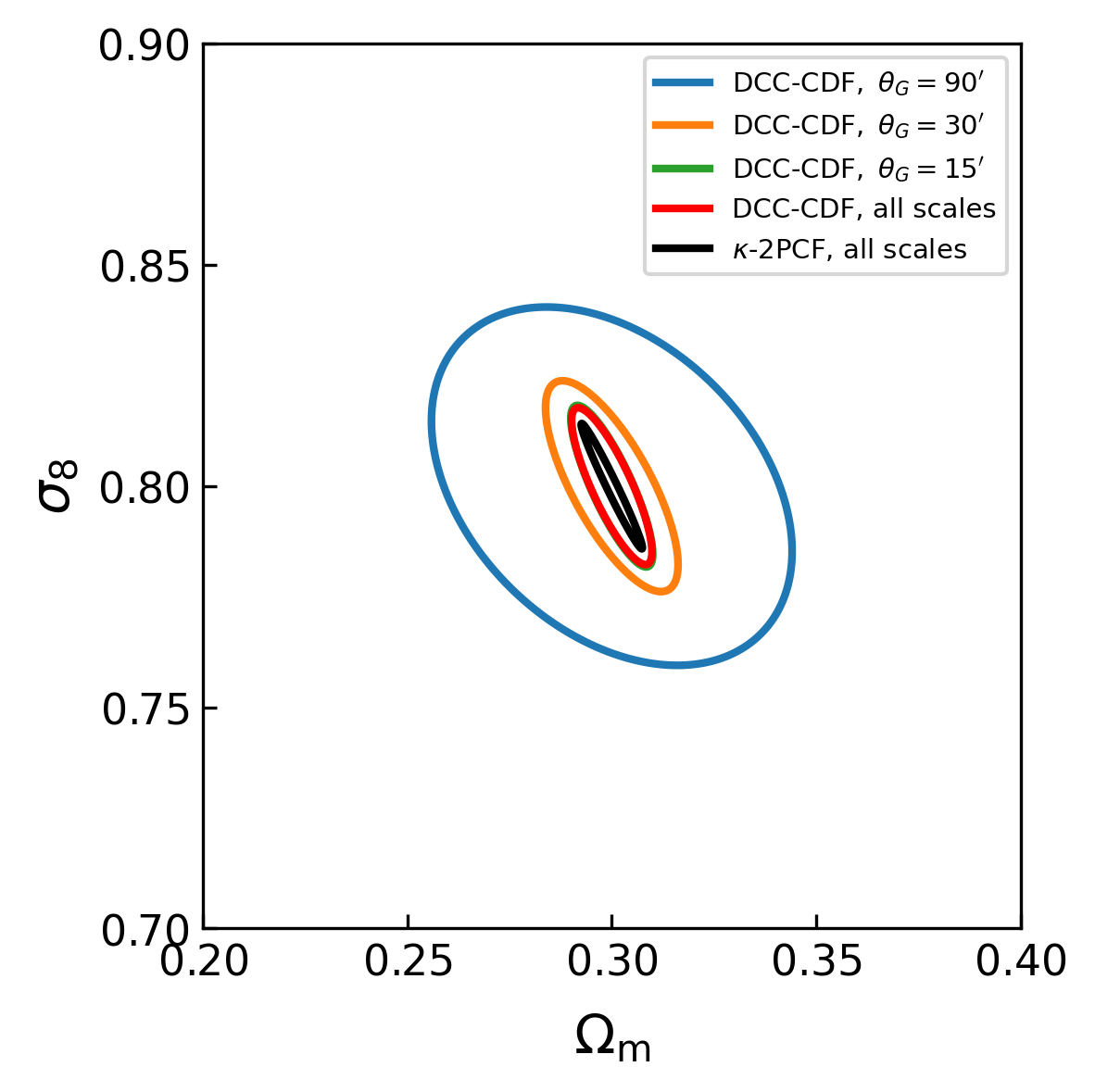}
    \includegraphics[width=0.48\textwidth]{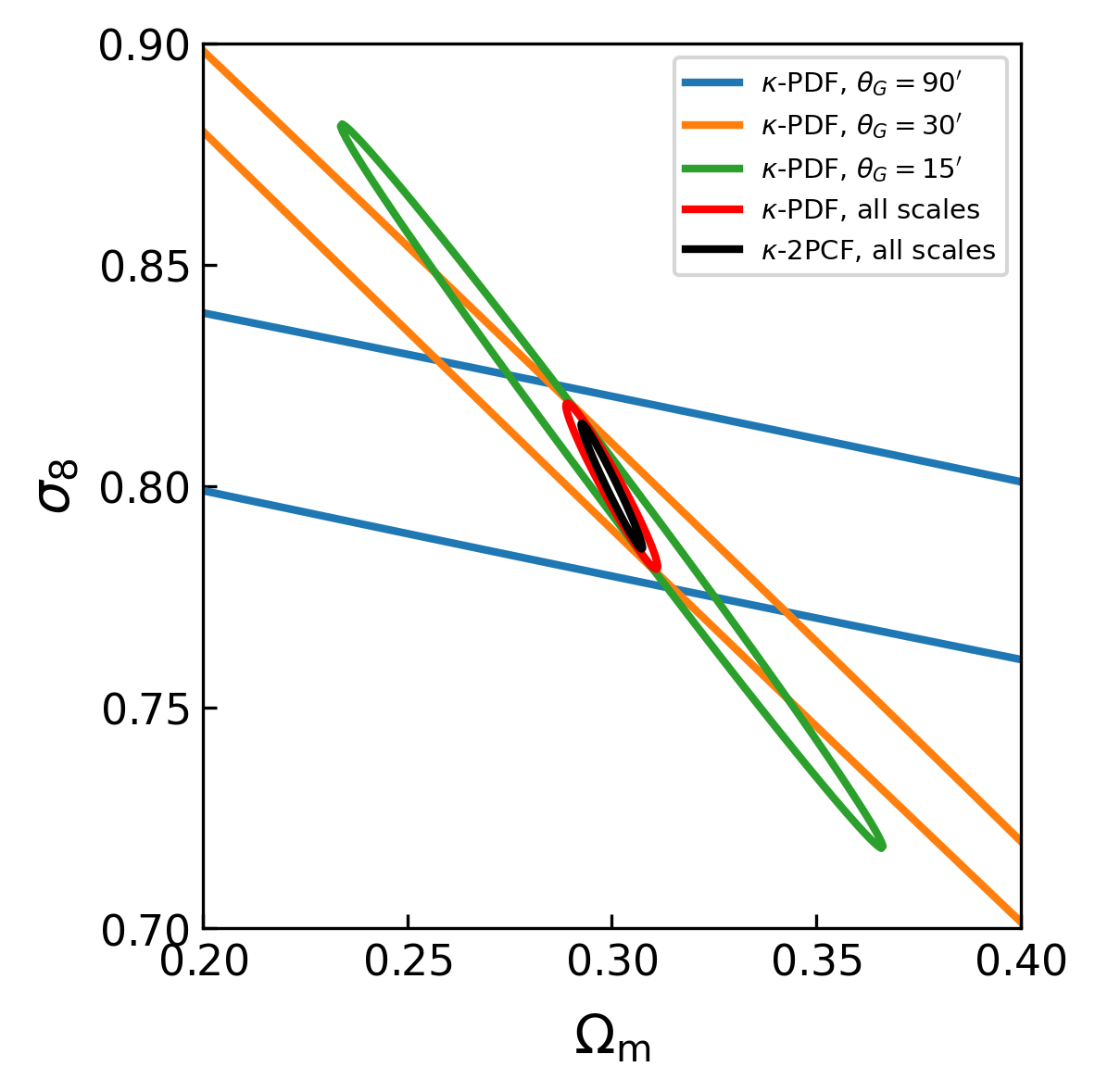}
    \caption{Forecasted constraints on $(\Omega_{\rm m}, \sigma_8)$ from different statistics in the Gaussian case. From top to bottom and left to right are D-2PCFs, kNN-CDFs of defects, DCC-CDFs, and overall convergence PDF respectively.
            The different colours represent different smoothing scales, while the black contour is the result from $\kappa$-2PCF for reference.
            }
    \label{fig:fisher_gaussian}
\end{figure*}

\begin{figure*}[htp]
    \centering
    \includegraphics[width=0.48\textwidth]{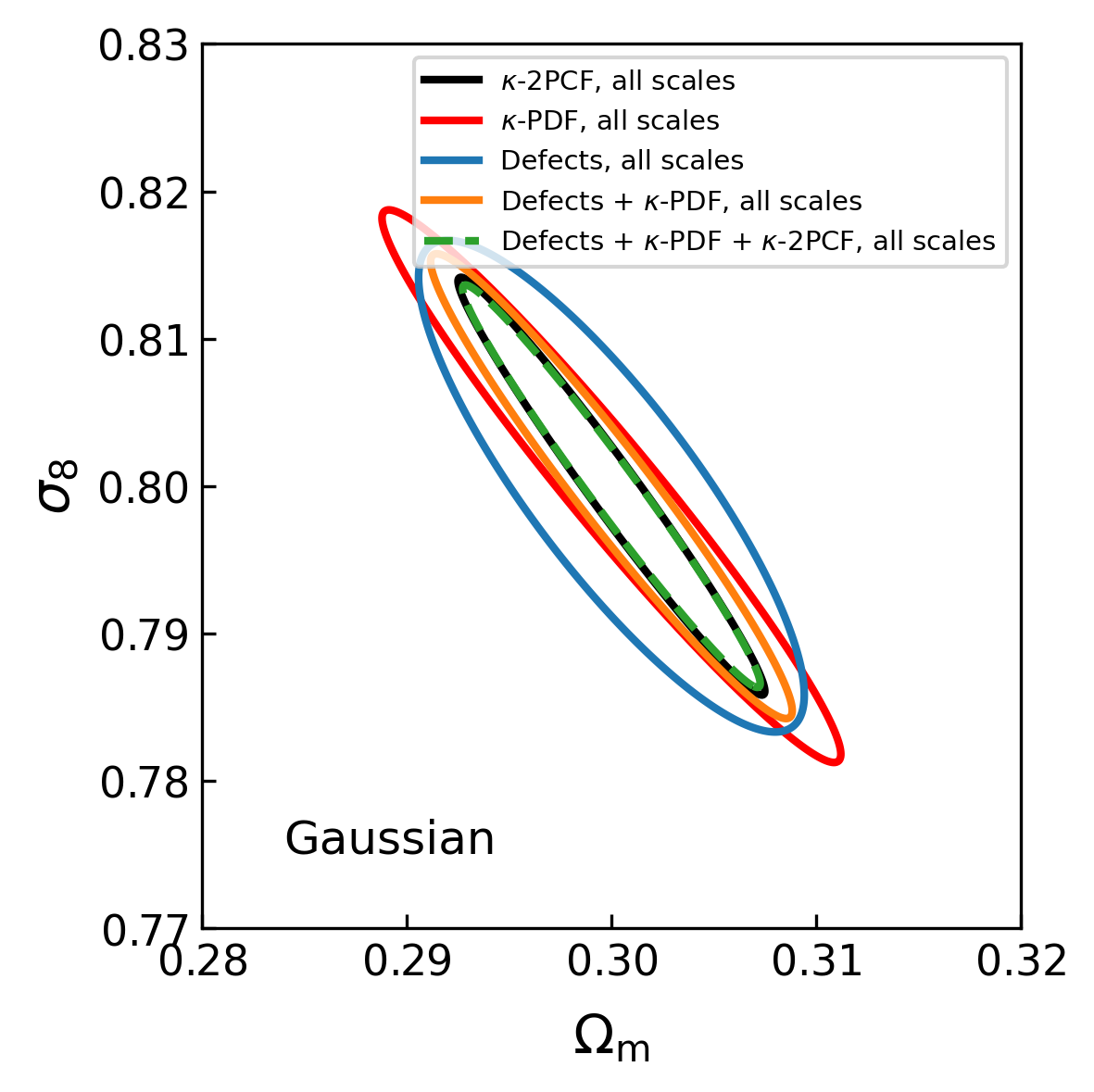}
    \includegraphics[width=0.48\textwidth]{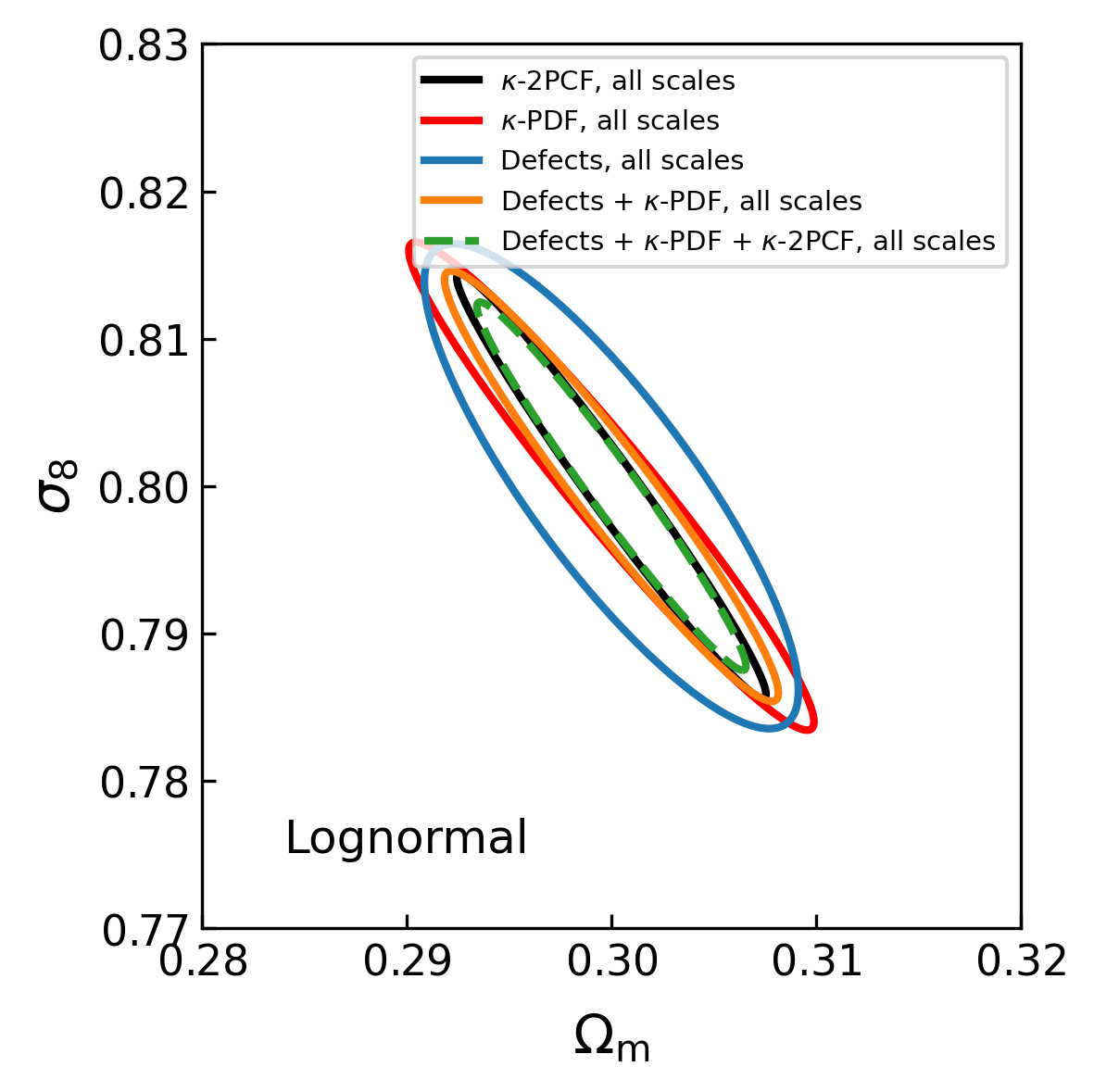}
    \caption{Forecasted constraints on $\Omega_{\rm m}$ and $\sigma_8$ from combinations of statistics, shown for the Gaussian case (left) and the lognormal case (right).
    Here and hereafter, ``Defects'' refers to the combination of all defect statistics.}
    \label{fig:fisher_lognormal}
\end{figure*}

With the Fisher derivatives at hand, we compute the Fisher matrix introduced in Sec.~\ref{sec:fisher_formalism} and forecast the constraints on $(\Omega_{\rm m}, \sigma_8)$ from different statistics in this case.
In Fig.~\ref{fig:fisher_gaussian}, we compare constraints on parameters from different statistics in the Gaussian setup.
In each panel, we show the results from different smoothing scales for each statistic, while the black contour is the result from $\kappa$-2PCF for reference.
Comparing the first two panels on the top row, we see the defect kNN-CDFs (upper-right) give noticably improved constraints on parameters compared to the 2PCF (upper-left) of defects, especially on $\Omega_{\rm m}$, which is consistent with our previous explanation that the kNN-CDFs contain higher-order information of the defect clustering\footnote{We remind the reader that defect clustering can be non-Gaussian even when the underlying field as a whole is Gaussian.} while the 2PCF only contains second-order information.
Comparing the two lower panels, we see the DCC-CDFs (bottom-left) can break the degeneracy between $\Omega_{\rm m}$ and $\sigma_8$ to some extent, while the overall convergence PDFs (bottom-right) need to be combined from different smoothing scales to break the degeneracy, which is consistent with previous research \cite{2023MNRAS.520.1721G, 2021MNRAS.505.2886B, 2012PhRvD..85j3513K}.
This arises from the fact that the DCC-CDFs contain both amplitude and phase information of the shear field, while the overall convergence PDF only contains amplitude information.

The other noticeable point is that none of the higher-order statistics can outperform the $\kappa$-2PCF in the Gaussian setup, which is expected as the 2PCF contains all information in a Gaussian field.
Moreover, as we can see from the defect statistics, combining different smoothing scales doesn't improve the constraints significantly, indicating that most of the information is contained in the smallest smoothing scale.
This allows us to use only one smoothing scale in the following noisy and Euclid-like cases.

In Fig.~\ref{fig:fisher_lognormal}, we compare parameter constraints obtained from different statistics in Gaussian and lognormal setups.
In the Gaussian case, the $\kappa$-2PCF provides the tightest parameter constraints among all statistics, as expected given that the 2PCF captures the complete information content of a Gaussian field. The combination of defect and $\kappa$-PDF statistics (orange) approaches the near-optimal constraining power exhibited by the $\kappa$-2PCF (black).
In the lognormal case, the $\kappa$-2PCF still gives the tightest constraints on parameters, but when combining the $\kappa$-2PCF with the higher-order statistics, the constraints are mildly improved by a factor of 10\%.
This indicates that the higher-order statistics contain complementary information to the $\kappa$-2PCF in non-Gaussian fields.
The constraints from the defect statistics are not improved significantly from the Gaussian to the lognormal case, which might be due to the fact that lognormal maps are not sufficiently non-Gaussian to fully exploit the information content of the defect statistics.

\subsection{Noisy Case}\label{sec:results_noisy}
As we have seen in the idealised case, the defect statistics show significant sensitivity to $\Omega_{\rm m}$ but little dependence on $\sigma_8$.
Here we investigate how the presence of noise in the maps will impact the cosmological sensitivities of defect statistics as it has been shown many times in the literature that higher-order statistics can be more sensitive to noise than the 2PCF \cite{2023A&A...675A.120E,2018MNRAS.480.5529G,2023MNRAS.520.1721G}.

In this case, we consider two defect statistics in the lognormal random maps: the kNN-CDFs of defects and the DCC-CDFs.
We do not include the D-2PCFs in this setup because we have already shown in the idealised case that the kNN-CDFs (Fig.~\ref{fig:fisher_gaussian}) are sufficient to capture the defect clustering information.

\begin{figure*}[htp]
    \centering
    \includegraphics[width=0.9\textwidth]{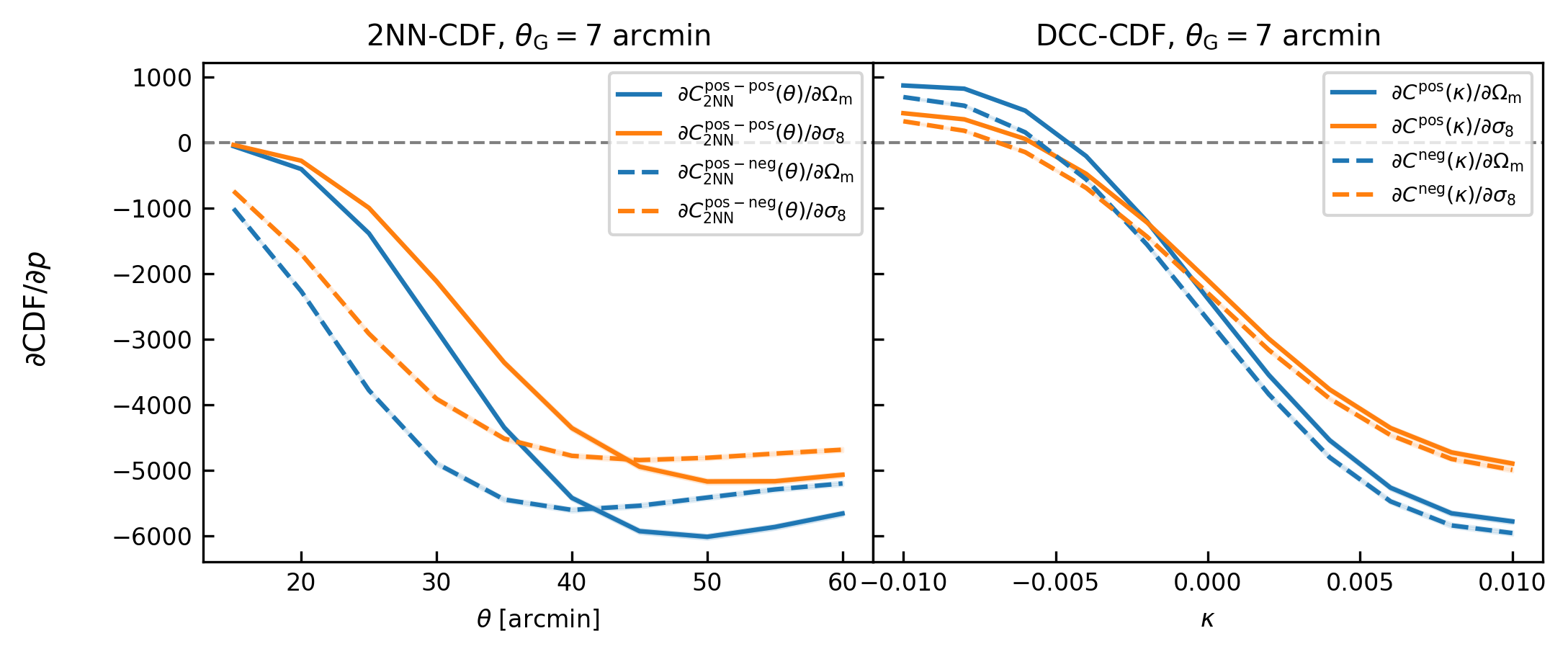}
    \caption{Derivatives of the 2NN-CDFs (left) and DCC-CDFs (right) of defects with respect to $\Omega_{\rm m}$ (blue) and $\sigma_8$ (orange) in the noisy case. The shaded regions (thinner than the width of the line) indicate the uncertainty on the mean across 100 realisations. Positive and negative defect results are distinguished by solid and dashed lines, respectively.}
    \label{fig:derivs_nns_dcdf_s7}
\end{figure*}

In Fig.~\ref{fig:derivs_nns_dcdf_s7}, we show the derivatives of the 2NN-CDFs and DCC-CDFs of defects with respect to $\Omega_{\rm m}$ and $\sigma_8$ in the noisy case.
We only show the 2NN-CDFs with respect to positive defects as examples, while other kNN statistics show similar behaviours as in the left panel of Fig.~\ref{fig:derivs_nns_dcdf_s7}.
We see that the derivatives with respect to both parameters are significant.
In particular, besides the sensitivity to $\Omega_{\rm m}$ already present in the idealised case, we also see a sensitivity to $\sigma_8$, which is not the case for the 2NN-CDFs in the idealised case (Fig.~\ref{fig:derivs_nns}).
Most interestingly, the derivatives are flipped in sign compared to the idealised case, which is due to the existence of noise in the maps: the noise raises the relative contribution of small scales and thus the defect number density, so that increasing $\Omega_{\rm m}$ or $\sigma_8$ now enhances the signal part (relatively large scale) of the power spectrum and decreases the defect number density instead.
Comparing the right panel with Fig.~\ref{fig:derivs_dcdf}, the DCC-CDFs decrease with both parameters in the noisy case, which is also due to the reduced defect numbers, whereas in the idealised case they increase with $\Omega_{\rm m}$ and their $\sigma_8$ dependence mainly follows that of the underlying convergence distribution.
We discuss these behaviours in more detail in Sec.~\ref{sec:discussion}.

\begin{figure*}[htp]
    \centering
    \includegraphics[width=0.48\textwidth]{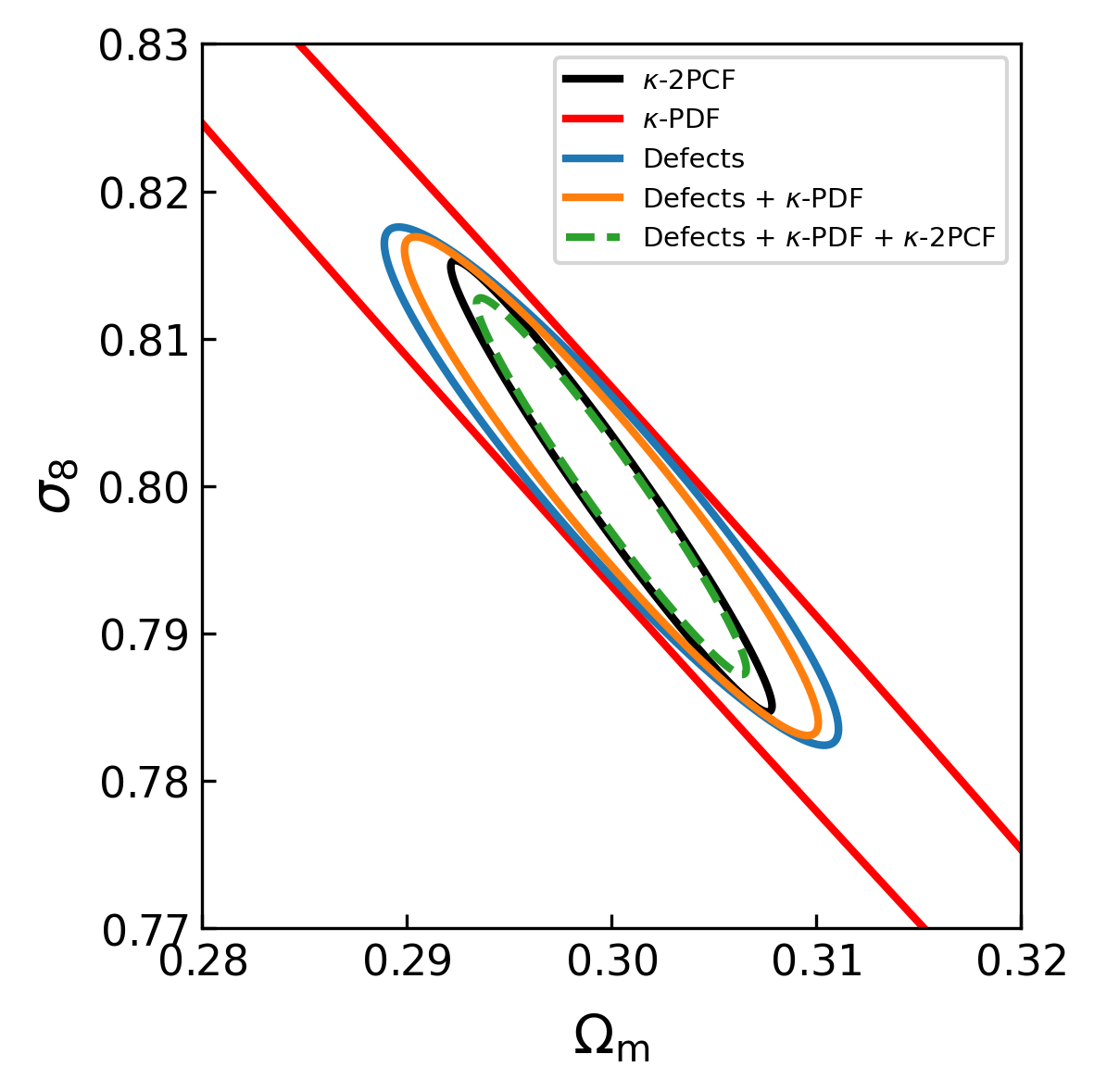}
    \caption{Forecasted constraints on $(\Omega_{\rm m}, \sigma_8)$ from different statistics in the noisy (lognormal) case.}
    \label{fig:fisher_s7}
\end{figure*}

The corresponding Fisher contours are shown in Fig.~\ref{fig:fisher_s7}.
We see that the constraints from both the kNN-CDFs and DCC-CDFs are significantly improved compared to the idealised case, especially for $\sigma_8$, which is consistent with the emergence of sensitivity to $\sigma_8$ in the derivatives.
Both kNN-CDFs and DCC-CDFs give constraints outperforming the $\kappa$-PDF, and comparable to the $\kappa$-2PCF.
Just like the idealised case, when combining the higher-order statistics with the $\kappa$-2PCF, the constraints are also improved by approximately 10\% for both parameters.
The reason why the constraints from higher-order statistics are still not tighter than the $\kappa$-2PCF might be due to the non-Gaussianity of the lognormal maps is not strong enough to hide information beyond the 2PCF, and the scales we consider are not small enough to fully exploit the non-Gaussian information content of the maps.
We thus proceed to the Euclid-like case to further explore the constraining power of defect statistics in more realistic and non-Gaussian maps.

\subsection{Euclid-like Case}\label{sec:results_euclid}
The Euclid-like case represents another step towards more realistic maps, using the N-body simulations from the Euclid HOWLS analysis \cite{2023A&A...675A.120E} and hence incorporating more small-scale information and more complex non-Gaussian features than the previous cases considered.
As with the noisy case, we focus on the kNN-CDFs of defects and DCC-CDFs in this setup, as they are sufficient to capture the defect information.

\begin{figure*}[htp]
    \centering
    \includegraphics[width=0.9\textwidth]{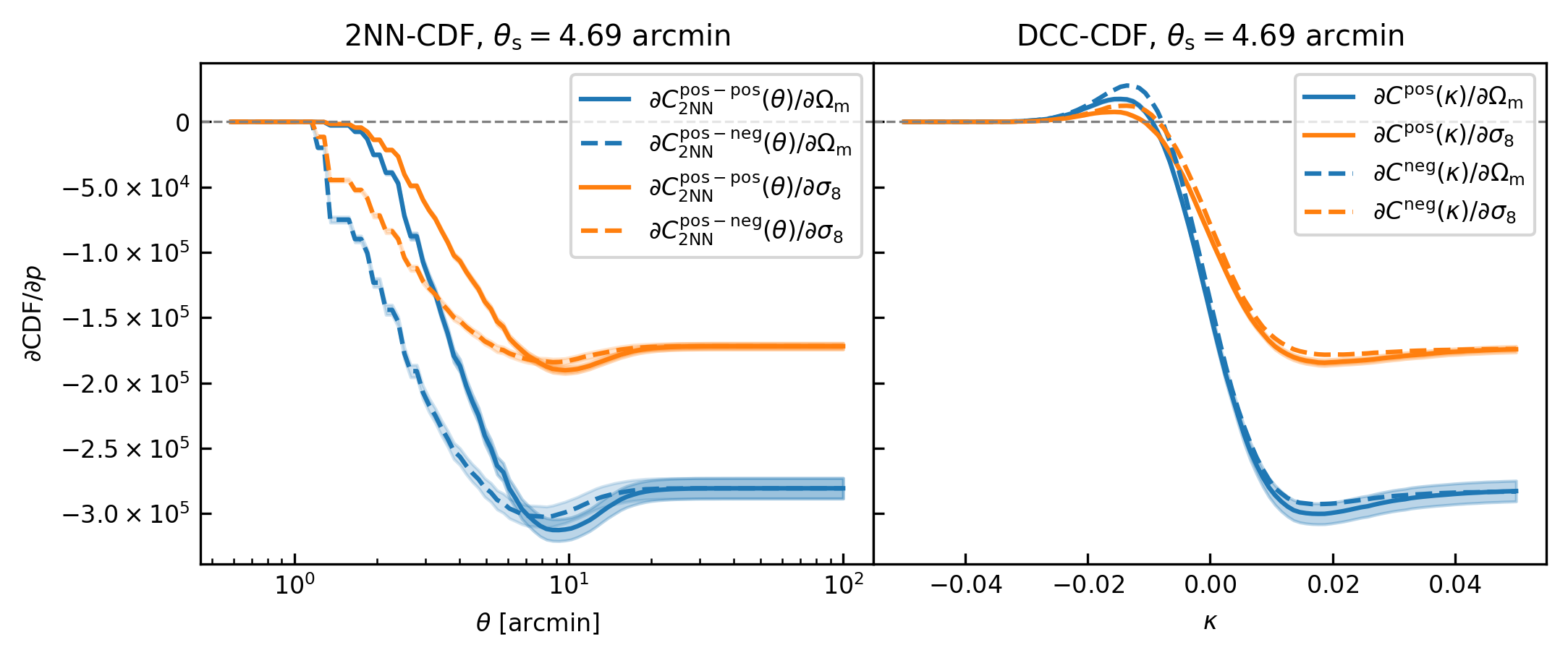}
    \caption{The same as Fig.~\ref{fig:derivs_nns_dcdf_s7} but for the Euclid-like maps.}
    \label{fig:howls_derivs_nns_cdf}
\end{figure*}

Fig.~\ref{fig:howls_derivs_nns_cdf} presents the derivatives of the 2NN-CDFs and DCC-CDFs with respect to $\Omega_{\rm m}$ and $\sigma_8$ in the Euclid-like case.
Similarly to Fig.~\ref{fig:derivs_nns_dcdf_s7}, the derivatives with respect to both parameters are significant, and they also show a flipped sign compared to the idealised case, which is again due to the presence of noise in the maps.
We explain this in more detail in Sec.~\ref{sec:discussion}.
We note there are some discrete jumps in the small-scale derivatives, which are due to discreteness separation of defects at small scales caused by pixelisation.
Our data selection criteria will remove those correlated bins, so they do not affect the final constraints.

\begin{figure*}[htp]
    \centering
    \includegraphics[width=0.48\textwidth]{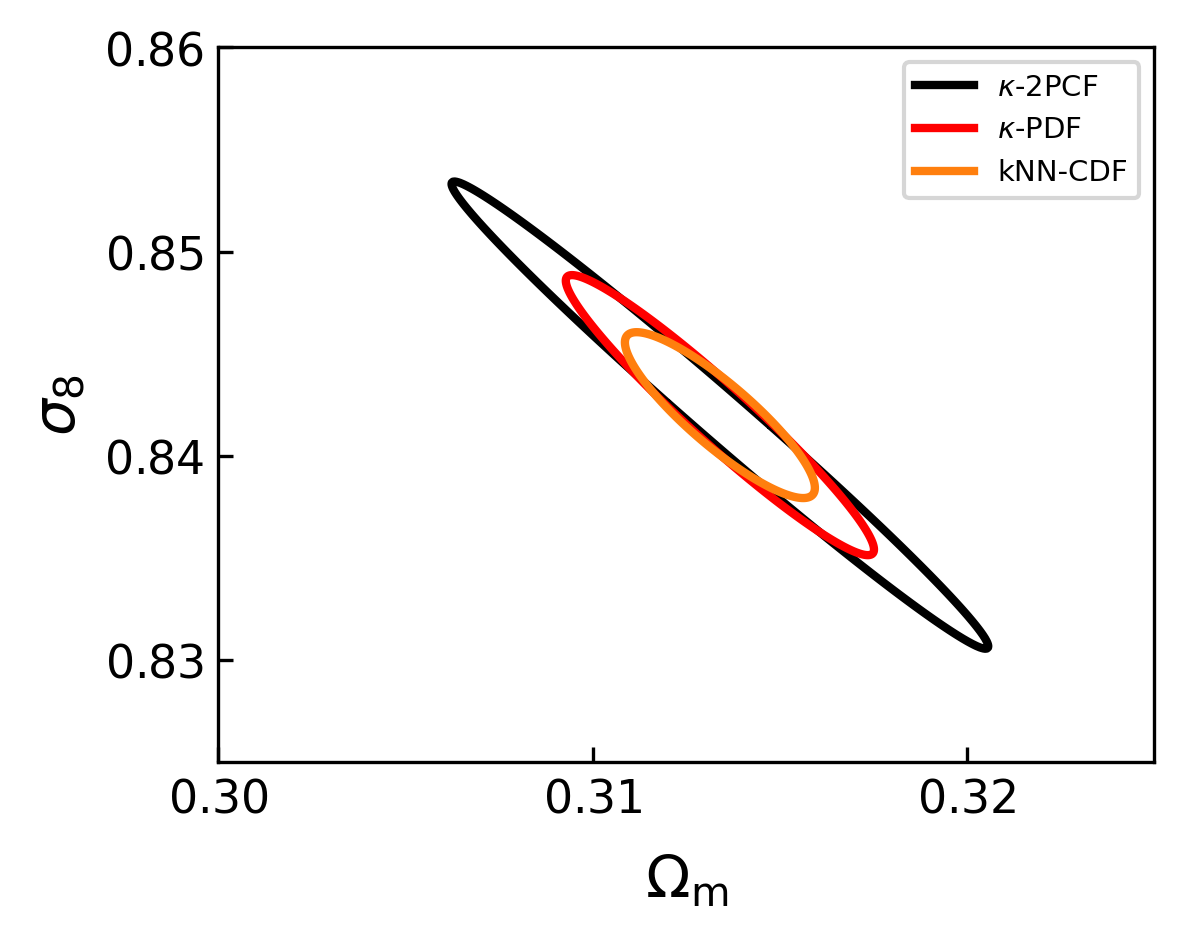}
    \includegraphics[width=0.48\textwidth]{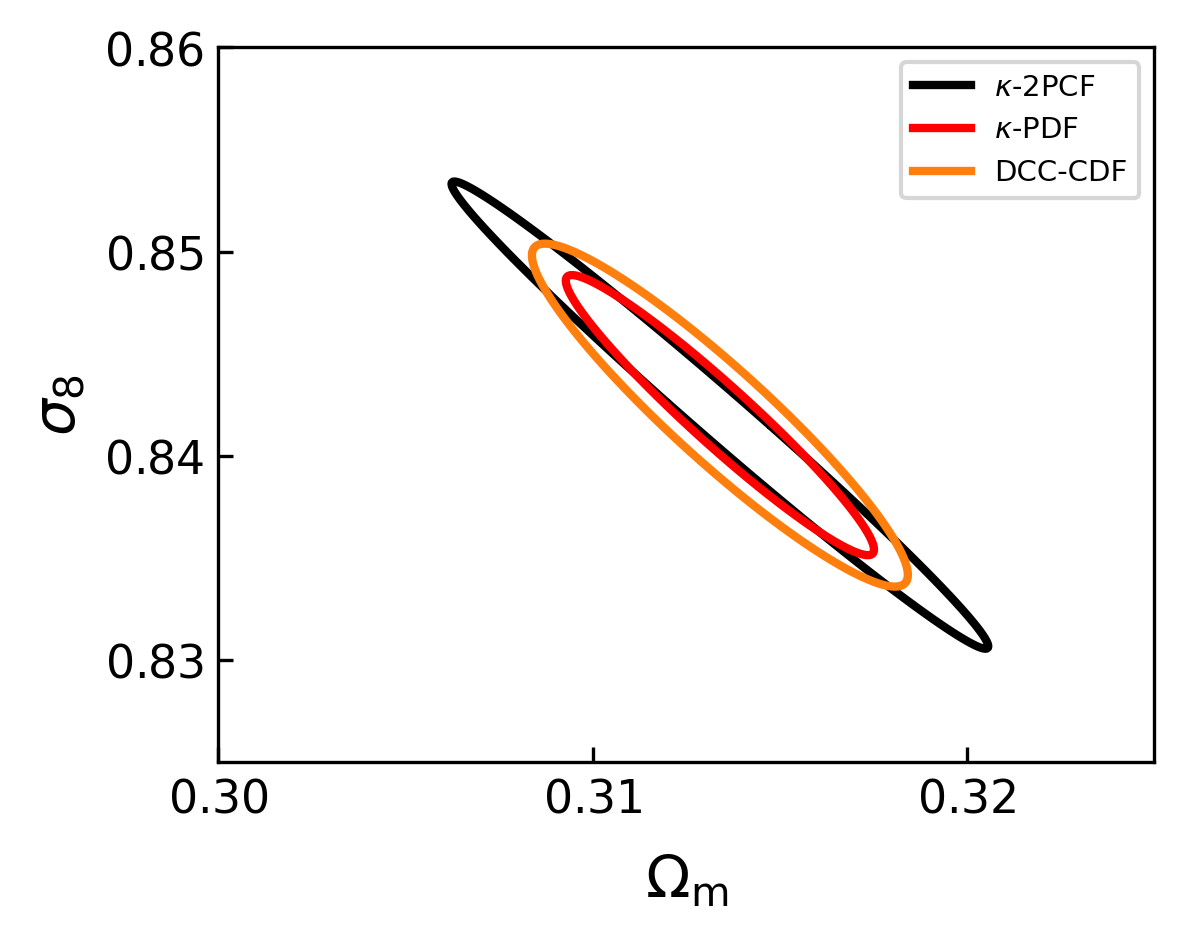}
    \includegraphics[width=0.48\textwidth]{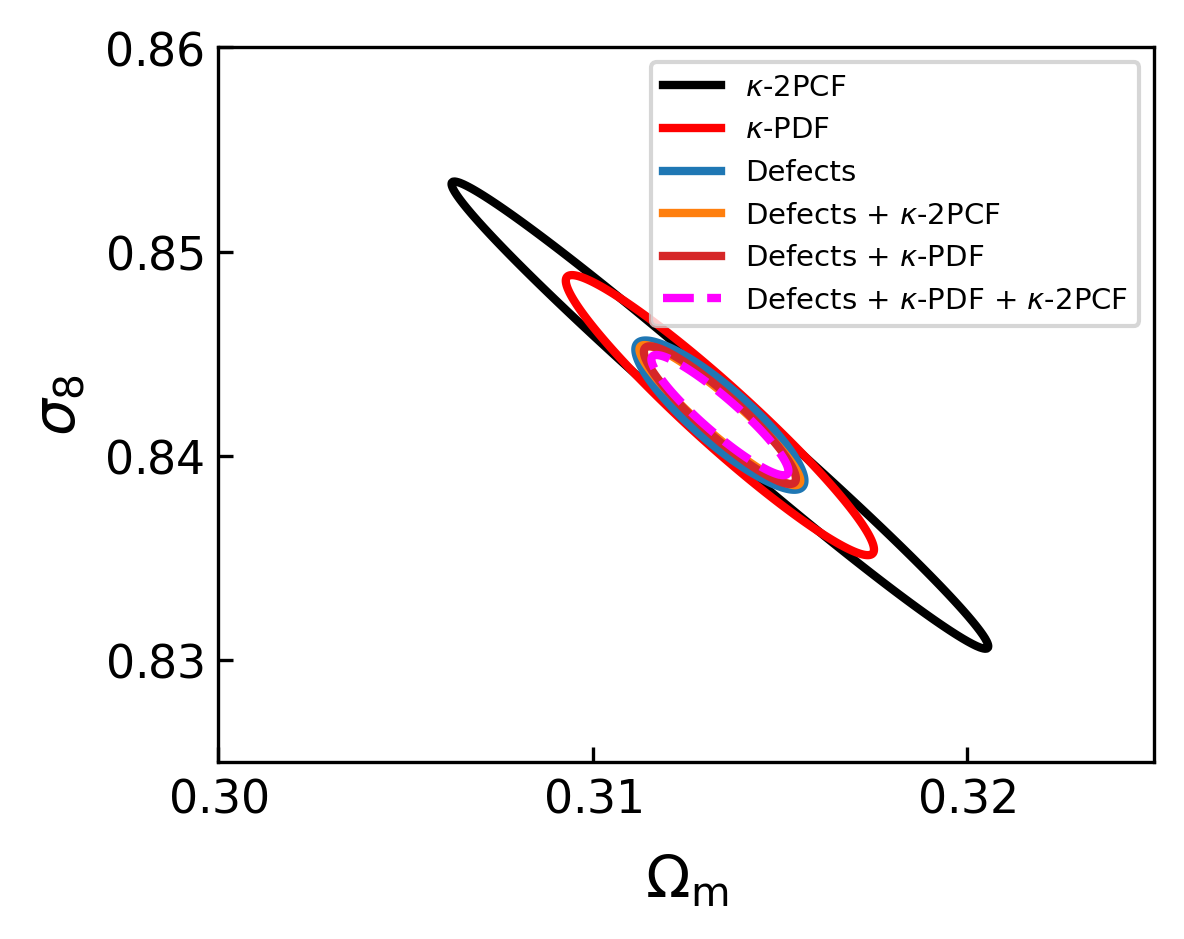}
    \caption{Forecasted constraints on $(\Omega_{\rm m}, \sigma_8)$ from different statistics in the Euclid-like Fisher analysis. From left to right are kNN-CDFs of defects and DCC-CDFs, while the bottom panel shows all defect statistics combined. For comparison, the constraints from the $\kappa$-2PCF (black) and the $\kappa$-PDF (red) are also shown.}
    \label{fig:fisher_howls}
\end{figure*}

\begin{figure}[htp]
    \centering
    \includegraphics[width=0.9\textwidth]{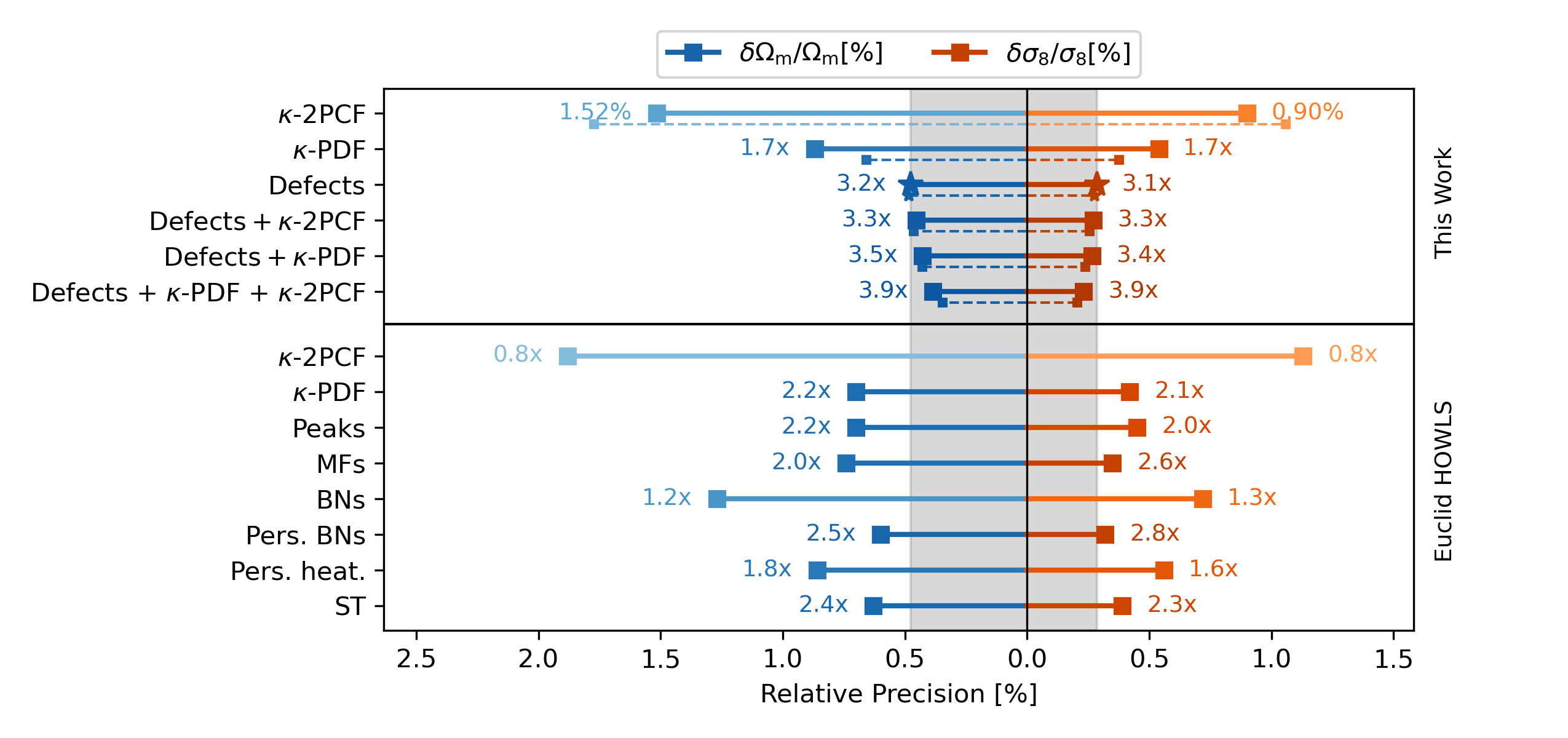}
    \caption{Forecasted relative precision of $1\sigma$ constraints on $\Omega_{\rm m}$ (blue bars) and $\sigma_8$ (orange) from different statistics in the Euclid-like Fisher analysis.
    The results in the upper panel are obtained in this work, while those in the lower panel are taken from the Euclid HOWLS analysis based on other statistics. 
    The solid bars in the upper panel are results from our fiducial analysis while the dashed bars are obtained using analysis settings consistent with the Euclid HOWLS configuration.
    The intensity of the colours represents the constraining power of different statistics; the darker the colour, the tighter the constraints.
    The numbers on the solid bars indicate the improvement factors of the constraints compared to the $\kappa$-2PCF following our settings (the numbers on the $\kappa$-2PCF are the exact relative precision of its constraints).
    The shaded region is the constraints from the defect statistics, which is the main focus of this work.
    }
    \label{fig:bars}
\end{figure}

The results of the Fisher analysis are shown in Fig.~\ref{fig:fisher_howls}, where we compare the forecasted constraints on $(\Omega_{\rm m}, \sigma_8)$ from defect statistics, also including the $\kappa$-PDF measured at the same smoothing scale ($\theta_{\rm s} = 4.69$ arcmin), and the $\kappa$-2PCF for reference.
For the defect kNN-CDFs, the 68\% constraints on $\Omega_{\rm m}$ and $\sigma_8$ are tightened by factors of 3.2 and 3.1, respectively, relative to the $\kappa$-2PCF, and they exhibit a consistent degeneracy direction with both the overall $\kappa$-PDF and the $\kappa$-2PCF. 
The DCC-CDFs, however, yield slightly weaker constraints on both parameters due to the limited number of defect samples in the $5\times5$ deg$^2$ footprint maps, which introduces substantial uncertainties in the data vectors. 
The bottom panel shows the constraints from all defect statistics (green) and their combinations with other probes: Defect + $\kappa$-2PCF (dark red), Defect + $\kappa$-PDF (light red), and Defect + $\kappa$-2PCF + $\kappa$-PDF (dashed black). Defect statistics alone provide the tightest constraints among the statistics considered, and their combination with other probes further tightens the confidence contours, demonstrating that defect statistics supply valuable and complementary cosmological information. 
This is in clear contrast to the Gaussian and noisy (lognormal) cases, where the defect statistics could at best match the $\kappa$-2PCF (Figs.~\ref{fig:fisher_gaussian} and \ref{fig:fisher_s7}). In the Gaussian case the 2PCF captures all the information of the field, while the lognormal maps are not sufficiently non-Gaussian for the defect statistics to add information on the probed scales; in the Euclid-like case, instead, the maps contain stronger non-Gaussian features and more small-scale information, which are efficiently captured by the defect statistics. 
When combined with $\kappa$-2PCF, the constraints are improved by a factor of 3.3 compared to the $\kappa$-2PCF alone, and including all higher-order statistics together with the $\kappa$-2PCF yields an enhancement of 3.9 for both parameters.
In Fig.~\ref{fig:bars}, we summarise the forecasted $\Omega_{\rm m}$ and $\sigma_8$ from all different statistics in the Euclid-like Fisher analysis in more details.

A possible physical interpretation of the strong constraining power of the
defect statistics can be obtained from the analytic expression for the
defect number density. As shown in Eq.~\ref{eq:defect_density}, the defect
density depends on the ratio of the spectral moments
$\xi_{\psi,0}^{(6)}/\xi_{\psi,0}^{(4)}$, which can equivalently be
expressed in terms of the convergence field as
$\langle|\nabla\kappa|^2\rangle/\langle\kappa^2\rangle$. The numerator
therefore introduces sensitivity to flexion, or equivalently to spatial
gradients of the convergence, in addition to the information contained in
the convergence amplitude. Because spatial derivatives introduce
additional powers of angular wavenumber, this ratio gives greater weight
to small-scale modes and hence to the nonlinear regime than statistics
based solely on the convergence field at second order. Defect statistics
can therefore provide an efficient way of extracting information from
the small-scale, nonlinear structure of the field. 

We also compare our results with the Euclid HOWLS analysis, which includes several higher-order statistics such as peak counts, Minkowski functionals, Betti numbers, persistent Betti numbers, persistent heatmaps, and scattering transform coefficients. We refer to the Euclid HOWLS analysis \cite{2023A&A...675A.120E} for detailed descriptions of these statistics and their corresponding parameter constraints.
In Fig.~\ref{fig:bars}, we present the relative constraints obtained from all statistics considered in this work, together with those reported in the Euclid HOWLS analysis. The solid bars in the upper panel correspond to the fiducial results obtained using our analysis settings.

We note, however, that our fiducial configuration does not fully follow the same procedure as the Euclid HOWLS analysis. In particular, we adopt different binning schemes and data selection criteria, as well as a different number and size of simulations for covariance estimation, all of which can affect the forecasted parameter constraints. To enable a more direct comparison, we therefore perform a re-analysis using settings consistent with the Euclid HOWLS configuration, for example by not splitting the SLICS maps into four subfields and by applying the same scale cuts to the $\kappa$-2PCF as in their analysis, while keeping other settings (such as the correlation and K-L divergence thresholds) unchanged. The corresponding results are shown as dashed bars in the figure.

As shown in Fig.~\ref{fig:bars}, the defect statistics outperform all other statistics in constraining $\Omega_{\rm m}$ and $\sigma_8$, highlighting the strong cosmological sensitivity of topological defects in weak-lensing convergence maps.
We also find that the constraints obtained from the defect statistics are only weakly affected by these changes in analysis settings, whereas the $\kappa$-PDF and $\kappa$-2PCF results become more consistent with those reported in the Euclid HOWLS analysis.

Finally, we comment on the qualitative difference between the constraints obtained in the Euclid-like case and those in the Gaussian and noisy cases, which is worth explaining before moving to the discussion.
The three setups differ both in the range of scales probed and in the amount of non-Gaussian information contained in the maps.
In the idealised case the noise-free maps are smoothed on scales of 15--90 arcmin, and the field is either Gaussian or only mildly non-Gaussian, so the $\kappa$-2PCF alone captures the cosmological information and none of the higher-order statistics can outperform it.
The noisy case probes smaller scales and the presence of noise makes the defect statistics sensitive to $\sigma_8$ as well, bringing them on par with the $\kappa$-2PCF, but the lognormal transformation of a Gaussian field does not generate sufficiently strong non-Gaussian features on the probed scales to allow the defect statistics to add information beyond the two-point statistics.
The Euclid-like maps, in contrast, are produced by N-body simulations and have a much higher resolution (0.59 arcmin), so they retain a large amount of information on small, genuinely non-Gaussian scales; this is precisely the regime that the defect statistics are designed to capture, which explains why they improve so dramatically over the $\kappa$-2PCF in this case.
We discuss the physics behind this behaviour in more detail in Sec.~\ref{sec:discussion}.

\subsection{Bridging the Three Cases}\label{sec:discussion}
As mentioned in Sec.~\ref{sec:results_noisy} and Sec.~\ref{sec:results_euclid}, the noise in the maps is the key factor that leads to the emergence of sensitivity to $\sigma_8$ in the defect statistics, and also causes the derivatives with respect to both parameters to flip sign compared to the idealised case.
The main consequence of adding noise to the maps is that it increases the number of defects, which can be theoretically expressed as an additive term to the convergence power spectrum
\begin{equation}
    C_\ell^{\kappa} \rightarrow C_\ell^{\kappa} + N_\ell^{\kappa}
\end{equation}
where $N_\ell^{\kappa}$ is the noise power spectrum, which is typically a constant related to the shape noise and the number density of observed sources
\begin{equation}
    N_\ell^{\kappa} = \frac{\sigma_\epsilon^2}{2 n_{\rm gal}},
\end{equation}
where $\sigma_\epsilon$ is the shape noise per component and $n_{\rm gal}$ is the number density of observed galaxies.

As computed in Eq.~\ref{eq:defect_density_theory} in App.~\ref{app:theories}, the defect number density is positively dependent on the relative contribution of the small scales in the power spectrum (because of an $\ell^2$ factor).
In most of the cases of cosmological interest, the power spectrum is a decreasing function of $\ell$, thus adding noise to the maps increases the relative contribution of small scales and thus increases the defect number density.
A pure noise field, consequently, will contain a higher defect number density than one containing cosmological signal.

Now the sensitivity to $\sigma_8$ emerges with the inclusion of noise because tuning $\sigma_8$ only changes the signal part of the power spectrum, while the noise part remains unchanged, which leads to a shape modification of the power spectrum and thus the statistics of the defects.
This can be understood more explicitly from Eq.~\ref{eq:defect_density}, where the defect number density depends only on the ratio $\xi^{(6)}_{\psi,0}/\xi^{(4)}_{\psi,0}$ of the derivatives of the lens potential correlation function.
In the noise-free case, changing $\sigma_8$ only rescales the amplitude of the shear correlation functions, which cancels in this ratio, so the defect number density is independent of $\sigma_8$; this is why we find no sensitivity to $\sigma_8$ for the defect statistics in the Gaussian case.
When noise is added, however, the observed power spectrum becomes $C_\ell^{\kappa} + N_\ell^{\kappa}$ as discussed above: the noise contribution is fixed and does not rescale with the amplitude of the signal, so the cancellation no longer takes place and the defect statistics become sensitive to $\sigma_8$.
The flipped sign of the derivatives can also be understood in a similar way.
When increasing $\Omega_{\rm m}$ or $\sigma_8$, the signal part of the power spectrum increases, which leads to a decrease in the relative contribution of small scales and thus a decrease in the defect number density, which is opposite to the behaviour in the idealised case where there is no noise and thus only tuning up $\Omega_{\rm m}$ can add relative contribution of small scales and increase the defect number density.

\begin{figure*}[t]
    \centering
    \includegraphics[width=0.47\textwidth]{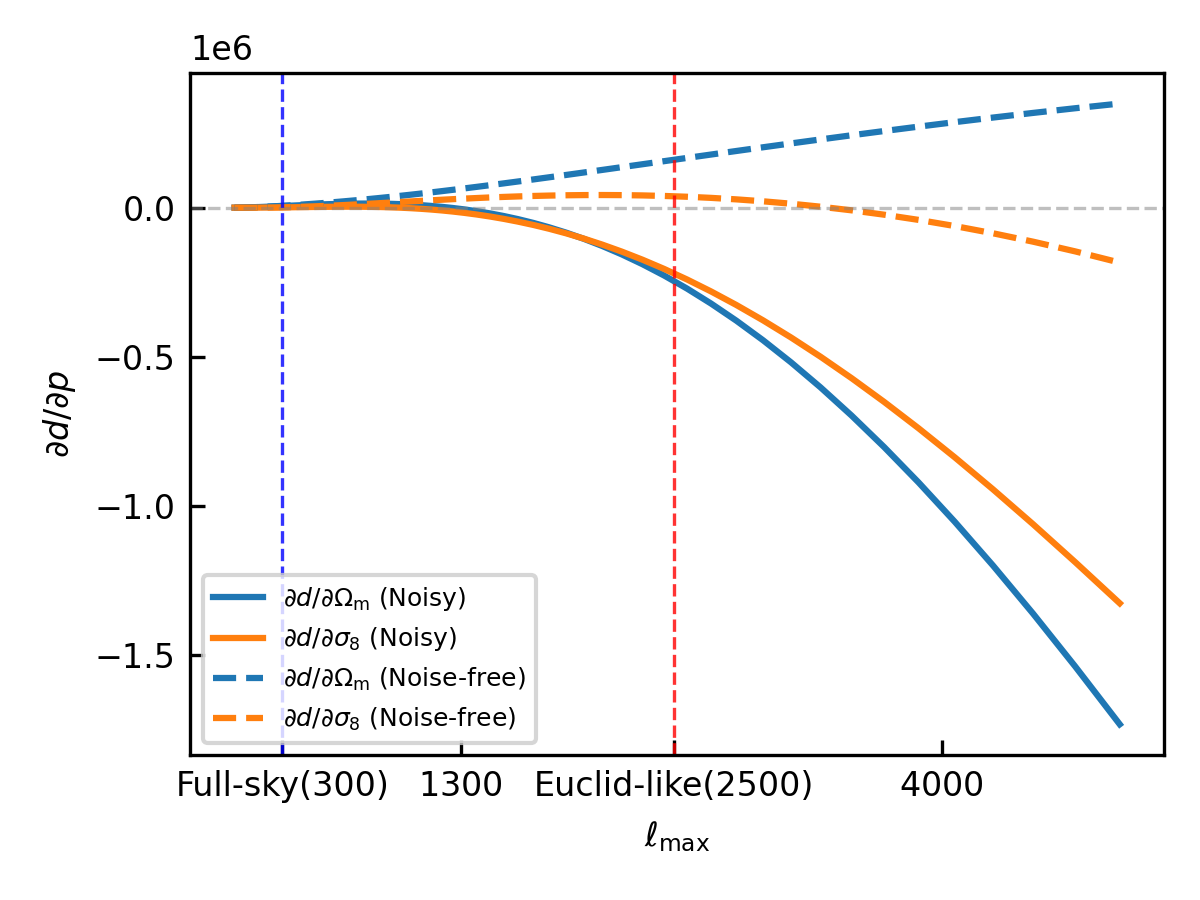}
    \includegraphics[width=0.48\textwidth]{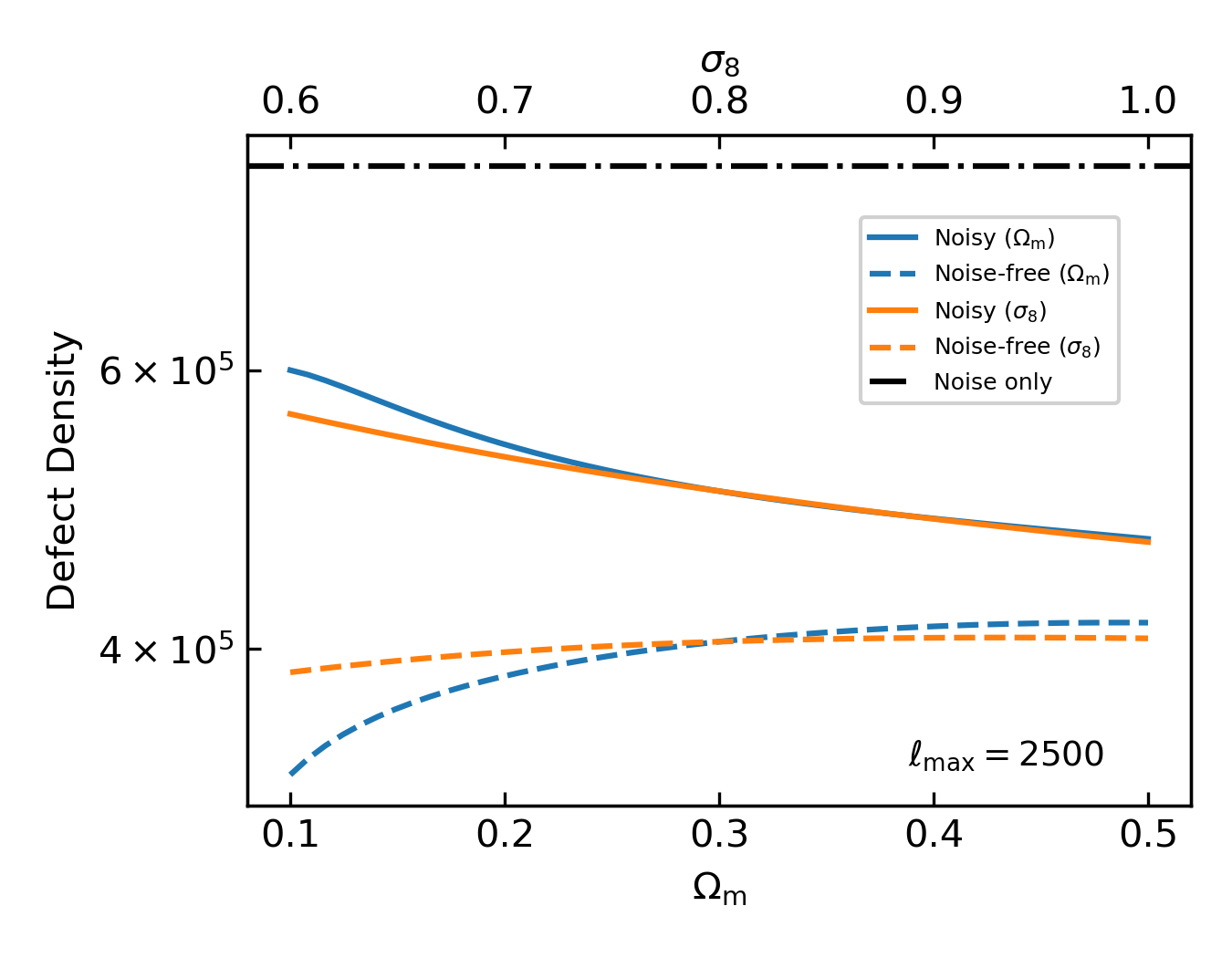}
    \caption{Left: The derivatives of the defect number density with respect to $\Omega_{\rm m}$ (blue) and $\sigma_8$ (orange) as a function of the cutoff scale $\ell_{\rm max}$ in both noise-free and noisy cases.
    The characteristic $\ell_{\rm max}$ values for our full sky maps and Euclid-like maps are denoted by the blue and red dashed vertical lines, respectively.
    Right: Theoretical predictions of the defect number density as a function of $\Omega_{\rm m}$ (blue) and $\sigma_8$ (orange) in the presence of noise with $\ell_{\rm max} = 2500$, and when one parameter is varying, the other is fixed to the fiducial value: ($\Omega_{\rm m}$,$\sigma_8$)=(0.3, 0.8).
    The solid/dashed lines are the predictions with/without noise.
    The defect number density with pure noise is shown as the dashed-dotted line for reference.}
    \label{fig:noise_defect_density}
\end{figure*}

This interpretation can be verified with analytical calculations of the defect number density in the presence of noise.
In the left panel of Fig.~\ref{fig:noise_defect_density}, we show the derivatives of the defect number density with respect to $\Omega_{\rm m}$ and $\sigma_8$ in both noise-free and noisy cases with different cutoff scales $\ell_{\rm max}$.
The derivatives are computed using the \texttt{jax-cosmo} package \cite{2023OJAp....6E..15C}.
We see that in the noise-free case, the derivative with respect to $\Omega_{\rm m}$ is positive while the derivative with respect to $\sigma_8$ is close to zero, which is consistent with our numerical results in the idealised case.
But as $\ell_{\rm max}$ increases, the derivative with respect to $\sigma_8$ becomes negative because, although increasing $\sigma_8$ enhances the high-$\ell$ power in the nonlinear regime, the fractional enhancement at the highest multipoles is weaker than at intermediate multipoles. This reduces their relative contribution to the spectral moments and thus lowers the defect number density. 
In the noisy case, the derivatives with respect to both parameters are negative, which is consistent with our interpretation above.

In the right panel Fig.~\ref{fig:noise_defect_density}, we show the theoretical expectations of the defect number density as a function of $\Omega_{\rm m}$ and $\sigma_8$ with and without noise, where the defect number density with pure noise is also shown for reference.
The figure shows that, although the defect number density increases when tuning up $\Omega_{\rm m}$ or $\sigma_8$ in the noise-free case, its overall amplitude is lower than in the noisy and pure-noise cases: adding noise to the maps raises the defect number density significantly above the noise-free case for all $\Omega_{\rm m}$ and $\sigma_8$ values, while it remains below the pure noise case.

\section{Conclusion}\label{sec:conclusion}
In this work, we present a new point of view to extract cosmological information from weak lensing fields, by analysing the topological defects in the cosmic shear maps.
This is the first time that topological defects in cosmic shear maps have been used as cosmological probes.

We first introduce the theoretical background of topological defects in 2D shear fields, and give the algorithm to identify defects in weak lensing shear maps.
We verify the algorithm by applying it to Gaussian random fields, and comparing the numerical results with the theoretical predictions of defect number density and 2PCFs.
We find an excellent agreement between the numerical and theoretical results.

We then go further to characterise the clustering and convergence environment of defects by measuring the 2PCFs, k-th nearest neighbour CDFs, and defect-conditioned convergence CDFs.
We find the properties for defects with different charges are quite different in their clustering and convergence dependency, suggesting that they may trace different features of the shear field and carry distinct cosmological information, which we will investigate further in future analyses.

We investigate the cosmological information of these defect statistics within a Fisher analysis framework. 
As a first step, we measure their derivatives with respect to $\Omega_{\rm m}$ and $\sigma_8$ to quantify their parameter sensitivities. 
We find that the clustering of defects is quite sensitive to $\Omega_{\rm m}$ but not $\sigma_8$ in linear idealised cases, whereas both parameters significantly affect the convergence environment of defects.
In noisy and nonlinear cases, both parameters show significant effects on all defect statistics.
This transition can be understood as the existence of noise in maps breaks the global amplitude tuning effect from $\sigma_8$ and thus induces a non-zero dependence of the defect statistics on $\sigma_8$.

Finally, we forecast the cosmological constraints from these defect statistics in three different cases, including idealised, noisy, and Euclid-like cases.
In the idealised case, none of the defect statistics outperforms the $\kappa$-2PCF, as expected for a purely Gaussian field, serving as a validation of our analysis pipeline. While in the lognormal setup, combining defect statistics with $\kappa$-PDF and $\kappa$-2PCF can noticeably tighten the constraints  (Fig.~\ref{fig:fisher_lognormal}). 
In the noisy case, the defect statistics alone can already outperform the $\kappa$-PDF and give comparable constraints to the $\kappa$-2PCF, and combining them with the $\kappa$-2PCF can further tighten the constraints (Fig.~\ref{fig:fisher_s7}).
In the Euclid-like case, we find that defect statistics significantly improve parameter constraints compared to the $\kappa$-2PCF, tightening the $\Omega_{\rm m}$ and $\sigma_8$ constraints by roughly a factor of 3, and by nearly a factor of 4 when combined with the $\kappa$-2PCF (Fig.~\ref{fig:fisher_howls} and Fig.~\ref{fig:bars}).
Compared to other higher-order statistics considered in the Euclid HOWLS analysis, the defect statistics also provide the tightest constraints on $\Omega_{\rm m}$ and $\sigma_8$.

This proof-of-concept analysis has firmly established topological defects as a promising weak lensing probe of cosmological information. In order for defects to become a competitive metric in cosmic shear analyses, further work is required.
First, the asymmetric distributions between positive and negative defects needs further investigation to understand its origin and cosmological implications.
Second, more realistic analyses must account for both astrophysical effects, such as baryonic feedback, source lens clustering and intrinsic alignments, and observational systematics, including photometric redshift uncertainties, multiplicative biases and survey masking.
Finally, tomographic analysis needs to be developed to fully exploit the potential of defect statistics in ongoing weak lensing surveys like Euclid.

\section{Data Availability}
The code used to generate the results in this paper is available at \url{https://github.com/AstroBai/topo_defects_gaussian} and \url{https://github.com/AstroBai/DefectHOWLS}.
The Euclid HOWLS simulations used in this work are available at \url{https://archive.lam.fr/GECO/HOWLS/}.

\begin{acknowledgments}
We'd like to thank Dr. Joachim Harnois-Deraps for valuable discussion and comments on this work.
We'd also like to thank Dr. Zhaoting Chen for helpful discussion on Fisher analysis.
We acknowledge the Euclid HOWLS team for releasing the data for our Euclid-like analysis.
We are grateful to Eric Tittley for managing the Institute for Astronomy's \textit{cuillin} supercomputer, on which this analysis was performed.
JB acknowledges support from the Science and Technology Facilities Council (STFC) studentship. 
BG acknowledges support from the UKRI Stephen Hawking Fellowship (grant reference EP/Y017137/1). 
ANT acknowledges support provided by the STFC.
We also acknowledge the use of generative AI tools, including \texttt{GPT}, \texttt{DeepSeek}, \texttt{GitHub Copilot}, and \texttt{OpenCode}, for assistance with code debugging and for editorial support in writing and polishing the manuscript.
\end{acknowledgments}

\appendix

\section{Topological Defects and Cosmic Flexion}\label{app:flexion}
In this section, we explore the connection between topological defects in shear fields and cosmic flexion, which is a higher-order weak lensing effect that describes the distortion of galaxy images beyond shear and convergence.
Following the formula for the defect number density, the determinant of the Jacobian matrix $\mathbf{J}$ of the mapping from the angular coordinates $\boldsymbol{\theta} = (\theta_1, \theta_2)$ to the shear field $\boldsymbol{\gamma} = (\gamma_1, \gamma_2)$ is a key quantity in determining the defect number density.
The determinant of the Jacobian matrix can be expressed as
\begin{equation}
    \mathcal{J} = \det \mathbf{J} = \frac{\partial \gamma_1}{\partial \theta_1} \frac{\partial \gamma_2}{\partial \theta_2} - \frac{\partial \gamma_1}{\partial \theta_2} \frac{\partial \gamma_2}{\partial \theta_1}.
\end{equation}

To connect the Jacobian to the cosmic flexion, we introduce the complex notation for the shear and flexion.
The spin raising and lowering operators are defined as
\begin{equation}
    \eth = \left(\frac{\partial}{\partial \theta_1} + i \frac{\partial}{\partial \theta_2}\right), \quad \bar{\eth} = \left(\frac{\partial}{\partial \theta_1} - i \frac{\partial}{\partial \theta_2}\right).
\end{equation}
With these operators, the shear and convergence fields can be expressed as
\begin{equation}
    \gamma = \frac{1}{2} \eth \eth \psi, \quad \kappa = \frac{1}{2} \bar{\eth} \eth \psi,
\end{equation}
where $\psi$ is the lens potential. We can see that shear is a spin-2 field, while convergence is a spin-0 field.
The first flexion $\mathcal{F}$ and the second flexion $\mathcal{G}$ are defined as \cite{2006MNRAS.365..414B, 2022PhRvD.105l3521A}
\begin{equation}
    \mathcal{F} = \frac{1}{2} \eth \bar{\eth} \eth \psi, \quad \mathcal{G} = \frac{1}{2} \eth \eth \eth \psi,
\end{equation}
from which we see that the first flexion $\mathcal{F}$ is a spin-1 field, while the second flexion $\mathcal{G}$ is a spin-3 field.

In terms of shear, the flexion can be written as
\begin{equation}
    \mathcal{F} = \bar{\eth} \gamma, \quad \mathcal{G} = \eth \gamma.
\end{equation}
Then we can compose the Jacobian determinant $\mathcal{J}$ in terms of the flexion to be
\begin{equation}
    \mathcal{J} = \frac{1}{4} \left( |\mathcal{F}|^2 - |\mathcal{G}|^2 \right).
\end{equation}  

This means that where the first flexion dominates, the Jacobian determinant is positive, leading to positive defects, while where the second flexion dominates, the Jacobian determinant is negative, leading to negative defects.
Moreover, the larger the difference between the first and second flexion, the higher the defect number density, which means that defects are more likely to be found in regions where the flexion is strong.

To understand how the first and second flexion correspond to positive and negative defects, we can construct two toy models of the shear field, one with only the first flexion and the other with only the second flexion.
We choose the lens potential for these two models as
\begin{equation}
    \psi_1 = r^3 \cos \theta, \quad \psi_2 = \frac{1}{3} r^3 \cos 3\theta \label{eq:flexion_models}
\end{equation}
where $(r, \theta)$ are the polar coordinates. One can prove these two models satisfy the conditions for the first and second flexion, respectively.
And the factor of $\frac{1}{3}$ in the second model is chosen to make the magnitudes of the first flexion in model 1 and the second flexion in model 2 equal ($|\mathcal{F}_1| = |\mathcal{G}_2| = 4$). 

We can also create another two models by combining the two models linearly, for example
\begin{equation}
    \psi_3 = \psi_1 + 0.5\psi_2, \quad \psi_4 = 0.5\psi_1 + \psi_2. \label{eq:flexion_models_2}
\end{equation}
These two models have both the first and second flexion, but the first flexion dominates in the first model while the second flexion dominates in the second model, thus we expect the first model to have only positive defects while the second model to have only negative defects.

Theoretically, we can directly compute the defect charge located at the origin for these models.
For example, for the first model, we can compute the shear field as
\begin{equation}
    \gamma = \frac{1}{2} \eth \eth \psi_1 = 2 r e^{i\theta},
\end{equation}
and then we can compute the defect charge at the origin as
\begin{align}
    m
    &= \frac{1}{2\pi} \oint \nabla \varphi \cdot d\mathbf{l} \\
    &= \frac{1}{2\pi} \oint \nabla \left( \frac{1}{2}\arg(\gamma) \right) \cdot d\mathbf{l} \\
    &= \frac{1}{2\pi} \oint \nabla \left( \frac{\theta}{2} \right) \cdot d\mathbf{l} \\
    &= \frac{1}{2},
\end{align}
where $\varphi = \frac{1}{2}\arg(\gamma) = \frac{1}{2}\tan^{-1}\left(\frac{\gamma_2}{\gamma_1}\right)$ is the shear orientation angle.
One can similarly compute the defect charge for the other three models, and we find that the first and third models have positive defect charges while the second and fourth models have negative charges.

\begin{figure*}[htp]
    \centering
    \includegraphics[width=0.7\textwidth]{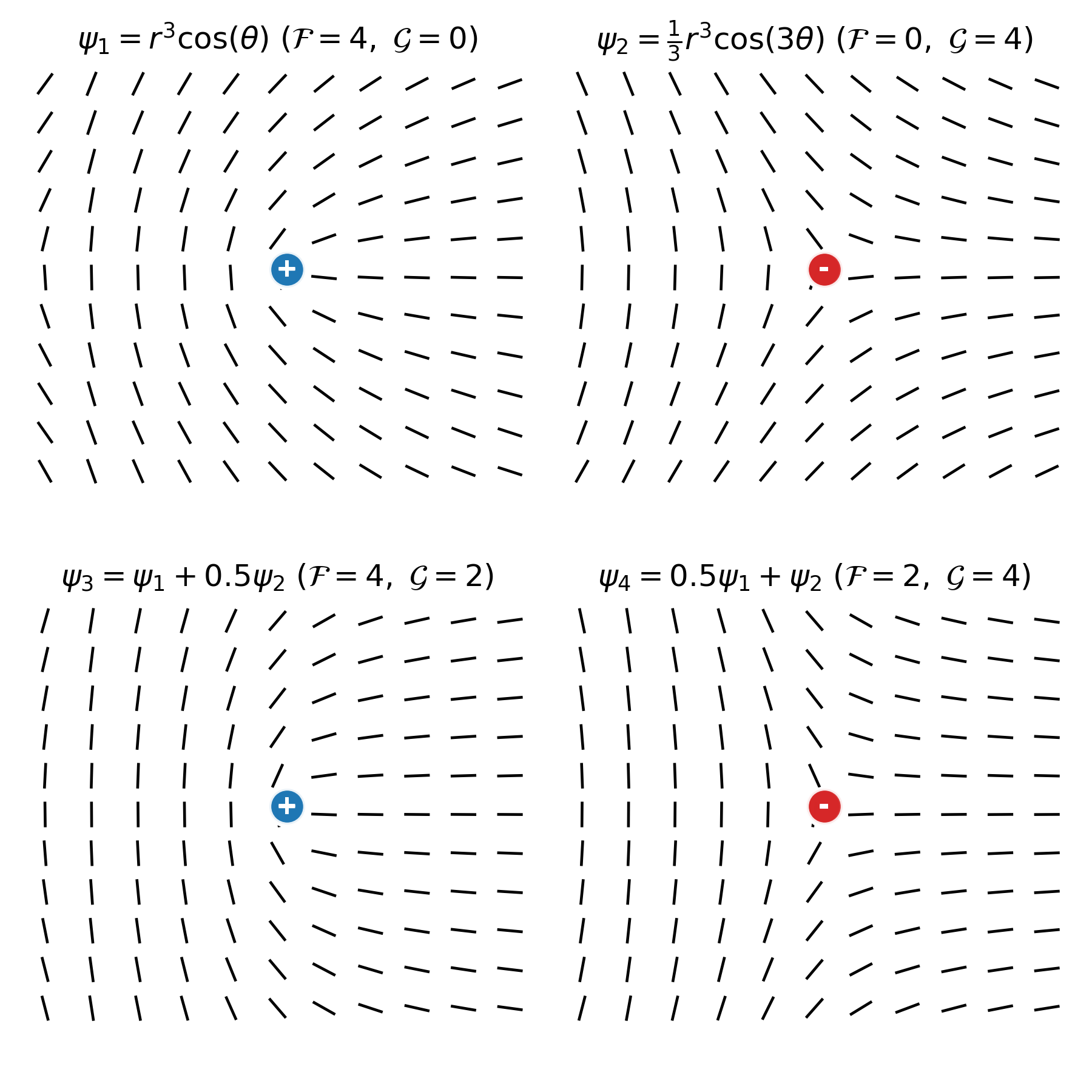}
    \caption{Shear fields from the four potential models, where the upper-left, upper-right, lower-left, and lower-right panels correspond to the first, second, third, and fourth models defined in Eq.~\ref{eq:flexion_models} and Eq.~\ref{eq:flexion_models_2}, respectively.
    The blue and red dots indicate the locations of positive and negative defects, respectively.}
    \label{fig:flexion_shear}
\end{figure*}

The shear fields for these 4 potential models are shown in Fig.~\ref{fig:flexion_shear}, where the upper-left, upper-right, lower-left, and lower-right panels correspond to the first, second, third, and fourth models, respectively.
The blue and red dots indicate the locations of positive and negative defects.
As observed from the figure, the first and third models, where the first flexion dominates, only contain positive defects, while the second and fourth models, where the second flexion dominates, only contain negative defects, which is consistent with our interpretation above.

\section{Theoretical Recipe for Defect Number Density}\label{app:theories}
In this section, we proceed to derive the theoretical predictions for the defect number density and 2PCF in Gaussian random fields, which are used to validate our numerical results in the idealised case.

Following Eq.~\ref{eq:defect_density}, the defect number density involves the 4-th and 6-th derivatives of the lens potential at $\theta=0$, in which only the Lengendre polynomials contain the $\theta$ dependence, thus we need to derive their derivatives at $\theta=0$.

The full expression of the $n=2m$-th (even order) derivative of $P_\ell(\cos\theta)$ at $\theta=0$ is
\begin{equation}
    \frac{\mathrm{d}^{2m}}{\mathrm{d}\theta^{2m}} P_\ell(\cos\theta) \bigg\vert_{\theta=0} = (-1)^m \frac{(2m)!}{(m!)^2} \frac{(\ell + m)!}{(\ell - m)!} \left(\frac{1}{4}\right)^{m}.
\end{equation}
For our case, we need the 4-th and 6-th derivatives, which are
\begin{equation}
    \begin{aligned}
    \frac{\mathrm{d}^4}{\mathrm{d}\theta^4} P_\ell(\cos\theta) \bigg\vert_{\theta=0} &= \frac{3}{8} (\ell + 2)(\ell + 1)\ell(\ell - 1),
    \\ \frac{\mathrm{d}^6}{\mathrm{d}\theta^6} P_\ell(\cos\theta) \bigg\vert_{\theta=0} &= -\frac{15}{16} (\ell + 3)(\ell + 2)(\ell + 1)\ell(\ell - 1)(\ell - 2).
    \end{aligned}
\end{equation}
Thus the defect number density is
\begin{equation}
    d = \frac{3}{10\pi} \times \frac{15}{16} \times \frac{8}{3} \times \frac{\sum_\ell (2\ell + 1)(\ell + 3)(\ell + 2)(\ell + 1)\ell(\ell - 1)(\ell - 2) C_\ell^{\psi}}{\sum_\ell (2\ell + 1)(\ell + 2)(\ell + 1)\ell(\ell - 1) C_\ell^{\psi}} + \frac{1}{\pi},
\end{equation}
which can be further simplified to
\begin{equation}
    d = \frac{3}{4\pi} \times \frac{\sum_\ell \frac{(2\ell + 1)(\ell+3)(\ell + 2)(\ell-1)(\ell - 2)}{\ell(\ell + 1)} C_\ell^{\kappa}}{\sum_\ell \frac{(2\ell + 1)(\ell + 2)(\ell - 1)}{\ell(\ell + 1)} C_\ell^{\kappa}} + \frac{1}{\pi}, \label{eq:defect_density_theory}
\end{equation}
where we have used the relation between the power spectrum of the lens potential and the convergence in Eq.~\ref{eq:Cl_psi}.


\bibliographystyle{JHEP}
\bibliography{biblio.bib}

@ARTICLE{2022PhRvD.105b3520A,
       author = {{Abbott}, T.~M.~C. and {Aguena}, M. and {Alarcon}, A. and {Allam}, S. and {Alves}, O. and {Amon}, A. and {Andrade-Oliveira}, F. and {Annis}, J. and {Avila}, S. and {Bacon}, D. and {Baxter}, E. and {Bechtol}, K. and {Becker}, M.~R. and {Bernstein}, G.~M. and {Bhargava}, S. and {Birrer}, S. and {Blazek}, J. and {Brandao-Souza}, A. and {Bridle}, S.~L. and {Brooks}, D. and {Buckley-Geer}, E. and {Burke}, D.~L. and {Camacho}, H. and {Campos}, A. and {Carnero Rosell}, A. and {Carrasco Kind}, M. and {Carretero}, J. and {Castander}, F.~J. and {Cawthon}, R. and {Chang}, C. and {Chen}, A. and {Chen}, R. and {Choi}, A. and {Conselice}, C. and {Cordero}, J. and {Costanzi}, M. and {Crocce}, M. and {da Costa}, L.~N. and {da Silva Pereira}, M.~E. and {Davis}, C. and {Davis}, T.~M. and {De Vicente}, J. and {DeRose}, J. and {Desai}, S. and {Di Valentino}, E. and {Diehl}, H.~T. and {Dietrich}, J.~P. and {Dodelson}, S. and {Doel}, P. and {Doux}, C. and {Drlica-Wagner}, A. and {Eckert}, K. and {Eifler}, T.~F. and {Elsner}, F. and {Elvin-Poole}, J. and {Everett}, S. and {Evrard}, A.~E. and {Fang}, X. and {Farahi}, A. and {Fernandez}, E. and {Ferrero}, I. and {Fert{\'e}}, A. and {Fosalba}, P. and {Friedrich}, O. and {Frieman}, J. and {Garc{\'\i}a-Bellido}, J. and {Gatti}, M. and {Gaztanaga}, E. and {Gerdes}, D.~W. and {Giannantonio}, T. and {Giannini}, G. and {Gruen}, D. and {Gruendl}, R.~A. and {Gschwend}, J. and {Gutierrez}, G. and {Harrison}, I. and {Hartley}, W.~G. and {Herner}, K. and {Hinton}, S.~R. and {Hollowood}, D.~L. and {Honscheid}, K. and {Hoyle}, B. and {Huff}, E.~M. and {Huterer}, D. and {Jain}, B. and {James}, D.~J. and {Jarvis}, M. and {Jeffrey}, N. and {Jeltema}, T. and {Kovacs}, A. and {Krause}, E. and {Kron}, R. and {Kuehn}, K. and {Kuropatkin}, N. and {Lahav}, O. and {Leget}, P.-F. and {Lemos}, P. and {Liddle}, A.~R. and {Lidman}, C. and {Lima}, M. and {Lin}, H. and {MacCrann}, N. and {Maia}, M.~A.~G. and {Marshall}, J.~L. and {Martini}, P. and {McCullough}, J. and {Melchior}, P. and {Mena-Fern{\'a}ndez}, J. and {Menanteau}, F. and {Miquel}, R. and {Mohr}, J.~J. and {Morgan}, R. and {Muir}, J. and {Myles}, J. and {Nadathur}, S. and {Navarro-Alsina}, A. and {Nichol}, R.~C. and {Ogando}, R.~L.~C. and {Omori}, Y. and {Palmese}, A. and {Pandey}, S. and {Park}, Y. and {Paz-Chinch{\'o}n}, F. and {Petravick}, D. and {Pieres}, A. and {Plazas Malag{\'o}n}, A.~A. and {Porredon}, A. and {Prat}, J. and {Raveri}, M. and {Rodriguez-Monroy}, M. and {Rollins}, R.~P. and {Romer}, A.~K. and {Roodman}, A. and {Rosenfeld}, R. and {Ross}, A.~J. and {Rykoff}, E.~S. and {Samuroff}, S. and {S{\'a}nchez}, C. and {Sanchez}, E. and {Sanchez}, J. and {Sanchez Cid}, D. and {Scarpine}, V. and {Schubnell}, M. and {Scolnic}, D. and {Secco}, L.~F. and {Serrano}, S. and {Sevilla-Noarbe}, I. and {Sheldon}, E. and {Shin}, T. and {Smith}, M. and {Soares-Santos}, M. and {Suchyta}, E. and {Swanson}, M.~E.~C. and {Tabbutt}, M. and {Tarle}, G. and {Thomas}, D. and {To}, C. and {Troja}, A. and {Troxel}, M.~A. and {Tucker}, D.~L. and {Tutusaus}, I. and {Varga}, T.~N. and {Walker}, A.~R. and {Weaverdyck}, N. and {Wechsler}, R. and {Weller}, J. and {Yanny}, B. and {Yin}, B. and {Zhang}, Y. and {Zuntz}, J. and {DES Collaboration}},
        title = "{Dark Energy Survey Year 3 results: Cosmological constraints from galaxy clustering and weak lensing}",
      journal = {Physics Review D},
         year = 2022,
        month = jan,
       volume = {105},
       number = {2},
          eid = {023520},
        pages = {023520},
          doi = {10.1103/PhysRevD.105.023520},
archivePrefix = {arXiv},
       eprint = {2105.13549},
 primaryClass = {astro-ph.CO},
       adsurl = {https://ui.adsabs.harvard.edu/abs/2022PhRvD.105b3520A}
}

@ARTICLE{2025A&A...703A.158W,
       author = {{Wright}, Angus H. and {St{\"o}lzner}, Benjamin and {Asgari}, Marika and {Bilicki}, Maciej and {Giblin}, Benjamin and {Heymans}, Catherine and {Hildebrandt}, Hendrik and {Hoekstra}, Henk and {Joachimi}, Benjamin and {Kuijken}, Konrad and {Li}, Shun-Sheng and {Reischke}, Robert and {von Wietersheim-Kramsta}, Maximilian and {Yoon}, Mijin and {Burger}, Pierre and {Chisari}, Nora Elisa and {de Jong}, Jelte and {Dvornik}, Andrej and {Georgiou}, Christos and {Harnois-D{\'e}raps}, Joachim and {Jalan}, Priyanka and {William}, Anjitha John and {Joudaki}, Shahab and {Lesci}, Giorgio Francesco and {Linke}, Laila and {Loureiro}, Arthur and {Mahony}, Constance and {Maturi}, Matteo and {Miller}, Lance and {Moscardini}, Lauro and {Napolitano}, Nicola R. and {Porth}, Lucas and {Radovich}, Mario and {Schneider}, Peter and {Tr{\"o}ster}, Tilman and {Valentijn}, Edwin and {Wittje}, Anna and {Yan}, Ziang and {Zhang}, Yun-Hao},
        title = "{KiDS-Legacy: Cosmological constraints from cosmic shear with the complete Kilo-Degree Survey}",
      journal = {Astronomy \& Astrophysics},
         year = 2025,
        month = nov,
       volume = {703},
          eid = {A158},
        pages = {A158},
          doi = {10.1051/0004-6361/202554908},
archivePrefix = {arXiv},
       eprint = {2503.19441},
 primaryClass = {astro-ph.CO},
       adsurl = {https://ui.adsabs.harvard.edu/abs/2025A&A...703A.158W}
}

@ARTICLE{2023PhRvD.108l3519D,
       author = {{Dalal}, Roohi and {Li}, Xiangchong and {Nicola}, Andrina and {Zuntz}, Joe and {Strauss}, Michael A. and {Sugiyama}, Sunao and {Zhang}, Tianqing and {Rau}, Markus M. and {Mandelbaum}, Rachel and {Takada}, Masahiro and {More}, Surhud and {Miyatake}, Hironao and {Kannawadi}, Arun and {Shirasaki}, Masato and {Taniguchi}, Takanori and {Takahashi}, Ryuichi and {Osato}, Ken and {Hamana}, Takashi and {Oguri}, Masamune and {Nishizawa}, Atsushi J. and {Malag{\'o}n}, Andr{\'e}s A. Plazas and {Sunayama}, Tomomi and {Alonso}, David and {Slosar}, An{\v{z}}e and {Luo}, Wentao and {Armstrong}, Robert and {Bosch}, James and {Hsieh}, Bau-Ching and {Komiyama}, Yutaka and {Lupton}, Robert H. and {Lust}, Nate B. and {MacArthur}, Lauren A. and {Miyazaki}, Satoshi and {Murayama}, Hitoshi and {Nishimichi}, Takahiro and {Okura}, Yuki and {Price}, Paul A. and {Tait}, Philip J. and {Tanaka}, Masayuki and {Wang}, Shiang-Yu},
        title = "{Hyper Suprime-Cam Year 3 results: Cosmology from cosmic shear power spectra}",
      journal = {Physics Review D},
         year = 2023,
        month = dec,
       volume = {108},
       number = {12},
          eid = {123519},
        pages = {123519},
          doi = {10.1103/PhysRevD.108.123519},
archivePrefix = {arXiv},
       eprint = {2304.00701},
 primaryClass = {astro-ph.CO},
       adsurl = {https://ui.adsabs.harvard.edu/abs/2023PhRvD.108l3519D}
}

@ARTICLE{2001PhR...340..291B,
       author = {{Bartelmann}, M. and {Schneider}, P.},
        title = "{Weak gravitational lensing}",
      journal = {Physics Reports},
         year = 2001,
        month = jan,
       volume = {340},
       number = {4-5},
        pages = {291-472},
          doi = {10.1016/S0370-1573(00)00082-X},
archivePrefix = {arXiv},
       eprint = {astro-ph/9912508},
 primaryClass = {astro-ph},
       adsurl = {https://ui.adsabs.harvard.edu/abs/2001PhR...340..291B}
}

@ARTICLE{2025A&A...697A...1E,
       author = {{Euclid Collaboration} and {Mellier}, Y. and {Abdurro'uf} and {Acevedo Barroso}, J.~A. and {Ach{\'u}carro}, A. and {Adamek}, J. and {Adam}, R. and {Addison}, G.~E. and {Aghanim}, N. and {Aguena}, M. and {Ajani}, V. and {Akrami}, Y. and {Al-Bahlawan}, A. and {Alavi}, A. and {Albuquerque}, I.~S. and {Alestas}, G. and {Alguero}, G. and {Allaoui}, A. and {Allen}, S.~W. and {Allevato}, V. and {Alonso-Tetilla}, A.~V. and {Altieri}, B. and {Alvarez-Candal}, A. and {Alvi}, S. and {Amara}, A. and {Amendola}, L. and {Amiaux}, J. and {Andika}, I.~T. and {Andreon}, S. and {Andrews}, A. and {Angora}, G. and {Angulo}, R.~E. and {Annibali}, F. and {Anselmi}, A. and {Anselmi}, S. and {Arcari}, S. and {Archidiacono}, M. and {Aric{\`o}}, G. and {Arnaud}, M. and {Arnouts}, S. and {Asgari}, M. and {Asorey}, J. and {Atayde}, L. and {Atek}, H. and {Atrio-Barandela}, F. and {Aubert}, M. and {Aubourg}, E. and {Auphan}, T. and {Auricchio}, N. and {Aussel}, B. and {Aussel}, H. and {Avelino}, P.~P. and {Avgoustidis}, A. and {Avila}, S. and {Awan}, S. and {Azzollini}, R. and {Baccigalupi}, C. and {Bachelet}, E. and {Bacon}, D. and {Baes}, M. and {Bagley}, M.~B. and {Bahr-Kalus}, B. and {Balaguera-Antolinez}, A. and {Balbinot}, E. and {Balcells}, M. and {Baldi}, M. and {Baldry}, I. and {Balestra}, A. and {Ballardini}, M. and {Ballester}, O. and {Balogh}, M. and {Ba{\~n}ados}, E. and {Barbier}, R. and {Bardelli}, S. and {Baron}, M. and {Barreiro}, T. and {Barrena}, R. and {Barriere}, J.-C. and {Barros}, B.~J. and {Barthelemy}, A. and {Bartolo}, N. and {Basset}, A. and {Battaglia}, P. and {Battisti}, A.~J. and {Baugh}, C.~M. and {Baumont}, L. and {Bazzanini}, L. and {Beaulieu}, J.-P. and {Beckmann}, V. and {Belikov}, A.~N. and {Bel}, J. and {Bellagamba}, F. and {Bella}, M. and {Bellini}, E. and {Benabed}, K. and {Bender}, R. and {Benevento}, G. and {Bennett}, C.~L. and {Benson}, K. and {Bergamini}, P. and {Bermejo-Climent}, J.~R. and {Bernardeau}, F. and {Bertacca}, D. and {Berthe}, M. and {Berthier}, J. and {Bethermin}, M. and {Beutler}, F. and {Bevillon}, C. and {Bhargava}, S. and {Bhatawdekar}, R. and {Bianchi}, D. and {Bisigello}, L. and {Biviano}, A. and {Blake}, R.~P. and {Blanchard}, A. and {Blazek}, J. and {Blot}, L. and {Bosco}, A. and {Bodendorf}, C. and {Boenke}, T. and {B{\"o}hringer}, H. and {Boldrini}, P. and {Bolzonella}, M. and {Bonchi}, A. and {Bonici}, M. and {Bonino}, D. and {Bonino}, L. and {Bonvin}, C. and {Bon}, W. and {Booth}, J.~T. and {Borgani}, S. and {Borlaff}, A.~S. and {Borsato}, E. and {Bose}, B. and {Botticella}, M.~T. and {Boucaud}, A. and {Bouche}, F. and {Boucher}, J.~S. and {Boutigny}, D. and {Bouvard}, T. and {Bouwens}, R. and {Bouy}, H. and {Bowler}, R.~A.~A. and {Bozza}, V. and {Bozzo}, E. and {Branchini}, E. and {Brando}, G. and {Brau-Nogue}, S. and {Brekke}, P. and {Bremer}, M.~N. and {Brescia}, M. and {Breton}, M.-A. and {Brinchmann}, J. and {Brinckmann}, T. and {Brockley-Blatt}, C. and {Brodwin}, M. and {Brouard}, L. and {Brown}, M.~L. and {Bruton}, S. and {Bucko}, J. and {Buddelmeijer}, H. and {Buenadicha}, G. and {Buitrago}, F. and {Burger}, P. and {Burigana}, C. and {Busillo}, V. and {Busonero}, D. and {Cabanac}, R. and {Cabayol-Garcia}, L. and {Cagliari}, M.~S. and {Caillat}, A. and {Caillat}, L. and {Calabrese}, M. and {Calabro}, A. and {Calderone}, G. and {Calura}, F. and {Camacho Quevedo}, B. and {Camera}, S. and {Campos}, L. and {Ca{\~n}as-Herrera}, G. and {Candini}, G.~P. and {Cantiello}, M. and {Capobianco}, V. and {Cappellaro}, E. and {Cappelluti}, N. and {Cappi}, A. and {Caputi}, K.~I. and {Cara}, C. and {Carbone}, C. and {Cardone}, V.~F. and {Carella}, E. and {Carlberg}, R.~G. and {Carle}, M. and {Carminati}, L. and {Caro}, F. and {Carrasco}, J.~M. and {Carretero}, J. and {Carrilho}, P. and {Carron Duque}, J. and {Carry}, B.},
        title = "{Euclid: I. Overview of the Euclid mission}",
      journal = {Astronomy \& Astrophysics},
         year = 2025,
        month = may,
       volume = {697},
          eid = {A1},
        pages = {A1},
          doi = {10.1051/0004-6361/202450810},
archivePrefix = {arXiv},
       eprint = {2405.13491},
 primaryClass = {astro-ph.CO},
       adsurl = {https://ui.adsabs.harvard.edu/abs/2025A&A...697A...1E}
}

@ARTICLE{2019ApJ...873..111I,
       author = {{Ivezi{\'c}}, {\v{Z}}eljko and {Kahn}, Steven M. and {Tyson}, J. Anthony and {Abel}, Bob and {Acosta}, Emily and {Allsman}, Robyn and {Alonso}, David and {AlSayyad}, Yusra and {Anderson}, Scott F. and {Andrew}, John and {Angel}, James Roger P. and {Angeli}, George Z. and {Ansari}, Reza and {Antilogus}, Pierre and {Araujo}, Constanza and {Armstrong}, Robert and {Arndt}, Kirk T. and {Astier}, Pierre and {Aubourg}, {\'E}ric and {Auza}, Nicole and {Axelrod}, Tim S. and {Bard}, Deborah J. and {Barr}, Jeff D. and {Barrau}, Aurelian and {Bartlett}, James G. and {Bauer}, Amanda E. and {Bauman}, Brian J. and {Baumont}, Sylvain and {Bechtol}, Ellen and {Bechtol}, Keith and {Becker}, Andrew C. and {Becla}, Jacek and {Beldica}, Cristina and {Bellavia}, Steve and {Bianco}, Federica B. and {Biswas}, Rahul and {Blanc}, Guillaume and {Blazek}, Jonathan and {Blandford}, Roger D. and {Bloom}, Josh S. and {Bogart}, Joanne and {Bond}, Tim W. and {Booth}, Michael T. and {Borgland}, Anders W. and {Borne}, Kirk and {Bosch}, James F. and {Boutigny}, Dominique and {Brackett}, Craig A. and {Bradshaw}, Andrew and {Brandt}, William Nielsen and {Brown}, Michael E. and {Bullock}, James S. and {Burchat}, Patricia and {Burke}, David L. and {Cagnoli}, Gianpietro and {Calabrese}, Daniel and {Callahan}, Shawn and {Callen}, Alice L. and {Carlin}, Jeffrey L. and {Carlson}, Erin L. and {Chandrasekharan}, Srinivasan and {Charles-Emerson}, Glenaver and {Chesley}, Steve and {Cheu}, Elliott C. and {Chiang}, Hsin-Fang and {Chiang}, James and {Chirino}, Carol and {Chow}, Derek and {Ciardi}, David R. and {Claver}, Charles F. and {Cohen-Tanugi}, Johann and {Cockrum}, Joseph J. and {Coles}, Rebecca and {Connolly}, Andrew J. and {Cook}, Kem H. and {Cooray}, Asantha and {Covey}, Kevin R. and {Cribbs}, Chris and {Cui}, Wei and {Cutri}, Roc and {Daly}, Philip N. and {Daniel}, Scott F. and {Daruich}, Felipe and {Daubard}, Guillaume and {Daues}, Greg and {Dawson}, William and {Delgado}, Francisco and {Dellapenna}, Alfred and {de Peyster}, Robert and {de Val-Borro}, Miguel and {Digel}, Seth W. and {Doherty}, Peter and {Dubois}, Richard and {Dubois-Felsmann}, Gregory P. and {Durech}, Josef and {Economou}, Frossie and {Eifler}, Tim and {Eracleous}, Michael and {Emmons}, Benjamin L. and {Fausti Neto}, Angelo and {Ferguson}, Henry and {Figueroa}, Enrique and {Fisher-Levine}, Merlin and {Focke}, Warren and {Foss}, Michael D. and {Frank}, James and {Freemon}, Michael D. and {Gangler}, Emmanuel and {Gawiser}, Eric and {Geary}, John C. and {Gee}, Perry and {Geha}, Marla and {Gessner}, Charles J.~B. and {Gibson}, Robert R. and {Gilmore}, D. Kirk and {Glanzman}, Thomas and {Glick}, William and {Goldina}, Tatiana and {Goldstein}, Daniel A. and {Goodenow}, Iain and {Graham}, Melissa L. and {Gressler}, William J. and {Gris}, Philippe and {Guy}, Leanne P. and {Guyonnet}, Augustin and {Haller}, Gunther and {Harris}, Ron and {Hascall}, Patrick A. and {Haupt}, Justine and {Hernandez}, Fabio and {Herrmann}, Sven and {Hileman}, Edward and {Hoblitt}, Joshua and {Hodgson}, John A. and {Hogan}, Craig and {Howard}, James D. and {Huang}, Dajun and {Huffer}, Michael E. and {Ingraham}, Patrick and {Innes}, Walter R. and {Jacoby}, Suzanne H. and {Jain}, Bhuvnesh and {Jammes}, Fabrice and {Jee}, M. James and {Jenness}, Tim and {Jernigan}, Garrett and {Jevremovi{\'c}}, Darko and {Johns}, Kenneth and {Johnson}, Anthony S. and {Johnson}, Margaret W.~G. and {Jones}, R. Lynne and {Juramy-Gilles}, Claire and {Juri{\'c}}, Mario and {Kalirai}, Jason S. and {Kallivayalil}, Nitya J. and {Kalmbach}, Bryce and {Kantor}, Jeffrey P. and {Karst}, Pierre and {Kasliwal}, Mansi M. and {Kelly}, Heather and {Kessler}, Richard and {Kinnison}, Veronica and {Kirkby}, David and {Knox}, Lloyd and {Kotov}, Ivan V. and {Krabbendam}, Victor L. and {Krughoff}, K. Simon and {Kub{\'a}nek}, Petr and {Kuczewski}, John and {Kulkarni}, Shri and {Ku}, John and {Kurita}, Nadine R. and {Lage}, Craig S. and {Lambert}, Ron and {Lange}, Travis and {Langton}, J. Brian and {Le Guillou}, Laurent and {Levine}, Deborah and {Liang}, Ming and {Lim}, Kian-Tat and {Lintott}, Chris J. and {Long}, Kevin E. and {Lopez}, Margaux and {Lotz}, Paul J. and {Lupton}, Robert H. and {Lust}, Nate B. and {MacArthur}, Lauren A. and {Mahabal}, Ashish and {Mandelbaum}, Rachel and {Markiewicz}, Thomas W. and {Marsh}, Darren S. and {Marshall}, Philip J. and {Marshall}, Stuart and {May}, Morgan and {McKercher}, Robert and {McQueen}, Michelle and {Meyers}, Joshua and {Migliore}, Myriam and {Miller}, Michelle and {Mills}, David J.},
        title = "{LSST: From Science Drivers to Reference Design and Anticipated Data Products}",
      journal = {The Astrophysical Journal},
         year = 2019,
        month = mar,
       volume = {873},
       number = {2},
          eid = {111},
        pages = {111},
          doi = {10.3847/1538-4357/ab042c},
archivePrefix = {arXiv},
       eprint = {0805.2366},
 primaryClass = {astro-ph},
       adsurl = {https://ui.adsabs.harvard.edu/abs/2019ApJ...873..111I}
}

@ARTICLE{2023MNRAS.520.1721G,
       author = {{Giblin}, Benjamin and {Cai}, Yan-Chuan and {Harnois-D{\'e}raps}, Joachim},
        title = "{Enhancing cosmic shear with the multiscale lensing probability density function}",
      journal = {Monthly Notices of the Royal Astronomical Society},
         year = 2023,
        month = apr,
       volume = {520},
       number = {2},
        pages = {1721-1737},
          doi = {10.1093/mnras/stad230},
archivePrefix = {arXiv},
       eprint = {2211.05708},
 primaryClass = {astro-ph.CO},
       adsurl = {https://ui.adsabs.harvard.edu/abs/2023MNRAS.520.1721G}
}

@ARTICLE{2023A&A...675A.120E,
       author = {{Euclid Collaboration} and {Ajani}, V. and {Baldi}, M. and {Barthelemy}, A. and {Boyle}, A. and {Burger}, P. and {Cardone}, V.~F. and {Cheng}, S. and {Codis}, S. and {Giocoli}, C. and {Harnois-D{\'e}raps}, J. and {Heydenreich}, S. and {Kansal}, V. and {Kilbinger}, M. and {Linke}, L. and {Llinares}, C. and {Martinet}, N. and {Parroni}, C. and {Peel}, A. and {Pires}, S. and {Porth}, L. and {Tereno}, I. and {Uhlemann}, C. and {Vicinanza}, M. and {Vinciguerra}, S. and {Aghanim}, N. and {Auricchio}, N. and {Bonino}, D. and {Branchini}, E. and {Brescia}, M. and {Brinchmann}, J. and {Camera}, S. and {Capobianco}, V. and {Carbone}, C. and {Carretero}, J. and {Castander}, F.~J. and {Castellano}, M. and {Cavuoti}, S. and {Cimatti}, A. and {Cledassou}, R. and {Congedo}, G. and {Conselice}, C.~J. and {Conversi}, L. and {Corcione}, L. and {Courbin}, F. and {Cropper}, M. and {Da Silva}, A. and {Degaudenzi}, H. and {Di Giorgio}, A.~M. and {Dinis}, J. and {Douspis}, M. and {Dubath}, F. and {Dupac}, X. and {Farrens}, S. and {Ferriol}, S. and {Fosalba}, P. and {Frailis}, M. and {Franceschi}, E. and {Galeotta}, S. and {Garilli}, B. and {Gillis}, B. and {Grazian}, A. and {Grupp}, F. and {Hoekstra}, H. and {Holmes}, W. and {Hornstrup}, A. and {Hudelot}, P. and {Jahnke}, K. and {Jhabvala}, M. and {K{\"u}mmel}, M. and {Kitching}, T. and {Kunz}, M. and {Kurki-Suonio}, H. and {Lilje}, P.~B. and {Lloro}, I. and {Maiorano}, E. and {Mansutti}, O. and {Marggraf}, O. and {Markovic}, K. and {Marulli}, F. and {Massey}, R. and {Mei}, S. and {Mellier}, Y. and {Meneghetti}, M. and {Moresco}, M. and {Moscardini}, L. and {Niemi}, S.-M. and {Nightingale}, J. and {Nutma}, T. and {Padilla}, C. and {Paltani}, S. and {Pedersen}, K. and {Pettorino}, V. and {Polenta}, G. and {Poncet}, M. and {Popa}, L.~A. and {Raison}, F. and {Renzi}, A. and {Rhodes}, J. and {Riccio}, G. and {Romelli}, E. and {Roncarelli}, M. and {Rossetti}, E. and {Saglia}, R. and {Sapone}, D. and {Sartoris}, B. and {Schneider}, P. and {Schrabback}, T. and {Secroun}, A. and {Seidel}, G. and {Serrano}, S. and {Sirignano}, C. and {Stanco}, L. and {Starck}, J.-L. and {Tallada-Cresp{\'\i}}, P. and {Taylor}, A.~N. and {Toledo-Moreo}, R. and {Torradeflot}, F. and {Tutusaus}, I. and {Valentijn}, E.~A. and {Valenziano}, L. and {Vassallo}, T. and {Wang}, Y. and {Weller}, J. and {Zamorani}, G. and {Zoubian}, J. and {Andreon}, S. and {Bardelli}, S. and {Boucaud}, A. and {Bozzo}, E. and {Colodro-Conde}, C. and {Di Ferdinando}, D. and {Fabbian}, G. and {Farina}, M. and {Graci{\'a}-Carpio}, J. and {Keih{\"a}nen}, E. and {Lindholm}, V. and {Maino}, D. and {Mauri}, N. and {Neissner}, C. and {Schirmer}, M. and {Scottez}, V. and {Zucca}, E. and {Akrami}, Y. and {Baccigalupi}, C. and {Balaguera-Antol{\'\i}nez}, A. and {Ballardini}, M. and {Bernardeau}, F. and {Biviano}, A. and {Blanchard}, A. and {Borgani}, S. and {Borlaff}, A.~S. and {Burigana}, C. and {Cabanac}, R. and {Cappi}, A. and {Carvalho}, C.~S. and {Casas}, S. and {Castignani}, G. and {Castro}, T. and {Chambers}, K.~C. and {Cooray}, A.~R. and {Coupon}, J. and {Courtois}, H.~M. and {Davini}, S. and {de la Torre}, S. and {De Lucia}, G. and {Desprez}, G. and {Dole}, H. and {Escartin}, J.~A. and {Escoffier}, S. and {Ferrero}, I. and {Finelli}, F. and {Ganga}, K. and {Garcia-Bellido}, J. and {George}, K. and {Giacomini}, F. and {Gozaliasl}, G. and {Hildebrandt}, H. and {Jimenez Mu{\~n}oz}, A. and {Joachimi}, B. and {Kajava}, J.~J.~E. and {Kirkpatrick}, C.~C. and {Legrand}, L. and {Loureiro}, A. and {Magliocchetti}, M. and {Maoli}, R. and {Marcin}, S. and {Martinelli}, M. and {Martins}, C.~J.~A.~P. and {Matthew}, S. and {Maurin}, L. and {Metcalf}, R.~B. and {Monaco}, P. and {Morgante}, G. and {Nadathur}, S. and {Nucita}, A.~A. and {Popa}, V. and {Potter}, D. and {Pourtsidou}, A. and {P{\"o}ntinen}, M.},
        title = "{Euclid preparation. XXVIII. Forecasts for ten different higher-order weak lensing statistics}",
      journal = {Astronomy \& Astrophysics},
         year = 2023,
        month = jul,
       volume = {675},
          eid = {A120},
        pages = {A120},
          doi = {10.1051/0004-6361/202346017},
archivePrefix = {arXiv},
       eprint = {2301.12890},
 primaryClass = {astro-ph.CO},
       adsurl = {https://ui.adsabs.harvard.edu/abs/2023A&A...675A.120E}
}

@ARTICLE{2012PhRvD..85j3513K,
       author = {{Kratochvil}, Jan M. and {Lim}, Eugene A. and {Wang}, Sheng and {Haiman}, Zolt{\'a}n and {May}, Morgan and {Huffenberger}, Kevin},
        title = "{Probing cosmology with weak lensing Minkowski functionals}",
      journal = {Physics Review D},
         year = 2012,
        month = may,
       volume = {85},
       number = {10},
          eid = {103513},
        pages = {103513},
          doi = {10.1103/PhysRevD.85.103513},
archivePrefix = {arXiv},
       eprint = {1109.6334},
 primaryClass = {astro-ph.CO},
       adsurl = {https://ui.adsabs.harvard.edu/abs/2012PhRvD..85j3513K}
}

@ARTICLE{2017A&A...599A..79P,
       author = {{Peel}, Austin and {Lin}, Chieh-An and {Lanusse}, Fran{\c{c}}ois and {Leonard}, Adrienne and {Starck}, Jean-Luc and {Kilbinger}, Martin},
        title = "{Cosmological constraints with weak-lensing peak counts and second-order statistics in a large-field survey}",
      journal = {Astronomy \& Astrophysics},
         year = 2017,
        month = mar,
       volume = {599},
          eid = {A79},
        pages = {A79},
          doi = {10.1051/0004-6361/201629928},
archivePrefix = {arXiv},
       eprint = {1612.02264},
 primaryClass = {astro-ph.CO},
       adsurl = {https://ui.adsabs.harvard.edu/abs/2017A&A...599A..79P}
}

@ARTICLE{2023arXiv231102449C,
       author = {{Carqueville}, Nils and {Del Zotto}, Michele and {Runkel}, Ingo},
        title = "{Topological defects}",
      journal = {arXiv e-prints},
         year = 2023,
        month = nov,
          eid = {arXiv:2311.02449},
        pages = {arXiv:2311.02449},
          doi = {10.48550/arXiv.2311.02449},
archivePrefix = {arXiv},
       eprint = {2311.02449},
 primaryClass = {math-ph},
       adsurl = {https://ui.adsabs.harvard.edu/abs/2023arXiv231102449C}
}

@ARTICLE{2005PhRvD..72d3004H,
       author = {{Huterer}, Dragan and {Vachaspati}, Tanmay},
        title = "{Distribution of singularities in the cosmic microwave background polarization}",
      journal = {Physics Review D},
         year = 2005,
        month = aug,
       volume = {72},
       number = {4},
          eid = {043004},
        pages = {043004},
          doi = {10.1103/PhysRevD.72.043004},
archivePrefix = {arXiv},
       eprint = {astro-ph/0405474},
 primaryClass = {astro-ph},
       adsurl = {https://ui.adsabs.harvard.edu/abs/2005PhRvD..72d3004H}
}

@ARTICLE{2009JCAP...09..034V,
       author = {{Vitelli}, Vincenzo and {Jain}, Bhuvnesh and {Kamien}, Randall D.},
        title = "{Topological defects in gravitational lensing shear fields}",
      journal = {Journal of Cosmology and Astroparticle Physics},
         year = 2009,
        month = sep,
       volume = {2009},
       number = {9},
          eid = {034},
        pages = {034},
          doi = {10.1088/1475-7516/2009/09/034},
archivePrefix = {arXiv},
       eprint = {0906.0124},
 primaryClass = {astro-ph.CO},
       adsurl = {https://ui.adsabs.harvard.edu/abs/2009JCAP...09..034V}
}

@ARTICLE{2003JPhA...36.6611D,
       author = {{Dennis}, M.~R.},
        title = "{Correlations and screening of topological charges in Gaussian random fields}",
      journal = {Journal of Physics A Mathematical General},
         year = 2003,
        month = jun,
       volume = {36},
       number = {24},
        pages = {6611-6628},
          doi = {10.1088/0305-4470/36/24/301},
archivePrefix = {arXiv},
       eprint = {math-ph/0302024},
 primaryClass = {math-ph},
       adsurl = {https://ui.adsabs.harvard.edu/abs/2003JPhA...36.6611D}
}

@ARTICLE{2023OJAp....6E..11T,
       author = {{Tessore}, Nicolas and {Loureiro}, Arthur and {Joachimi}, Benjamin and {von Wietersheim-Kramsta}, Maximilian and {Jeffrey}, Niall},
        title = "{GLASS: Generator for Large Scale Structure}",
      journal = {The Open Journal of Astrophysics},
         year = 2023,
        month = mar,
       volume = {6},
          eid = {11},
        pages = {11},
          doi = {10.21105/astro.2302.01942},
archivePrefix = {arXiv},
       eprint = {2302.01942},
 primaryClass = {astro-ph.CO},
       adsurl = {https://ui.adsabs.harvard.edu/abs/2023OJAp....6E..11T}
}

@ARTICLE{2018MNRAS.481.2813G,
       author = {{Giocoli}, Carlo and {Baldi}, Marco and {Moscardini}, Lauro},
        title = "{Weak lensing light-cones in modified gravity simulations with and without massive neutrinos}",
      journal = {Monthly Notices of the Royal Astronomical Society},
         year = 2018,
        month = dec,
       volume = {481},
       number = {2},
        pages = {2813-2828},
          doi = {10.1093/mnras/sty2465},
archivePrefix = {arXiv},
       eprint = {1806.04681},
 primaryClass = {astro-ph.CO},
       adsurl = {https://ui.adsabs.harvard.edu/abs/2018MNRAS.481.2813G}
}

@ARTICLE{2018MNRAS.481.1337H,
       author = {{Harnois-D{\'e}raps}, J. and {Amon}, A. and {Choi}, A. and {Demchenko}, V. and {Heymans}, C. and {Kannawadi}, A. and {Nakajima}, R. and {Sirks}, E. and {van Waerbeke}, L. and {Cai}, Yan-Chuan and {Giblin}, B. and {Hildebrandt}, H. and {Hoekstra}, H. and {Miller}, L. and {Tr{\"o}ster}, T.},
        title = "{Cosmological simulations for combined-probe analyses: covariance and neighbour-exclusion bias}",
      journal = {Monthly Notices of the Royal Astronomical Society},
         year = 2018,
        month = nov,
       volume = {481},
       number = {1},
        pages = {1337-1367},
          doi = {10.1093/mnras/sty2319},
archivePrefix = {arXiv},
       eprint = {1805.04511},
 primaryClass = {astro-ph.CO},
       adsurl = {https://ui.adsabs.harvard.edu/abs/2018MNRAS.481.1337H}
}

@ARTICLE{2021MNRAS.500.5479B,
       author = {{Banerjee}, Arka and {Abel}, Tom},
        title = "{Nearest neighbour distributions: New statistical measures for cosmological clustering}",
      journal = {Monthly Notices of the Royal Astronomical Society},
         year = 2021,
        month = jan,
       volume = {500},
       number = {4},
        pages = {5479-5499},
          doi = {10.1093/mnras/staa3604},
archivePrefix = {arXiv},
       eprint = {2007.13342},
 primaryClass = {astro-ph.CO},
       adsurl = {https://ui.adsabs.harvard.edu/abs/2021MNRAS.500.5479B}
}

@ARTICLE{2007A&A...464..399H,
       author = {{Hartlap}, J. and {Simon}, P. and {Schneider}, P.},
        title = "{Why your model parameter confidences might be too optimistic. Unbiased estimation of the inverse covariance matrix}",
      journal = {Astronomy \& Astrophysics},
         year = 2007,
        month = mar,
       volume = {464},
       number = {1},
        pages = {399-404},
          doi = {10.1051/0004-6361:20066170},
archivePrefix = {arXiv},
       eprint = {astro-ph/0608064},
 primaryClass = {astro-ph},
       adsurl = {https://ui.adsabs.harvard.edu/abs/2007A&A...464..399H}
}

@ARTICLE{2005ApJ...622..759G,
       author = {{G{\'o}rski}, K.~M. and {Hivon}, E. and {Banday}, A.~J. and {Wandelt}, B.~D. and {Hansen}, F.~K. and {Reinecke}, M. and {Bartelmann}, M.},
        title = "{HEALPix: A Framework for High-Resolution Discretization and Fast Analysis of Data Distributed on the Sphere}",
      journal = {The Astrophysical Journal},
         year = 2005,
        month = apr,
       volume = {622},
       number = {2},
        pages = {759-771},
          doi = {10.1086/427976},
archivePrefix = {arXiv},
       eprint = {astro-ph/0409513},
 primaryClass = {astro-ph},
       adsurl = {https://ui.adsabs.harvard.edu/abs/2005ApJ...622..759G}
}

@software{2015ascl.soft08007J,
       author = {{Jarvis}, Mike},
        title = "{TreeCorr: Two-point correlation functions}",
 howpublished = {Astrophysics Source Code Library, record ascl:1508.007},
         year = 2015,
        month = aug,
          eid = {ascl:1508.007},
archivePrefix = {ascl},
       eprint = {1508.007},
       adsurl = {https://ui.adsabs.harvard.edu/abs/2015ascl.soft08007J}
}

@ARTICLE{2020NatMe..17..261V,
       author = {{Virtanen}, Pauli and {Gommers}, Ralf and {Oliphant}, Travis E. and {Haberland}, Matt and {Reddy}, Tyler and {Cournapeau}, David and {Burovski}, Evgeni and {Peterson}, Pearu and {Weckesser}, Warren and {Bright}, Jonathan and {van der Walt}, St{\'e}fan J. and {Brett}, Matthew and {Wilson}, Joshua and {Millman}, K. Jarrod and {Mayorov}, Nikolay and {Nelson}, Andrew R.~J. and {Jones}, Eric and {Kern}, Robert and {Larson}, Eric and {Carey}, C.~J. and {Polat}, {\.I}lhan and {Feng}, Yu and {Moore}, Eric W. and {VanderPlas}, Jake and {Laxalde}, Denis and {Perktold}, Josef and {Cimrman}, Robert and {Henriksen}, Ian and {Quintero}, E.~A. and {Harris}, Charles R. and {Archibald}, Anne M. and {Ribeiro}, Ant{\^o}nio H. and {Pedregosa}, Fabian and {van Mulbregt}, Paul and {SciPy 1.  0 Contributors}},
        title = "{SciPy 1.0: fundamental algorithms for scientific computing in Python}",
      journal = {Nature Medicine},
         year = 2020,
        month = feb,
       volume = {17},
        pages = {261-272},
          doi = {10.1038/s41592-019-0686-2},
archivePrefix = {arXiv},
       eprint = {1907.10121},
 primaryClass = {cs.MS},
       adsurl = {https://ui.adsabs.harvard.edu/abs/2020NatMe..17..261V}
}

@ARTICLE{2011JMLR...12.2825P,
       author = {{Pedregosa}, Fabian and {Varoquaux}, Ga{\"e}l and {Gramfort}, Alexandre and {Michel}, Vincent and {Thirion}, Bertrand and {Grisel}, Olivier and {Blondel}, Mathieu and {M{\"u}ller}, Andreas and {Nothman}, Joel and {Louppe}, Gilles and {Prettenhofer}, Peter and {Weiss}, Ron and {Dubourg}, Vincent and {Vanderplas}, Jake and {Passos}, Alexandre and {Cournapeau}, David and {Brucher}, Matthieu and {Perrot}, Matthieu and {Duchesnay}, {\'E}douard},
        title = "{Scikit-learn: Machine Learning in Python}",
      journal = {Journal of Machine Learning Research},
         year = 2011,
        month = oct,
       volume = {12},
        pages = {2825-2830},
          doi = {10.48550/arXiv.1201.0490},
archivePrefix = {arXiv},
       eprint = {1201.0490},
 primaryClass = {cs.LG},
       adsurl = {https://ui.adsabs.harvard.edu/abs/2011JMLR...12.2825P}
}

@ARTICLE{2025JCAP...02..021A,
       author = {{Adame}, A.~G. and {Aguilar}, J. and {Ahlen}, S. and {Alam}, S. and {Alexander}, D.~M. and {Alvarez}, M. and {Alves}, O. and {Anand}, A. and {Andrade}, U. and {Armengaud}, E. and {Avila}, S. and {Aviles}, A. and {Awan}, H. and {Bahr-Kalus}, B. and {Bailey}, S. and {Baltay}, C. and {Bault}, A. and {Behera}, J. and {BenZvi}, S. and {Bera}, A. and {Beutler}, F. and {Bianchi}, D. and {Blake}, C. and {Blum}, R. and {Brieden}, S. and {Brodzeller}, A. and {Brooks}, D. and {Buckley-Geer}, E. and {Burtin}, E. and {Calderon}, R. and {Canning}, R. and {Carnero Rosell}, A. and {Cereskaite}, R. and {Cervantes-Cota}, J.~L. and {Chabanier}, S. and {Chaussidon}, E. and {Chaves-Montero}, J. and {Chen}, S. and {Chen}, X. and {Claybaugh}, T. and {Cole}, S. and {Cuceu}, A. and {Davis}, T.~M. and {Dawson}, K. and {de la Macorra}, A. and {de Mattia}, A. and {Deiosso}, N. and {Dey}, A. and {Dey}, B. and {Ding}, Z. and {Doel}, P. and {Edelstein}, J. and {Eftekharzadeh}, S. and {Eisenstein}, D.~J. and {Elliott}, A. and {Fagrelius}, P. and {Fanning}, K. and {Ferraro}, S. and {Ereza}, J. and {Findlay}, N. and {Flaugher}, B. and {Font-Ribera}, A. and {Forero-S{\'a}nchez}, D. and {Forero-Romero}, J.~E. and {Frenk}, C.~S. and {Garcia-Quintero}, C. and {Gazta{\~n}aga}, E. and {Gil-Mar{\'\i}n}, H. and {Gontcho a Gontcho}, S. and {Gonzalez-Morales}, A.~X. and {Gonzalez-Perez}, V. and {Gordon}, C. and {Green}, D. and {Gruen}, D. and {Gsponer}, R. and {Gutierrez}, G. and {Guy}, J. and {Hadzhiyska}, B. and {Hahn}, C. and {Hanif}, M.~M.~S. and {Herrera-Alcantar}, H.~K. and {Honscheid}, K. and {Howlett}, C. and {Huterer}, D. and {Ir{\v{s}}i{\v{c}}}, V. and {Ishak}, M. and {Juneau}, S. and {Kara{\c{c}}ayl{\i}}, N.~G. and {Kehoe}, R. and {Kent}, S. and {Kirkby}, D. and {Kremin}, A. and {Krolewski}, A. and {Lai}, Y. and {Lan}, T.-W. and {Landriau}, M. and {Lang}, D. and {Lasker}, J. and {Le Goff}, J.~M. and {Le Guillou}, L. and {Leauthaud}, A. and {Levi}, M.~E. and {Li}, T.~S. and {Linder}, E. and {Lodha}, K. and {Magneville}, C. and {Manera}, M. and {Margala}, D. and {Martini}, P. and {Maus}, M. and {McDonald}, P. and {Medina-Varela}, L. and {Meisner}, A. and {Mena-Fern{\'a}ndez}, J. and {Miquel}, R. and {Moon}, J. and {Moore}, S. and {Moustakas}, J. and {Mueller}, E. and {Mu{\~n}oz-Guti{\'e}rrez}, A. and {Myers}, A.~D. and {Nadathur}, S. and {Napolitano}, L. and {Neveux}, R. and {Newman}, J.~A. and {Nguyen}, N.~M. and {Nie}, J. and {Niz}, G. and {Noriega}, H.~E. and {Padmanabhan}, N. and {Paillas}, E. and {Palanque-Delabrouille}, N. and {Pan}, J. and {Penmetsa}, S. and {Percival}, W.~J. and {Pieri}, M.~M. and {Pinon}, M. and {Poppett}, C. and {Porredon}, A. and {Prada}, F. and {P{\'e}rez-Fern{\'a}ndez}, A. and {P{\'e}rez-R{\`a}fols}, I. and {Rabinowitz}, D. and {Raichoor}, A. and {Ram{\'\i}rez-P{\'e}rez}, C. and {Ramirez-Solano}, S. and {Rashkovetskyi}, M. and {Ravoux}, C. and {Rezaie}, M. and {Rich}, J. and {Rocher}, A. and {Rockosi}, C. and {Roe}, N.~A. and {Rosado-Marin}, A. and {Ross}, A.~J. and {Rossi}, G. and {Ruggeri}, R. and {Ruhlmann-Kleider}, V. and {Samushia}, L. and {Sanchez}, E. and {Saulder}, C. and {Schlafly}, E.~F. and {Schlegel}, D. and {Schubnell}, M. and {Seo}, H. and {Shafieloo}, A. and {Sharples}, R. and {Silber}, J. and {Slosar}, A. and {Smith}, A. and {Sprayberry}, D. and {Tan}, T. and {Tarl{\'e}}, G. and {Taylor}, P. and {Trusov}, S. and {Ure{\~n}a-L{\'o}pez}, L.~A. and {Vaisakh}, R. and {Valcin}, D. and {Valdes}, F. and {Vargas-Maga{\~n}a}, M. and {Verde}, L. and {Walther}, M. and {Wang}, B. and {Wang}, M.~S. and {Weaver}, B.~A. and {Weaverdyck}, N. and {Wechsler}, R.~H. and {Weinberg}, D.~H. and {White}, M. and {Yu}, J. and {Yu}, Y. and {Yuan}, S. and {Y{\`e}che}, C. and {Zaborowski}, E.~A. and {Zarrouk}, P. and {Zhang}, H. and {Zhao}, C. and {Zhao}, R. and {Zhou}, R. and {Zhuang}, T.},
        title = "{DESI 2024 VI: cosmological constraints from the measurements of baryon acoustic oscillations}",
      journal = {Journal of Cosmology and Astroparticle Physics},
         year = 2025,
        month = feb,
       volume = {2025},
       number = {2},
          eid = {021},
        pages = {021},
          doi = {10.1088/1475-7516/2025/02/021},
archivePrefix = {arXiv},
       eprint = {2404.03002},
 primaryClass = {astro-ph.CO},
       adsurl = {https://ui.adsabs.harvard.edu/abs/2025JCAP...02..021A}
}

@ARTICLE{2025PhRvD.112h3515A,
       author = {{Abdul Karim}, M. and {Aguilar}, J. and {Ahlen}, S. and {Alam}, S. and {Allen}, L. and {Prieto}, C. Allende and {Alves}, O. and {Anand}, A. and {Andrade}, U. and {Armengaud}, E. and {Aviles}, A. and {Bailey}, S. and {Baltay}, C. and {Bansal}, P. and {Bault}, A. and {Behera}, J. and {BenZvi}, S. and {Bianchi}, D. and {Blake}, C. and {Brieden}, S. and {Brodzeller}, A. and {Brooks}, D. and {Buckley-Geer}, E. and {Burtin}, E. and {Calderon}, R. and {Canning}, R. and {Rosell}, A. Carnero and {Carrilho}, P. and {Casas}, L. and {Castander}, F.~J. and {Charles}, M. and {Chaussidon}, E. and {Chaves-Montero}, J. and {Chebat}, D. and {Chen}, X. and {Claybaugh}, T. and {Cole}, S. and {Cooper}, A.~P. and {Cuceu}, A. and {Dawson}, K.~S. and {de la Macorra}, A. and {de Mattia}, A. and {Deiosso}, N. and {Della Costa}, J. and {Demina}, R. and {Dey}, A. and {Dey}, B. and {Ding}, Z. and {Doel}, P. and {Edelstein}, J. and {Eisenstein}, D.~J. and {Elbers}, W. and {Fagrelius}, P. and {Fanning}, K. and {Fern{\'a}ndez-Garc{\'\i}a}, E. and {Ferraro}, S. and {Font-Ribera}, A. and {Forero-Romero}, J.~E. and {Frenk}, C.~S. and {Garcia-Quintero}, C. and {Garrison}, L.~H. and {Gazta{\~n}aga}, E. and {Gil-Mar{\'\i}n}, H. and {Gontcho A Gontcho}, S. and {Gonzalez}, D. and {Gonzalez-Morales}, A.~X. and {Gordon}, C. and {Green}, D. and {Gutierrez}, G. and {Guy}, J. and {Hadzhiyska}, B. and {Hahn}, C. and {He}, S. and {Herbold}, M. and {Herrera-Alcantar}, H.~K. and {Ho}, M.-F. and {Honscheid}, K. and {Howlett}, C. and {Huterer}, D. and {Ishak}, M. and {Juneau}, S. and {Kamble}, N.~V. and {Kara{\c{c}}ayl{\i}}, N.~G. and {Kehoe}, R. and {Kent}, S. and {Kim}, A.~G. and {Kirkby}, D. and {Kisner}, T. and {Koposov}, S.~E. and {Kremin}, A. and {Krolewski}, A. and {Lahav}, O. and {Lamman}, C. and {Landriau}, M. and {Lang}, D. and {Lasker}, J. and {Le Goff}, J.~M. and {Le Guillou}, L. and {Leauthaud}, A. and {Levi}, M.~E. and {Li}, Q. and {Li}, T.~S. and {Lodha}, K. and {Lokken}, M. and {Lozano-Rodr{\'\i}guez}, F. and {Magneville}, C. and {Manera}, M. and {Martini}, P. and {Matthewson}, W.~L. and {Meisner}, A. and {Mena-Fern{\'a}ndez}, J. and {Menegas}, A. and {Mergulh{\~a}o}, T. and {Miquel}, R. and {Moustakas}, J. and {Mu{\~n}oz-Guti{\'e}rrez}, A. and {Mu{\~n}oz-Santos}, D. and {Myers}, A.~D. and {Nadathur}, S. and {Naidoo}, K. and {Napolitano}, L. and {Newman}, J.~A. and {Niz}, G. and {Noriega}, H.~E. and {Paillas}, E. and {Palanque-Delabrouille}, N. and {Pan}, J. and {Peacock}, J.~A. and {Pellejero Ibanez}, M. and {Percival}, W.~J. and {P{\'e}rez-Fern{\'a}ndez}, A. and {P{\'e}rez-R{\`a}fols}, I. and {Pieri}, M.~M. and {Poppett}, C. and {Prada}, F. and {Rabinowitz}, D. and {Raichoor}, A. and {Ram{\'\i}rez-P{\'e}rez}, C. and {Rashkovetskyi}, M. and {Ravoux}, C. and {Rich}, J. and {Rocher}, A. and {Rockosi}, C. and {Rohlf}, J. and {Rom{\'a}n-Herrera}, J.~O. and {Ross}, A.~J. and {Rossi}, G. and {Ruggeri}, R. and {Ruhlmann-Kleider}, V. and {Samushia}, L. and {Sanchez}, E. and {Sanders}, N. and {Schlegel}, D. and {Schubnell}, M. and {Seo}, H. and {Shafieloo}, A. and {Sharples}, R. and {Silber}, J. and {Sinigaglia}, F. and {Sprayberry}, D. and {Tan}, T. and {Tarl{\'e}}, G. and {Taylor}, P. and {Turner}, W. and {Ure{\~n}a-L{\'o}pez}, L.~A. and {Vaisakh}, R. and {Valdes}, F. and {Valogiannis}, G. and {Vargas-Maga{\~n}a}, M. and {Verde}, L. and {Walther}, M. and {Weaver}, B.~A. and {Weinberg}, D.~H. and {White}, M. and {Wolfson}, M. and {Y{\`e}che}, C. and {Yu}, J. and {Zaborowski}, E.~A. and {Zarrouk}, P. and {Zhai}, Z. and {Zhang}, H. and {Zhao}, C. and {Zhao}, G.~B. and {Zhou}, R. and {Zou}, H. and {DESI Collaboration}},
        title = "{DESI DR2 results. II. Measurements of baryon acoustic oscillations and cosmological constraints}",
      journal = {Physics Review D},
         year = 2025,
        month = oct,
       volume = {112},
       number = {8},
          eid = {083515},
        pages = {083515},
          doi = {10.1103/tr6y-kpc6},
archivePrefix = {arXiv},
       eprint = {2503.14738},
 primaryClass = {astro-ph.CO},
       adsurl = {https://ui.adsabs.harvard.edu/abs/2025PhRvD.112h3515A}
}

@ARTICLE{2021CQGra..38o3001D,
       author = {{Di Valentino}, Eleonora and {Mena}, Olga and {Pan}, Supriya and {Visinelli}, Luca and {Yang}, Weiqiang and {Melchiorri}, Alessandro and {Mota}, David F. and {Riess}, Adam G. and {Silk}, Joseph},
        title = "{In the realm of the Hubble tension-a review of solutions}",
      journal = {Classical and Quantum Gravity},
         year = 2021,
        month = jul,
       volume = {38},
       number = {15},
          eid = {153001},
        pages = {153001},
          doi = {10.1088/1361-6382/ac086d},
archivePrefix = {arXiv},
       eprint = {2103.01183},
 primaryClass = {astro-ph.CO},
       adsurl = {https://ui.adsabs.harvard.edu/abs/2021CQGra..38o3001D}
}

@ARTICLE{2025PhRvD.111h3510N,
       author = {{Novaes}, Camila P. and {Thiele}, Leander and {Armijo}, Joaquin and {Cheng}, Sihao and {Cowell}, Jessica A. and {Marques}, Gabriela A. and {Ferreira}, Elisa G.~M. and {Shirasaki}, Masato and {Osato}, Ken and {Liu}, Jia},
        title = "{Cosmology from HSC Y1 weak lensing data with combined higher-order statistics and simulation-based inference}",
      journal = {Physics Review D},
         year = 2025,
        month = apr,
       volume = {111},
       number = {8},
          eid = {083510},
        pages = {083510},
          doi = {10.1103/PhysRevD.111.083510},
archivePrefix = {arXiv},
       eprint = {2409.01301},
 primaryClass = {astro-ph.CO},
       adsurl = {https://ui.adsabs.harvard.edu/abs/2025PhRvD.111h3510N}
}

@ARTICLE{2024MNRAS.533.3209F,
       author = {{Ferlito}, Fulvio and {Davies}, Christopher T. and {Springel}, Volker and {Reinecke}, Martin and {Greco}, Alessandro and {Delgado}, Ana Maria and {White}, Simon D.~M. and {Hern{\'a}ndez-Aguayo}, C{\'e}sar and {Bose}, Sownak and {Hernquist}, Lars},
        title = "{Ray-tracing versus Born approximation in full-sky weak lensing simulations of the MillenniumTNG project}",
      journal = {Monthly Notices of the Royal Astronomical Society},
         year = 2024,
        month = sep,
       volume = {533},
       number = {3},
        pages = {3209-3221},
          doi = {10.1093/mnras/stae2019},
archivePrefix = {arXiv},
       eprint = {2406.08540},
 primaryClass = {astro-ph.CO},
       adsurl = {https://ui.adsabs.harvard.edu/abs/2024MNRAS.533.3209F}
}

@ARTICLE{1993ApJ...412...64L,
       author = {{Landy}, Stephen D. and {Szalay}, Alexander S.},
        title = "{Bias and Variance of Angular Correlation Functions}",
      journal = {The Astrophysical Journal},
         year = 1993,
        month = jul,
       volume = {412},
        pages = {64},
          doi = {10.1086/172900},
       adsurl = {https://ui.adsabs.harvard.edu/abs/1993ApJ...412...64L}
}

@ARTICLE{2023OJAp....6E..15C,
       author = {{Campagne}, Jean-Eric and {Lanusse}, Fran{\c{c}}ois and {Zuntz}, Joe and {Boucaud}, Alexandre and {Casas}, Santiago and {Karamanis}, Minas and {Kirkby}, David and {Lanzieri}, Denise and {Peel}, Austin and {Li}, Yin},
        title = "{JAX-COSMO: An End-to-End Differentiable and GPU Accelerated Cosmology Library}",
      journal = {The Open Journal of Astrophysics},
         year = 2023,
        month = apr,
       volume = {6},
          eid = {15},
        pages = {15},
          doi = {10.21105/astro.2302.05163},
archivePrefix = {arXiv},
       eprint = {2302.05163},
 primaryClass = {astro-ph.CO},
       adsurl = {https://ui.adsabs.harvard.edu/abs/2023OJAp....6E..15C}
}

@article{poincare1881memoire,
  author    = {H. Poincar{\'e}},
  title     = {Mémoire sur les courbes définies par une équation différentielle (I)},
  journal   = {Journal de Mathématiques Pures et Appliquées},
  year      = {1881},
  volume    = {7},
  pages     = {375--422},
}

@article{hopf1926vektorfelder,
  author    = {Heinz Hopf},
  title     = {Vektorfelder in n-dimensionalen Mannigfaltigkeiten},
  journal   = {Mathematische Annalen},
  year      = {1926},
  volume    = {96},
  pages     = {209--221},
}

@ARTICLE{2014ApJ...784...31L,
       author = {{Liu}, Xiangkun and {Wang}, Qiao and {Pan}, Chuzhong and {Fan}, Zuhui},
        title = "{Mask Effects on Cosmological Studies with Weak-lensing Peak Statistics}",
      journal = {The Astrophysical Journal},
         year = 2014,
        month = mar,
       volume = {784},
       number = {1},
          eid = {31},
        pages = {31},
          doi = {10.1088/0004-637X/784/1/31},
archivePrefix = {arXiv},
       eprint = {1304.2873},
 primaryClass = {astro-ph.CO},
       adsurl = {https://ui.adsabs.harvard.edu/abs/2014ApJ...784...31L}
}

@ARTICLE{2013ApJ...774..111S,
       author = {{Shirasaki}, Masato and {Yoshida}, Naoki and {Hamana}, Takashi},
        title = "{Effect of Masked Regions on Weak-lensing Statistics}",
      journal = {The Astrophysical Journal},
         year = 2013,
        month = sep,
       volume = {774},
       number = {2},
          eid = {111},
        pages = {111},
          doi = {10.1088/0004-637X/774/2/111},
archivePrefix = {arXiv},
       eprint = {1304.2164},
 primaryClass = {astro-ph.CO},
       adsurl = {https://ui.adsabs.harvard.edu/abs/2013ApJ...774..111S}
}

@ARTICLE{2023MNRAS.521.1152R,
       author = {{Rizzato}, Matteo and {Sellentin}, Elena},
        title = "{Extremely expensive likelihoods: a variational-Bayes solution for precision cosmology}",
      journal = {Monthly Notices of the Royal Astronomical Society},
         year = 2023,
        month = may,
       volume = {521},
       number = {1},
        pages = {1152-1161},
          doi = {10.1093/mnras/stad638},
archivePrefix = {arXiv},
       eprint = {2203.05009},
 primaryClass = {astro-ph.CO},
       adsurl = {https://ui.adsabs.harvard.edu/abs/2023MNRAS.521.1152R}
}

@article{10.1145/2168773.2168777,
author = {Mor\'{e}, Jorge J. and Wild, Stefan M.},
title = {Estimating Derivatives of Noisy Simulations},
year = {2012},
issue_date = {April 2012},
publisher = {Association for Computing Machinery},
address = {New York, NY, USA},
volume = {38},
number = {3},
issn = {0098-3500},
url = {https://doi.org/10.1145/2168773.2168777},
doi = {10.1145/2168773.2168777},
journal = {ACM Trans. Math. Softw.},
month = apr,
articleno = {19},
numpages = {21}
}

@ARTICLE{2019ApJS..242....2C,
       author = {{Chisari}, Nora Elisa and {Alonso}, David and {Krause}, Elisabeth and {Leonard}, C. Danielle and {Bull}, Philip and {Neveu}, J{\'e}r{\'e}my and {Villarreal}, Antonia Sierra and {Singh}, Sukhdeep and {McClintock}, Thomas and {Ellison}, John and {Du}, Zilong and {Zuntz}, Joe and {Mead}, Alexander and {Joudaki}, Shahab and {Lorenz}, Christiane S. and {Tr{\"o}ster}, Tilman and {Sanchez}, Javier and {Lanusse}, Francois and {Ishak}, Mustapha and {Hlozek}, Ren{\'e}e and {Blazek}, Jonathan and {Campagne}, Jean-Eric and {Almoubayyed}, Husni and {Eifler}, Tim and {Kirby}, Matthew and {Kirkby}, David and {Plaszczynski}, St{\'e}phane and {Slosar}, An{\v{z}}e and {Vrastil}, Michal and {Wagoner}, Erika L. and {LSST Dark Energy Science Collaboration}},
        title = "{Core Cosmology Library: Precision Cosmological Predictions for LSST}",
      journal = {The Astrophysical Journal Supplement Series},
         year = 2019,
        month = may,
       volume = {242},
       number = {1},
          eid = {2},
        pages = {2},
          doi = {10.3847/1538-4365/ab1658},
archivePrefix = {arXiv},
       eprint = {1812.05995},
 primaryClass = {astro-ph.CO},
       adsurl = {https://ui.adsabs.harvard.edu/abs/2019ApJS..242....2C}
}

@ARTICLE{2006JCAP...10..013P,
       author = {{Perotto}, Laurence and {Lesgourgues}, Julien and {Hannestad}, Steen and {Tu}, Huitzu and {Y Y Wong}, Yvonne},
        title = "{Probing cosmological parameters with the CMB: forecasts from Monte Carlo simulations}",
      journal = {Journal of Cosmology and Astroparticle Physics},
         year = 2006,
        month = oct,
       volume = {2006},
       number = {10},
          eid = {013},
        pages = {013},
          doi = {10.1088/1475-7516/2006/10/013},
archivePrefix = {arXiv},
       eprint = {astro-ph/0606227},
 primaryClass = {astro-ph},
       adsurl = {https://ui.adsabs.harvard.edu/abs/2006JCAP...10..013P}
}

@ARTICLE{2021MNRAS.506.1623H,
       author = {{Harnois-D{\'e}raps}, Joachim and {Martinet}, Nicolas and {Castro}, Tiago and {Dolag}, Klaus and {Giblin}, Benjamin and {Heymans}, Catherine and {Hildebrandt}, Hendrik and {Xia}, Qianli},
        title = "{Cosmic shear cosmology beyond two-point statistics: a combined peak count and correlation function analysis of DES-Y1}",
      journal = {Monthly Notices of the Royal Astronomical Society},
         year = 2021,
        month = sep,
       volume = {506},
       number = {2},
        pages = {1623-1650},
          doi = {10.1093/mnras/stab1623},
archivePrefix = {arXiv},
       eprint = {2012.02777},
 primaryClass = {astro-ph.CO},
       adsurl = {https://ui.adsabs.harvard.edu/abs/2021MNRAS.506.1623H}
}

@ARTICLE{2024MNRAS.534.3305H,
       author = {{Harnois-D{\'e}raps}, Joachim and {Heydenreich}, Sven and {Giblin}, Benjamin and {Martinet}, Nicolas and {Tr{\"o}ster}, Tilman and {Asgari}, Marika and {Burger}, Pierre and {Castro}, Tiago and {Dolag}, Klaus and {Heymans}, Catherine and {Hildebrandt}, Hendrik and {Joachimi}, Benjamin and {Wright}, Angus H.},
        title = "{KiDS-1000 and DES-Y1 combined: cosmology from peak count statistics}",
      journal = {Monthly Notices of the Royal Astronomical Society},
         year = 2024,
        month = nov,
       volume = {534},
       number = {4},
        pages = {3305-3330},
          doi = {10.1093/mnras/stae2249},
archivePrefix = {arXiv},
       eprint = {2405.10312},
 primaryClass = {astro-ph.CO},
       adsurl = {https://ui.adsabs.harvard.edu/abs/2024MNRAS.534.3305H}
}

@ARTICLE{2021MNRAS.505.2886B,
       author = {{Boyle}, Aoife and {Uhlemann}, Cora and {Friedrich}, Oliver and {Barthelemy}, Alexandre and {Codis}, Sandrine and {Bernardeau}, Francis and {Giocoli}, Carlo and {Baldi}, Marco},
        title = "{Nuw CDM cosmology from the weak-lensing convergence PDF}",
      journal = {Monthly Notices of the Royal Astronomical Society},
         year = 2021,
        month = aug,
       volume = {505},
       number = {2},
        pages = {2886-2902},
          doi = {10.1093/mnras/stab1381},
archivePrefix = {arXiv},
       eprint = {2012.07771},
 primaryClass = {astro-ph.CO},
       adsurl = {https://ui.adsabs.harvard.edu/abs/2021MNRAS.505.2886B}
}

@software{2011ascl.soft02026L,
       author = {{Lewis}, Antony and {Challinor}, Anthony},
        title = "{CAMB: Code for Anisotropies in the Microwave Background}",
 howpublished = {Astrophysics Source Code Library, record ascl:1102.026},
         year = 2011,
        month = feb,
          eid = {ascl:1102.026},
archivePrefix = {ascl},
       eprint = {1102.026},
       adsurl = {https://ui.adsabs.harvard.edu/abs/2011ascl.soft02026L}
}

@ARTICLE{2000ApJ...538..473L,
       author = {{Lewis}, Antony and {Challinor}, Anthony and {Lasenby}, Anthony},
        title = "{Efficient Computation of Cosmic Microwave Background Anisotropies in Closed Friedmann-Robertson-Walker Models}",
      journal = {The Astrophysical Journal},
         year = 2000,
        month = aug,
       volume = {538},
       number = {2},
        pages = {473-476},
          doi = {10.1086/309179},
archivePrefix = {arXiv},
       eprint = {astro-ph/9911177},
 primaryClass = {astro-ph},
       adsurl = {https://ui.adsabs.harvard.edu/abs/2000ApJ...538..473L}
}

@ARTICLE{2013MNRAS.432.1928T,
       author = {{Taylor}, Andy and {Joachimi}, Benjamin and {Kitching}, Thomas},
        title = "{Putting the precision in precision cosmology: How accurate should your data covariance matrix be?}",
      journal = {Monthly Notices of the Royal Astronomical Society},
         year = 2013,
        month = jul,
       volume = {432},
       number = {3},
        pages = {1928-1946},
          doi = {10.1093/mnras/stt270},
archivePrefix = {arXiv},
       eprint = {1212.4359},
 primaryClass = {astro-ph.CO},
       adsurl = {https://ui.adsabs.harvard.edu/abs/2013MNRAS.432.1928T}
}

@ARTICLE{2020Natur.585..357H,
       author = {{Harris}, Charles R. and {Millman}, K. Jarrod and {van der Walt}, St{\'e}fan J. and {Gommers}, Ralf and {Virtanen}, Pauli and {Cournapeau}, David and {Wieser}, Eric and {Taylor}, Julian and {Berg}, Sebastian and {Smith}, Nathaniel J. and {Kern}, Robert and {Picus}, Matti and {Hoyer}, Stephan and {van Kerkwijk}, Marten H. and {Brett}, Matthew and {Haldane}, Allan and {del R{\'\i}o}, Jaime Fern{\'a}ndez and {Wiebe}, Mark and {Peterson}, Pearu and {G{\'e}rard-Marchant}, Pierre and {Sheppard}, Kevin and {Reddy}, Tyler and {Weckesser}, Warren and {Abbasi}, Hameer and {Gohlke}, Christoph and {Oliphant}, Travis E.},
        title = "{Array programming with NumPy}",
      journal = {Nature},
         year = 2020,
        month = sep,
       volume = {585},
       number = {7825},
        pages = {357-362},
          doi = {10.1038/s41586-020-2649-2},
archivePrefix = {arXiv},
       eprint = {2006.10256},
 primaryClass = {cs.MS},
       adsurl = {https://ui.adsabs.harvard.edu/abs/2020Natur.585..357H}
}

@article{MVBerry_1977,
doi = {10.1088/0305-4470/10/11/009},
url = {https://doi.org/10.1088/0305-4470/10/11/009},
year = {1977},
month = {nov},
publisher = {},
volume = {10},
number = {11},
pages = {1809},
author = {M V Berry and J H Hannay},
title = {Umbilic points on Gaussian random surfaces},
journal = {Journal of Physics A: Mathematical and General}
}

@ARTICLE{2001RSPSA.457..141B,
       author = {{Berry}, M.~V. and {Dennis}, M.~R.},
        title = "{Polarization singularities in isotropic random vector waves}",
      journal = {Proceedings of the Royal Society of London Series A},
         year = 2001,
        month = jan,
       volume = {457},
       number = {2005},
        pages = {141-155},
          doi = {10.1098/rspa.2000.0660},
       adsurl = {https://ui.adsabs.harvard.edu/abs/2001RSPSA.457..141B}
}

@ARTICLE{2023EPJST.232.1813F,
       author = {{Fumeron}, S{\'e}bastien and {Berche}, Bertrand},
        title = "{Introduction to topological defects: from liquid crystals to particle physics}",
      journal = {European Physical Journal Special Topics},
         year = 2023,
        month = sep,
       volume = {232},
       number = {11},
        pages = {1813-1833},
          doi = {10.1140/epjs/s11734-023-00803-x},
archivePrefix = {arXiv},
       eprint = {2209.07743},
 primaryClass = {cond-mat.soft},
       adsurl = {https://ui.adsabs.harvard.edu/abs/2023EPJST.232.1813F}
}

@ARTICLE{2004NJPh....6..162B,
       author = {{Berry}, M.~V. and {Dennis}, M.~R. and {Lee}, Jr., R.~L.},
        title = "{Polarization singularities in the clear sky}",
      journal = {New Journal of Physics},
         year = 2004,
        month = nov,
       volume = {6},
       number = {1},
        pages = {162},
          doi = {10.1088/1367-2630/6/1/162},
       adsurl = {https://ui.adsabs.harvard.edu/abs/2004NJPh....6..162B}
}

@ARTICLE{2019arXiv191204232S,
       author = {{Schoenholz}, Samuel S. and {Cubuk}, Ekin D.},
        title = "{JAX, M.D.: A Framework for Differentiable Physics}",
      journal = {arXiv e-prints},
         year = 2019,
        month = dec,
          eid = {arXiv:1912.04232},
        pages = {arXiv:1912.04232},
          doi = {10.48550/arXiv.1912.04232},
archivePrefix = {arXiv},
       eprint = {1912.04232},
 primaryClass = {physics.comp-ph},
       adsurl = {https://ui.adsabs.harvard.edu/abs/2019arXiv191204232S}
}

@ARTICLE{2022PhRvD.105l3521A,
       author = {{Arena}, Evan J. and {Goldberg}, David M. and {Bacon}, David J.},
        title = "{Cosmic flexion}",
      journal = {Physics Review D},
         year = 2022,
        month = jun,
       volume = {105},
       number = {12},
          eid = {123521},
        pages = {123521},
          doi = {10.1103/PhysRevD.105.123521},
archivePrefix = {arXiv},
       eprint = {2203.12036},
 primaryClass = {astro-ph.CO},
       adsurl = {https://ui.adsabs.harvard.edu/abs/2022PhRvD.105l3521A}
}

@ARTICLE{2006MNRAS.365..414B,
       author = {{Bacon}, D.~J. and {Goldberg}, D.~M. and {Rowe}, B.~T.~P. and {Taylor}, A.~N.},
        title = "{Weak gravitational flexion}",
      journal = {Monthly Notices of the Royal Astronomical Society},
         year = 2006,
        month = jan,
       volume = {365},
       number = {2},
        pages = {414-428},
          doi = {10.1111/j.1365-2966.2005.09624.x},
archivePrefix = {arXiv},
       eprint = {astro-ph/0504478},
 primaryClass = {astro-ph},
       adsurl = {https://ui.adsabs.harvard.edu/abs/2006MNRAS.365..414B}
}

@article{Kullback:1951zyt,
    author = "Kullback, S. and Leibler, R. A.",
    title = "{On Information and Sufficiency}",
    doi = "10.1214/aoms/1177729694",
    journal = "The Annals of Mathematical Statistics",
    volume = "22",
    number = "1",
    pages = "79--86",
    year = "1951"
}

@ARTICLE{2018MNRAS.480.5529G,
       author = {{Giblin}, Benjamin and {Heymans}, Catherine and {Harnois-D{\'e}raps}, Joachim and {Simpson}, Fergus and {Dietrich}, J{\"o}rg P. and {Van Waerbeke}, Ludovic and {Amon}, Alexandra and {Asgari}, Marika and {Erben}, Thomas and {Hildebrandt}, Hendrik and {Joachimi}, Benjamin and {Kuijken}, Konrad and {Martinet}, Nicolas and {Schneider}, Peter and {Tr{\"o}ster}, Tilman},
        title = "{KiDS-450: enhancing cosmic shear with clipping transformations}",
      journal = {Monthly Notices of the Royal Astronomical Society},
         year = 2018,
        month = nov,
       volume = {480},
       number = {4},
        pages = {5529-5549},
          doi = {10.1093/mnras/sty2271},
archivePrefix = {arXiv},
       eprint = {1805.12084},
 primaryClass = {astro-ph.CO},
       adsurl = {https://ui.adsabs.harvard.edu/abs/2018MNRAS.480.5529G}
}

@article{Tegmark1996KarhunenLoveEP,
  title={Karhunen-Lo{\`e}ve Eigenvalue Problems in Cosmology: How Should We Tackle Large Data Sets?},
  author={Max Tegmark and Andy N. Taylor and Alan F. Heavens},
  journal={The Astrophysical Journal},
  year={1996},
  volume={480},
  pages={22 - 35},
  url={https://api.semanticscholar.org/CorpusID:119066184}
}






\end{document}